\documentclass[a4paper,11pt,twoside,openright]{book}
\usepackage[indexonlyfirst]{glossaries}
\usepackage[noadjust]{cite}
\usepackage[american]{babel}
\usepackage[usenames,dvipsnames,svgnames,table,xcdraw]{xcolor}
\usepackage[utf8]{inputenc}
\usepackage[T1,T2A]{fontenc}
\usepackage{accents}
\usepackage{amsfonts}
\usepackage{amsmath, amsfonts, amsthm, amssymb}
\numberwithin{equation}{chapter}
\usepackage[toc,page,title,titletoc,header]{appendix}
\usepackage{bm}
\usepackage{cancel}
\usepackage{caption}
\usepackage{catchfile}
\usepackage{cite}
\usepackage{color}
\usepackage{commath}
\usepackage{comment}
\usepackage{lipsum}
\usepackage{csquotes}
\usepackage{datetime}
\usepackage{dsfont}
\usepackage{braket}
\usepackage{enumitem}
\usepackage{epsfig}
\usepackage{fancyhdr}
\usepackage{fix-cm}
\usepackage{float}
\usepackage{graphicx}
\usepackage{ifpdf}
\usepackage{latexsym}
\usepackage{letltxmacro}
\usepackage{mathrsfs}
\usepackage{mathtools}
\usepackage{multirow}
\usepackage{phaistos}
\usepackage{rotating}
\usepackage{slashed}
\usepackage{stackrel}
\usepackage{subcaption}
\usepackage{tcolorbox}
\tcbuselibrary{skins, breakable}
\usepackage{tikz}
\usepackage{tikz-cd}
\usetikzlibrary{decorations.pathmorphing}
\usetikzlibrary{decorations.markings}
\usepackage{titlesec}
\usepackage{units}
\usepackage{url}
\usepackage{wrapfig}
\usepackage{xspace}
\usepackage{tikz}
\usepackage{mathtools}
\usepackage{fontawesome}
\usetikzlibrary{backgrounds}
\makeatletter

\tikzset{%
  fancy quotes/.style={
    text width=\fq@width pt,
    align=justify,
    inner sep=1em,
    anchor=north west,
    minimum width=\linewidth,
  },
  fancy quotes width/.initial={.8\linewidth},
  fancy quotes marks/.style={
    scale=8,
    text=white,
    inner sep=0pt,
  },
  fancy quotes opening/.style={
    fancy quotes marks,
  },
  fancy quotes closing/.style={
    fancy quotes marks,
  },
  fancy quotes background/.style={
    show background rectangle,
    inner frame xsep=0pt,
    background rectangle/.style={
      fill=gray!15,
      rounded corners=10mm,
    },
  }
}

\newenvironment{fancyquotes}[1][]{%
\noindent
\tikzpicture[fancy quotes background]
\node[fancy quotes opening,anchor=north west] (fq@ul) at (0,0) {``};
\tikz@scan@one@point\pgfutil@firstofone(fq@ul.east)
\pgfmathsetmacro{\fq@width}{\linewidth - 2*\pgf@x}
\node[fancy quotes,#1] (fq@txt) at (fq@ul.north west) \bgroup}
{\egroup;
\node[overlay,fancy quotes closing,anchor=east] at (fq@txt.south east) {''} ;
\endtikzpicture}
\makeatother

\newcommand{\bea}{\begin{align}}
\newcommand{\beq}{\begin{equation}}

\newcommand{\bmat}{\begin{pmatrix}}

\newcommand{\ea}{\end{align}}
\newcommand{\eeq}{\end{equation}}
\newcommand{\emat}{\end{pmatrix}}

\newcommand{\comm}[1]{}
\newcommand{\sgn}{\text{sgn}}
\newcommand{\getenv}[2][]{%
  \CatchFileEdef{\temp}{"|kpsewhich --var-value #2"}{}%
  \if\relax\detokenize{#1}\relax\temp\else\let#1\temp\fi}

\def\dd{\text{d}}

\newcommand{\tr}{{\text{Tr}}}
\newcommand{\Area}{{\text{Area}}}

\newcommand{\LL}{\ell_{\Theta}}
\renewcommand{\to}{\rightarrow}

\newcommand{\southcorner}{\mathbin{\rotatebox[origin=c]{315}{$\lrcorner$}}}
\def\CB{\mathcal{B}}
\def\CC{\mathcal{C}}
\def\CD{\mathcal{D}}

\def\CH{\mathcal{H}}

\def\CM{\mathcal{M}}
\def\CN{\mathcal{N}}
\def\CO{\mathcal{O}}
\def\CP{\mathcal{P}}

\def\CS{\mathcal{S}}
\def\CT{\mathcal{T}}

\def\CV{\mathcal{V}}
\def\CW{\mathcal{W}}

\getenv[\USER]{USER}

\graphicspath{{Figures/}}
\graphicspath{{./Chapters/Chapter1/Figures/}     
              {./Chapters/Chapter2/Figures/}
              {./Chapters/Chapter3/Figures/}
              {./Chapters/Chapter4/Figures/}
              {./Chapters/Chapter5/Figures/}
              {./Chapters/Chapter6/Figures/}
              {./Chapters/Chapter7/Figures/}
              {./Chapters/Chapter8/Figures/}
              {./ThesisFigs/}}
\RequirePackage[colorlinks=true,urlcolor=black,anchorcolor=black,citecolor=green,filecolor=black,
               linkcolor=blue,menucolor=black,linktocpage=true,pdfproducer=medialab,pagebackref=true]{hyperref}
\hypersetup{
  colorlinks   = true, 
  urlcolor     = blue, 
  linkcolor    = blue, 
  citecolor   = blue 
}
\usepackage[capitalise]{cleveref}

\newcommand{\theAuthor}{Pablo Qu\`ilez}
\newcommand{\authorEmail}{pablo.quilez@uam.es}
\newcommand{\myTitle}{Blaaaa}
    \pdfcatalog { /PageMode (/UseOutlines)
                  /OpenAction (fitbh)  }

\usepackage{chngcntr}
\usepackage{etoolbox}

\AtBeginEnvironment{subappendices}{%
\chapter*{Appendices}
\addcontentsline{toc}{chapter}{Appendices}
\counterwithin{figure}{section}
\counterwithin{table}{section}
}

\AtEndEnvironment{subappendices}{%
\counterwithout{figure}{section}
\counterwithout{table}{section}
}

\makeatletter
\let\ps@plain\ps@empty
\makeatother

\makeatletter
\g@addto@macro\bfseries\boldmath
\makeatother

\makeatletter
\def \cleardoublepage {\clearpage \if@twoside
\ifodd \c@page
\else
\null\thispagestyle{empty}\clearpage
\fi
\fi}
\makeatother

\newcommand{\chsize}[1]
{
  \ifnum#1<10
    30
  \else
    48
  \fi
}

\titleformat{\chapter}[block]
 {\Huge\bfseries}
 {\raisebox{-\height}{\sffamily\normalsize\MakeUppercase{\chaptertitlename}}%
  \space\raisebox{-\height}{\bigchapternumber}\space }
 {0pt}
 { \printtitle}

\newlength\pretitlewidth
\newcommand\bigchapternumber{
\resizebox{\chsize{\thechapter}pt}
{!}{\mdseries\thechapter}}
\newcommand{\printtitle}[1]{%
  \settowidth{\pretitlewidth}{%
    {\sffamily\scriptsize\MakeUppercase{\chaptertitlename}}\space
    {\bigchapternumber}\space
  }%
  \parbox[t]{\dimexpr\textwidth-\pretitlewidth}{%
    \linespread{.9}\selectfont
    \hrule depth 1pt
    \vspace{1.5ex}
    \raggedright\bfseries #1
  }%
}
\begin{document}
\setlength{\abovedisplayskip}{6pt}
\setlength{\belowdisplayskip}{6pt}

\title{Thesis}

\begin{titlepage}
	\centering
	\includegraphics[width=4cm]{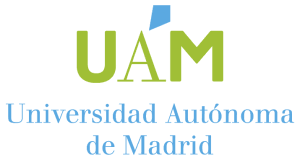} \hspace{1cm}\includegraphics[width=4cm]{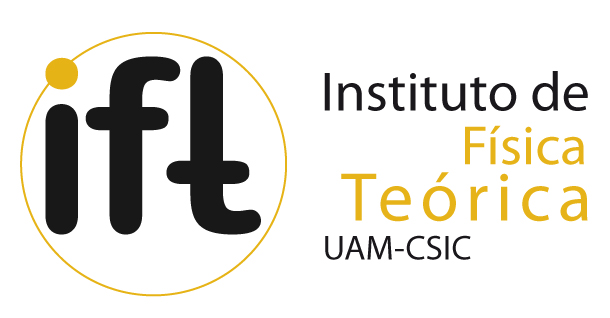}\par\vspace{1cm}
	{\scshape\LARGE Universidad Autónoma de Madrid \par}
	\vspace{1cm}
	{\scshape\Large Facultad de Ciencias \\ Departamento de Física Teórica\par}
	\vspace{1.5cm}
	{\huge\bfseries Quantum Complexity \\ and Holography      \par}
	\vspace{2cm}
	Memoria de tesis doctoral realizada por \\
	{\Large\itshape Javier Martín García\par}
	\vfill
	Dirigida por\par
	Dr.~José Luis Fernández Barbón

	\vfill

	{\large  4 de septiembre de 2020 }
\end{titlepage}




\pagenumbering{arabic}
\newpage
\newpage\null\thispagestyle{empty}\newpage
\setcounter{page}{1}
\section*{Abstract}

This thesis develops recent work \cite{BarbonMartinHyperbolic, noncomp, terminals, locking} on the so called Volume-Complexity and Action-Complexity conjectures. According to this family of proposals, geometric quantities can be defined in some holographic gravitational theories that can be mapped with the concept of quantum complexity for states in a dual quantum-mechanical theory.

In this work, we review the original motivations for the use of quantum-information theory in the search of a theory of quantum gravity, and argue in favour of holographic complexity as a promising new tool that could play a key role in the elucidation of the properties of black holes.

After this introduction, we devote some time to the study of `exotic' thermodynamical systems of diverse origin, confronting the conjectures with expectations and seeking for new behaviours of holographic complexity that could help us understand or refine the existing proposals.

Next, we turn our attention to the study of holographic complexity for singular spacetimes, defining slightly modified versions of the conjecture that are well adapted to singularities and searching for universal behaviours of complexity dynamics within these setups.

Finally, we finish with some speculations about the relation between holographic complexity and older characterization criteria for singularities in general relativity.

\newpage\null\thispagestyle{empty}\newpage
 
 \section*{Agradecimientos}
En primer lugar, me gustaría agradecer a Pepe su apoyo, su guía y todo lo que me ha enseñado a lo largo de estos años sobre física, sobre su historia y sobre todo por enseñarme a cómo hacerla, prestando atención a todo el conjunto y anteponiendo aprender a producir sin sentido.

A los compañeros del IFT por infinidad de discusiones en la cocina y los cafés de las que he aprendido tanto y por los momentos compartidos fuera del trabajo. No os puedo mencionar a todos, pero vosotros ya sabéis que estáis aquí. Reservo un par de menciones especiales: a Danky por todas esas tardes de discusiones de física o de todo lo demás y de risas a la búsqueda de los tramposos de MMII, pero sobre todo por su amistad; y a Martín por su amistad y por haber sido un compañero tan cojonudo.

A Isabel, Rebeca, Susana y las dos Mónicas por ser siempre tan agradables conmigo y hacer la vida tan fácil a alguien tan desastre con los papeleos como yo. 

Si he tenido vida fuera de ese edificio en la otra punta de Madrid ha sido gracias a JB. Venir de fuera a la gran ciudad sin conocer a nadie es un poco abrumador pero él me hizo sentir como si fuese de allí de toda la vida. 

A mi familia: Marieta, Tini, Luis y Ana, por haberme apoyado siempre en mi educación y con este doctorado. Incluso cuando yo mismo tuve dudas de si esto era para mí, vosotros lo tuvisteis claro. Nada de esto hubiese sido posible sin vosotros.

Una de las cosas que más te puede cambiar la vida es un buen profesor, y de no ser por la pasión que me transmitieron en sus clases Marc, Alfredo o Miguel Ángel es muy probable que no me hubiese dado cuenta de que la física teórica es la mejor rama de la ciencia que hay. A Miguel Ángel en particular tengo que agradecer además su guía y su apoyo incondicional durante estos años.

Por último, me gustaría agradecer a sistema público de educación el haberme dado, desde que tenía tres años y hasta ahora, la posibilidad  de llegar hasta aquí, así como a ArXiv.org por apostar por un modelo de conocimiento científico gratuito y abierto para todos.

 \newpage\null\thispagestyle{empty}
\vspace*{\fill}
 \begin{flushright}
  \emph{A Liam y a Marta}
 \end{flushright}
 \vspace*{\fill}
 
  \newpage\null\thispagestyle{empty}
\vspace*{\fill}
 \begin{flushright}
  \emph{ }
 \end{flushright}
 \vspace*{\fill}

  \newpage\null\thispagestyle{empty}
\vspace*{\fill}
 \begin{flushright}
  \emph{``Die Wahrheit ist konkret''}\\
  B.\CYRI.\CYRU.
 \end{flushright}
 \vspace*{\fill}

\newpage \null\thispagestyle{empty}

\cleardoublepage\phantomsection\pdfbookmark{Contents}{contents}
\tableofcontents
\newpage\null\thispagestyle{empty}

\part{ \sc{Foundations}}

\chapter{Gravity and Quantum Information}
\label{ch:capitulo1}
\interfootnotelinepenalty=100
\section{Quantum gravity and the holographic principle}

For over a century, quantum mechanics has shown to be arguably the most successful theoretical framework for a precise description of the fundamental interactions in nature, a triumph that can fairly be judged both from an experimental and a theoretical point of view. On the one hand, the Standard Model of particle physics provides us with tools to test a huge variety of phenomena, spanning scales separated by dozens of orders of magnitude and giving some of the most precise predictions in the history of science, with discrepancies with experiment that can be as small as one part in 100 millions for some quantities \cite{PDG}. On the other hand, the powerful use of mathematical symmetries and the constraining power of quantum field theory as a framework has made possible a unified picture of particle interactions arising from a small set of fundamental principles and parameters.

On a somewhat parallel path, the classical theory of general relativity has standed for over a century as the most solid description of the gravitational interaction, a compelling success that has recently been exhibited yet again with the direct detection of two of its most striking predictions: the existence of gravitational waves and black holes \cite{LIGO,EHT}.

If we were to consider gravity as a classical field and to the extent that we may regard naturalness problems as -- perhaps-- not so fundamental, the Standard Model could very well be the ``ultimate'' microscopic theory of particle interactions. Such first assumption, however should make us fell uneasy. From a merely aesthetic point of view, settling for the existence of fundamentally classical interactions would entail a dual description of nature, putting an end to the successfull reductionist paradigm of unification \cite{Weinbergdreams}. Most importantly, there are good reasons to argue that such possibility cannot be consistent and that we must include gravity into the framework of quantum mechanics. Indeed, the notion of a quantized theory of gravity is almost as old as general relativity itself and even prior to the establishment of the very principles of quantum mechanics. Already in 1916, Einstein pointed out (cf. \cite{Einsteingw}) that the existence of tiny gravitational waves produced by the electrons in atoms lead to the same problems that doomed the classical electromagnetic picture in the Rutherford atom model, implying therefore that a quantum version of general relativity would be just as needed as one for the Maxwell theory.

\begin{fancyquotes}
\emph{Nevertheless, due to the inneratomic movement of electrons, atoms would have to radiate not only electromagnetic but also gravitational energy, if only in tiny amounts. As this is hardly true in nature, it appears that the quantum theory would have to modify not only Maxwellian electrodynamics, but also the new theory of gravitation.}
\flushright{--Albert Einstein (1916)}
\end{fancyquotes}

A simple look at the Einstein field equations
\begin{equation}
R_{\mu \nu} - \dfrac{1}{2}R g_{\mu \nu} + \Lambda g_{\mu \nu}= 8\pi G T_{\mu \nu}
\end{equation}
already evidences the necessity to reconcile the classically geometric picture at the left hand side with the fundamentally quantum nature of matter in the right hand side. In fact, as argued in \cite{EppleyHannah} it is easy to come up with \textit{gedankenexperiments} in which the existence of a classical gravitational field causes violations of the Heissenberg principle (in a nutshell, one could determine position and momentum of a particle simultaneously by simply scattering a gravity wave with arbitrarily high frequency and low momentum). Furthermore, naive semiclassical versions of the Einstein equations have shown to be inconsistent or are experimentally excluded \cite{Pageqg}.

The program to find a quantum theory of the gravitational field, however, has not been exactly a piece of cake and it is still very far from being completed. In particular, the problem is suspected to require not only sophisticated technical tools but most importantly fundamentally new concepts that may challenge the present principles of physics; namely Lorentz invariance, unitarity and locality. Nonetheless, several approaches have coexisted during the last century (see \cite{RovelliQGreview, CarlipQGreport} for reviews) and it is fair to say that we have learned great deal about them, at least at the level of identifying the key obstacles to its resolution.

A partial success, initiated by Fierz and Pauli in the 30's and developed in the 60's by DeWitt \cite{FierzPauli, DeWittCovariant, DeWittApplications}, came with the perturbative description of gravity in terms of an interactive quantum field teory of gravitons. Although always hunted by the phantom of non-renormalizability \cite{tHooftRenormalization, GoroffSagnotti}, such description was shown to be perfectly consistent within the modern approach of effective theories \cite{WeinbergEFT, BurgessGR}, allowing for the calculation of some well-defined quantities such as quantum corrections to the Newtonian potential \cite{DonoghueQuantumCorrections}. Furthermore, it was soon shown (cf. \cite{Kraichnan, Gupta, WeinbergGravitons, Deser, BoulwareDeser, FeynmanQuantumGravity, FeynmanLecturesGravitation}) that general relativity is not only a consistent low energy effective quantum field theory but in fact the \emph{unique} one for a massless spin-2 field, a result that further suggests that the ultimate fate of gravity is to come up as an inevitable consequence of quantum mechanics. 

At energies of the order of the Planck mass, however, gravitons become strongly coupled and such effective description is of no use as all scattering amplitudes become non-unitary. There are, in short, two main attitudes toward this issue. The first one assumes that the metric degrees of freedom are all there is to be above the Planck mass and that unitarity violation is just an artifact of perturbation theory. The hope in this side is that UV divergencies might be fixed by suitable non-perturbative methods, somehow rendering gravity a finite theory. Although some partial resummations of Feynman diagrams succeded in the obtention of finite quantities, we do not know a general procedure to resum the perturbative series and this path has lead to little success. The second one -- inspired by other historical triumphs such as electroweak completion of Fermi's theory -- admits that the metric is only a good description at low energies that needs to be completed with new weakly coupled degrees of freedom at the Planck scale. Within field theory this seems to be hopeless \footnote{The main attemps to do this are in the framework of supergravity, where most theories have been proved to be non-renormalizable. There exists however the hope that the maximally supersymmetric $\mathcal{N}=8$ supergravity is finite \cite{Bern}.} and a major achievement was accomplished by the broader framework of string theory. Indeed, string theory was able to provide a suitable regularization for the graviton theory, yielding a unitary amplitude for the scattering of both open and closed strings at arbitrarily high energies \cite{Veneziano, Virasoro}.

Nonetheless, the existence of non-perturbative objects in both QFT and classical gravity suggests that string theory cannot be yet the end of the story. As a matter of fact, a generic prediction of general relativity eventually brings the UV problem back to the IR since production of black holes becomes dominant in the very high energy regime of any scattering process, rendering string theory a useful description only for a transient around $E \sim M_P$. Furthermore, the rich mesh of weak/strong-coupling  dualities that was found in the 90's for the different string theories seems to indicate the existence of an underlying fully non-perturbative theory of quantum gravity, out of which string theory would emerge as an effective weak-coupling description. The quest to find such completion however takes us somehow to the beginning, as there are very few clues pointing towards the answer or even the principles that are to be imposed in this task.

Be that as it may, all is not lost, as there is a piece of IR gravitational physics that has been so far an extraordinarily insightful window into the UV properties of quantum gravity: the physics of black holes. 

In 1972, it was realized by Jakob Bekenstein \cite{Bekensteinentropy, Bekensteinsecondlaw} that gravity posed some problems to the postulates of thermodynamics. In particular, for some spacetimes arbitrarily large volumes can fit in finite regions of space, thus seemingly allowing for the storage of an infinite entropy on a finite energy system. Such objects would violate the second law of thermodynamics (or rather the implicit assumption in it that the maximum entropy is to be bounded from above at fixed energy) as one could design a process to decrease the entropy of the universe by using this system as thermodynamic dump. Black holes are examples of such systems and hence their existence in nature implies one of the following options: either the laws of thermodynamics do not apply to gravity or black holes possess an entropy that cannot be accounted as usual as an extensive quantity. Opting for the the second option, Bekenstein argued that a better and finite notion of entropy could be assigned to black holes, i.e. one that would be proportional to the area of the event horizon measured in Planck units
\begin{equation}
\label{bhentropy}
S_{b.h.}=\eta \; \dfrac{A}{l_P^2}\,,
\end{equation}
where $\eta$ is an order one dimensionless parameter. Supported by recent proofs \cite {HawkingAreatheorem,HawkingAreatheorem2}  that no physical (classical) process could decrease such quantity, a generalization of the second law of thermodynamics 
\begin{equation}
\Delta S_{\text{out}} + \Delta S_{b.h.} \geq 0\, ,
\end{equation}
was in turn proposed to hold (see \cite{WallTenproofs} for a review of the current status of the proof of such \emph{Generalized Second Law} in the context of semiclassical gravity), solving the conundrum with the Maxwell-demonic character of black holes. Away from heuristic arguments, a rigorous derivation of \eqref{BHentropy} seems to require a complete knowledge of the quantum gravity theory, able to fix the value of the parameter $\eta$. 

\begin{fancyquotes}
\emph{[...]it would be somewhat pretentious to attempt to calculate the precise value of the constant $\eta$ without a full understanding of the quantum reality which underlies a ``classical'' black hole.}
\flushright{--Jacob Bekenstein (1973)}
\end{fancyquotes}
Somewhat surprisingly, no detailed quantum gravity description is actually needed to fix the proportionality coefficient $\eta$, which can be found after a clever interpretation of two important results that can be obtained within the semiclassical theory. The first one is a simple mechanical law stating the relation between infinitesimal changes in the mass of a black hole and the subsequent change in the area of its event horizon 
\begin{equation}
\label{firstlaw}
\delta M = \dfrac{\kappa}{8 \pi G}\, \delta A\,,
\end{equation}
where $\kappa$ stands for for the surface gravity along the horizon \cite{Smarr, FourLaws}. The second one, famously achieved by Stephen Hawking in 1975 \cite{HawkingRadiation, HawkingExplosions} is the realization that quantum fields propagating along the horizon of a black hole are thermally populated, making possible to assign a temperature to the system
\begin{equation}
T_H= \dfrac{\kappa}{2\pi}\,,
\end{equation}
and strongly suggesting a thermodynamic interpretation of \eqref{firstlaw}, which resembles the first law of thermodynamics for the gravitational ADM energy $M$. Identificating the Hawking temperature in \eqref{firstlaw} fixes the coefficient $\eta = 1/4$, and defines the Bekenstein-Hawking entropy of a black hole
\begin{equation}
\label{BHentropy}
S_{BH}=\dfrac{A}{4G}\,.
\end{equation}
While offering salvation to the principles of thermodynamics in the presence of gravity, the Bekenstein-Hawking entropy raises a number of puzzles of its own. The most obvious one concerns the statistical interpretation of $S_{BH}$, which should be recoverable from a precise counting of black hole microstates in some fully-fledged theory of quantum gravity. A quarter century would have to pass for a counting of that sort to be achieved, culminating with a successful microscopical description of maximally supersymmetric extremal black holes within the framework of string theory \cite{StromingerVafa, CallanMaldacena}. The generic (non-supersymmetric) case however still lacks a proof of that sort, a task that again seems to require non-perturbative tools out of the realm of string theory. 

Away from precise constructions, however, the fact that the black hole entropy scales as the area of the horizon suggests a picture of gravitational degrees of freedom located at this surface, in turn \textit{holographically} describing the physics in the interior of the black hole. Rather than considering black holes as exotic states posessing this property, the `clumping' nature of gravity forces us to consider them as actually very generic since any sufficiently energetic distribution of matter in a finite region of spacetime is doomed to undergo gravitational collapse \cite{GibbonsPerry}. 

These considerations suggest that the Bekenstein-Hawking entropy sets a bound (cf. \cite{BekensteinBound, BoussoBound}) on the very spectrum of any quantum theory of gravity, truncating thus the usual extensivity that is characteristic of ordinary QFTs. In this picture, we might consider a local QFT as a vastly redundant description of the gravitational physics, only appropriate to the description of some low-energy subspace of sufficiently `diluted' systems. For the high energy part of the spectrum on the other hand, dominated by black holes, a description based on `boundary' degrees of freedom seems more natural in the light of the Bekenstein-Hawking formula. As most of the states of the ensemble are actually black holes, this picture suggests that such boundary degrees of freedom are actually the fundamental ones, with the local \emph{bulk} gravitational theory emerging as a merely effective description for the diluted subspace.

The last two paragraphs conform the heuristic foundation of the so called \emph{holographic principle} \cite{tHooftHolography, SusskindHolography} which has conformed for the last two decades the main strategy in the quest to find a theory of quantum gravity. In order to put some meat on these bones we still need to provide a couple of elements for the conjecture (or at least an example which realizes it) to hold. 

The first one is the presence of some sort of boundary, or holographic screen in which to `place' our fundamental degrees of freedom. Certainly, our universe lacks such a screen but even coming up with a toy-model one in which to cage a piece of if is not an obvious task since no ordinary matter can be made impermeable to gravitons.

The second one of course is a suitable boundary Hamiltonian able to do the job. Even forgetting about the details it is obvious that such Hamiltonian cannot be a textbook example for an ordinary `caged universe' due its rather exotic density of states. In particular,  from \eqref{BHentropy} we see that such density asymptotes to
\begin{equation}
\Omega(E) \sim {\rm exp} \,\left(G^{\frac{1}{d-2}} \,E^{\frac{d-1}{d-2}}\right)\, , 
\end{equation}
which comprises a rather strong growth with energy that is not achieved by any known local QFT or even string theory. 

Fortunately, an example was soon found by Juan Maldacena that adresses these two points (cf. \cite{MaldacenaAdSCFT, WittenAdSCFT}) in a simple but yet successful manner. The key point is to use a box that is built up by gravity itself: Anti-de Sitter spacetime. As it is well know, this solution with negative cosmological constant $\Lambda <0$ effectively provides a confining harmonic oscillator potential able to reflect back both matter and gravitons in a finite amount of time, at the same time that allows for arbitrarily large stable black holes to live inside. Incidentally, the presence of the cosmological constant changes qualitatively the shape of the black hole density of states, which now behaves as
\begin{equation}
\Omega(E)_{AdS_{d+1}} \sim {\rm exp} \,\left( \dfrac{\ell^{\frac{d-1}{d}}}{G^{1/d}}\left(\ell\, E\right)^{\frac{d-1}{d}} \right)\, , 
\end{equation}
where $\ell \sim \Lambda^{-1}$ is the characteristic curvature radius of AdS. This is now a recognizable density of states of an ordinary local QFT in $d$ dimensions. Particularly, it matches with the expectation for a Conformal Field Theory with $N_*$ species in a box of size $\ell$  as long as
\begin{equation}
\label{Ndef}
N_* = \dfrac{\ell^{d-1}}{G}\, ,
\end{equation}
and suggests a dual picture of the gravitational theory being described (or rather fully non-perturbatively \emph{defined}) by a suitable CFT with no gravity and located in the boundary of AdS. This picture, supported by the fact that the isometry group of $AdS_{d+1}$ spacetime is no other that the $d-$dimensional conformal group $SO(d,2)$, conforms the statement of the celebrated AdS/CFT correspondence.

In short, the idea is that each state of the CFT system encodes all the information about a corresponding state in the dual gravitational system. For example, the vacuum state typically corresponds to the empty Anti-de Sitter spacetime, whereas small perturbations around the vacuum induce respectively small perturbations in the spacetime, for example in the form of gravitational waves. Typical high energy states in the boundary theory, on the other hand, usually correspond to very massive black holes.

Since its proposal in 1997, a huge plethora of non-trivial tests have piled up evidence supporting the validity of the conjecture, specially in its more precise and studied version involving $\mathcal{N}=4$ SYM on the CFT side and type IIB string theory on $AdS_5\times {\bf S}^5$ on the gravity one. Nevertheless, we still lack a proof for the correspondence, and even assuming its true, there remain a number of fundamental (partially) unanswered questions such as

\begin{itemize}
\item What is the detailed map between the theories? Given some quantity in the gravitational theory, what is the quantity in the CFT and how do we compute it?
\item What are the conditions for a CFT to have a smooth gravity dual?
\item How do the spacetime geometry, gravity and approximate bulk locality emerge from the CFT physics?
\end{itemize}

The answer to the first point is what is known as ``the dictionary'': a (yet incomplete) list of CFT expression for the relevant bulk quantities and vice-versa. A basic piece of the dictionary is the statement that the Hilbert space is by definition the same one at both sides, just as it is the Hamiltonian and the symmetry generators of $SO(d,2)$. Accordingly, quantities that only depend on these features, like the thermal partition function or the free energy also have a somehow trivial translation into the dual theory language. A different deal comes about when we want to talk about local bulk operators. Certainly, the very notion of locality better be dead in a bona-fide theory of quantum gravity, but we should be able to recover it in the approximate regime in which the gravitational EFT holds. Fortunately, we do have an answer for that and we may obtain relations between, say, a scalar primary operator $\mathcal{O}$ in the CFT and a corresponding bulk scalar field $\phi$ close to the boundary
\begin{equation}
\lim\limits_{r\rightarrow \infty} r^{\Delta_{\mathcal{O}}} \phi (r, t, \Omega ) = \mathcal{O}(t,\Omega)\, ,
\end{equation}
where $\Delta_{\mathcal{O}}$ is nothing but the scaling dimension of the CFT operator and the bulk field has a mass given by
\begin{equation}
\Delta_{\mathcal{O}} = \dfrac{d}{2} + \sqrt{\frac{d^2}{4}+m^2\ell^2}\, .
\end{equation}
For fields further inside the spacetime geometry, a general `bulk reconstruction' procedure (see \cite{Harlowbulk} for a review) has been developed which makes us able to compute the bulk local fields as a linear combination of CFT primaries, i.e. symbolically
\begin{equation}
\phi(x) = \int\limits_{\mathcal{B}} dy\,  K(y,x)\mathcal{O}(y)
\end{equation}
where $K(y,x)$ is a suitable kernel function and $\mathcal{B}$ is a region supported in the boundary. 

Regarding the second point, some necessary conditions have been argued \cite{PenedonesAdSCFT,PapadodimasAdSCFT} for a CFT to have the right spectrum of states but the discussion is far to be settled. We can however deduce easily the most important one from \eqref{Ndef} where we can see that the existence of a classical `Einsteinian' gravity regime of spacetime requires a large number ($N_* \gg 1$) of degrees of freedom in the boundary theory.  A more careful look at the problem shows that for the theory to be able to describe also the low-energy `diluted' subspace, a ``gap'' must be present in the spectrum of conformal weights \footnote{The so-called \textit{state-operator correspondence} in radial quantization allows us to talk interchangeably about states of energy $\Delta$ for the theory on $\mathbf{R}\times\mathbf{S}^{d-1}$ and operators of conformal dimension $\Delta$ on $\mathbf{R}^{d}$. }, where an $\mathcal{O}(N^0)$ number of light fields must survive the bulk weak coupling limit below certain threshold $\Delta_*$. The highly constraining power of conformal invariance makes fulfilling these requirements a non-trivial task and only a handful of supersymmetric theories are known to do the job. Most of the work done in the context of AdS/CFT assumes though that such appropriate large $N$ limit exists and that we can calculate every quantity in a corresponding t'Hooft expansion. Evidence from the better-known examples suggests that a large t'Hooft coupling is required as well.

\begin{figure}[t]
$$\hspace{4cm} \includegraphics[width=10cm]{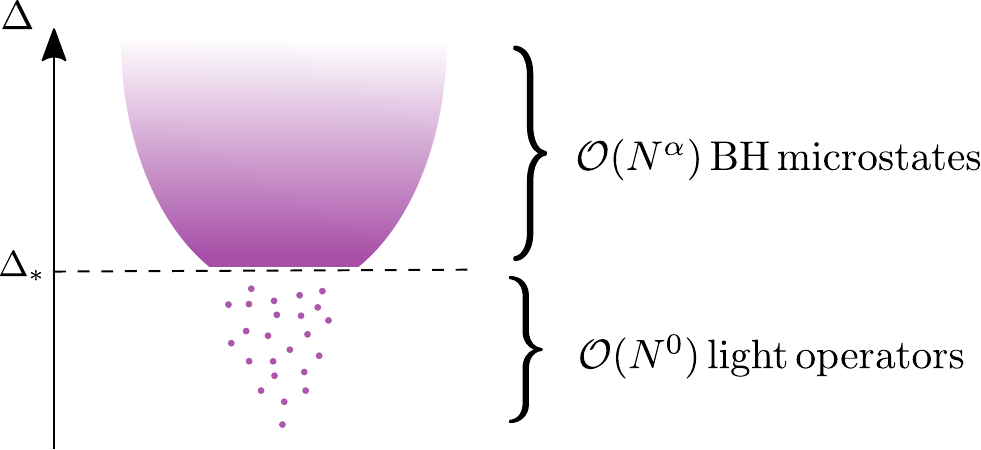} $$   
\begin{center}
\caption{\emph{Spectrum of conformal weights for the operators of a holographic CFT. The requirement that there is a finite number of weakly interacting bulk fields below the Planck mass as $G \rightarrow 0$ gets translated into a finite ($N$-independent) number of light operators below certain $\Delta_* \sim N^\gamma $ as $N \rightarrow \infty$. Above this threshold a huge tower of heavy operators with $\Delta$'s scaling with $N$ represent the black hole microstates.}}
\label{fig:spectrum}
 \end{center}
\end{figure} 

At last, the third point is really about a special chapter of the dictionary, but a rather deep one for which the usual language of QFT seems to be of little help. In the last decade however it has become clear that a better understanding of these questions can be achieved from the perspective of quantum information theory. In particular, evidence strongly suggests that the structure of quantum entanglement of the QFT boundary states plays a significant role in the geometrical structure of the higher-dimensional spacetime. Quantities that are natural in the quantum information theory, such as the different measures of entropy, become simple geometrical properties in the gravitational side. Along this thesis, we will explore such quantities, focusing on a particularly special measure of state entanglement that we call \emph{complexity}.

\section{Entanglement builds space}

After this short introduction to AdS/CFT, it seems like the right time to go back and ask ourselves whether we learned in this context anything new regarding the interpretation of the Bekenstein-Hawking entropy. As we saw, the thermal partition function as well as other thermodynamical quantities exactly match between both theories suggesting that the stable black holes in AdS are nothing but high-energy thermal states of the CFT. The answer is then automatic: the way to count the black hole microstates is actually to count the thermal CFT microstates in some ensemble, which we do know how to compute \footnote{Actually, when working at strong coupling, the free energy can be calculated up to an order-one numerical prefactor that is hard to get except for the simpler case of AdS$_3$/CFT$_2$, where modular invariance in the CFT simplifies the calculations. In higher dimensional cases, AdS/CFT would actually give a prediction for the value of that prefactor in the large $N$ limit so we may consider this as a smoking gun for the conjecture in case any lattice technique is able to calculate that number independently for, say, $\mathcal{N}=4$ SYM.}. As a result, we obtain a new piece of the dictionary relating entropy on the ordinary quantum system with the area of a black hole horizon in the bulk. As it turns out, this relation between geometry and entropy is just the tip of the iceberg and admits a beautiful generalization. In order to get there, we need first to understand a little more about the meaning of entropy in a quantum theory.

\begin{figure}[h]
$$ \includegraphics[width=11cm]{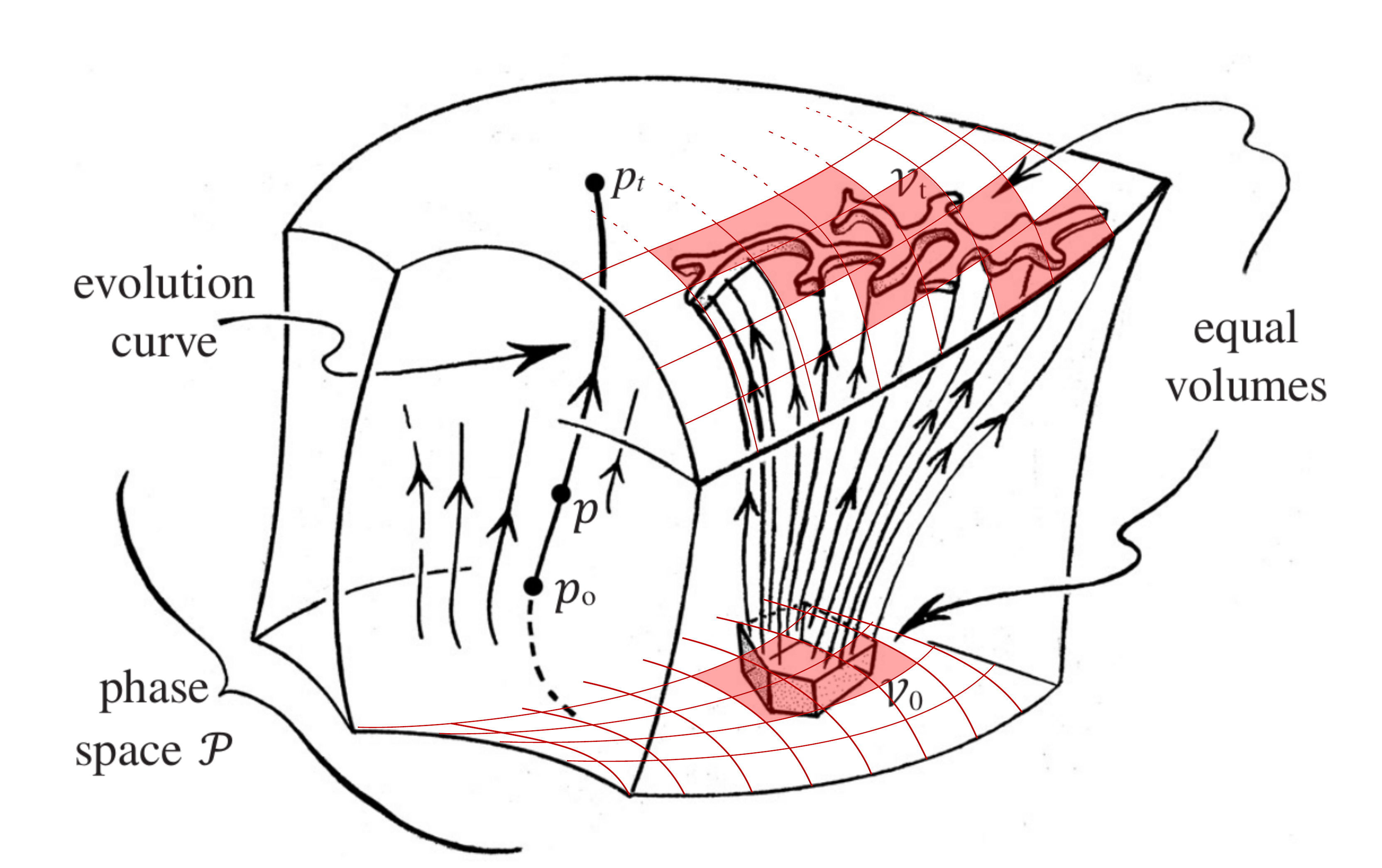} $$   
\begin{center}
\caption{\emph{Classical evolution in phase space of a probability distribution $\rho(t)$. Liouville theorem guarantees that the volume $\mathcal{V}(\rho)$ is conserved. Chaotic evolution however tends to spread $\rho$ over $\CP$ creating `dendritic' structures. As a result, any measure of volume involving an arbitrary coarse-graining tends to increase with time. Heuristically this is the statement of the second law of thermodynamics.}}
\label{fig:secondlaw}
 \end{center}
\end{figure} 

Already in classical statistical mechanics, there are a number of ways in which to define entropy, all of them essentially accounting for some coarse-grained notion of volume in phase space (see Figure \ref{fig:secondlaw}). In this context, a useful definition is that of Gibbs/Shannon, which can be defined for an arbitrary `mixed' ensemble $\lbrace \psi_i, p_i\rbrace  $ of microstates and probabilities, assigning to the mixture the entropy
\begin{equation}
\label{Gibbs}
S = - \sum\limits_{i} p_i \log p_i \, .
\end{equation}
In quantum mechanics, this granularity of phase space is somehow already built in the theory, with `volumes' that are now associated with the dimensionality of a Hilbert space, providing a direct way to count states without the need of any arbitrary coarse-graining scale. A quantum version of the the Gibbs/Shannon entropy with the same functional form \eqref{Gibbs} can be defined for a quantum ensemble of states $\lbrace \ket{\psi_i}, p_i\rbrace  $. The result is the so-called \emph{Von-Neumann entropy}, which we may rewrite
\begin{equation}
S = - \tr \left( \rho \log \rho \right) \, ,
\end{equation}
with $\rho = \sum\limits_i p_i \ket{\psi_i}\bra{\psi_i}$ the so called \emph{density matrix} containing all information about the mixed state \footnote{Despite the fact that Von-Neumann entropy is fine-grained in its definition, we are of course allowed to coarse-grain it as well for practical purposes. In this case, rather than a tiling of phase space such procedure requires a choice of coarse-grained observables $\CO_i$. Next , we must consider the set of all possible density matrices $\lbrace\tilde{\rho}_j\rbrace$ able to give the same result for those observables $\tr[\tilde{\rho}_j \CO_i]=\tr[\rho \CO_i]$, and we define the coarse-grained entropy by maximizing $S(\tilde{\rho_j})$ over this set.  Of course $S_{VN} \leq S_{\rm{coarse}}$ with the latter morally recovering the classical notion of thermal entropy.} . The Von-Neumann entropy has a number of remarkable properties that makes it richer than its classical cousin. Among them, we will focus on those that are linked to its behaviour under bipartitions of the Hilbert space or, in other words, its properties as a quantifier of \emph{entanglement}.

Consider a quantum system in a a pure\footnote{Everything here can also be discussed when the initial state is not pure, $\rho \neq \ket{\Psi} \bra{\Psi}$, but it only gets the conclusions messier by piling up classical uncertainty on top of the quantum one and obscuring the differences between entanglement and `mixing'. } state $\ket{\Psi} \in \mathcal{H}$ and take a subsystem $A$ of it. In that case, the Hilbert space can be decomposed as
\begin{equation}
\CH = \CH_A \otimes \CH_{\bar{A}}
\end{equation}
where $\bar{A}$ stands for the complementary region of $A$. This Hilbert space is spanned by states of the form $\ket{\phi_A} \otimes \ket{\phi_{\bar{A}}}$, where $\ket{\phi_A}$  and $\ket {\phi_{\bar{A}}}$ are respectively complete bases for $\CH_A$ and $\CH_{\bar{A}}$. Now, a legitimate question is: can we somehow isolate the `piece' of $\ket{\Psi}$ that describes only what is happening on subsystem $A$? Since $A$ could very well represent our laboratory and $\bar{A}$ the rest of the universe, this question seems a fairly important one if we want to do physics in $A$. A first naive try would be to look for some state $\ket{\psi_A} \in \CH_A$ able to do the job, i.e. demanding expectation values of operators $\CO_A$ in $A$ to match
\begin{equation}
\bra{\Psi} \CO_A \otimes \mathbb{I} \ket{\Psi} = \bra{\psi_A} \CO_A \ket{\psi_A}  \, .
\end{equation}
Such $\ket{\psi_A}$ however does not exist in general unless $\ket{\Psi}$ is in what is known as a \emph{product state} $\ket{\Psi}= \ket{\psi_A} \otimes \ket{\psi_{\bar{A}}}$, and recovering the left hand side typically requires an ensemble $\lbrace\ket{\psi_A^i}, p_i\rbrace$ such that
\begin{equation}
\bra{\Psi} \CO_A \otimes \mathbb{I} \ket{\Psi} = \sum\limits_i p _i \bra{\psi_A^i} \CO_A \ket{\psi_A^i}  \, .
\end{equation}
In other words, the failure of $\ket{\Psi}$ to being able to be represented as a simple product requires the state on $A$ to be mixed $\rho_A =\sum\limits_i p _i  \ket{\psi_A^i}  \bra{\psi_A^i}$ where the corresponding \emph{reduced density matrix} can be calculated by means of a partial trace over the complementary Hilbert space
\begin{equation}
\rho_A = \tr_{\bar A}\,  \rho \, .
\end{equation}
Recapping, we see that bipartite systems in quantum mechanics have a remarkable property with no classical analogue, i.e. they can be \emph{entangled}, meaning that the density matrices describing their subsystems can be mixed even when the total state is pure. In other words: even when the state of a system is completely known, in general there is no way of `zooming' into a subsystem without losing some information about the state there.

As it turns out, the structure of entanglement in the spectrum of states for quantum field theories encodes all kinds of interesting features about the theory itself, some of which may be hard to diagnose with standard local operators. Operators however do `feel' the presence of entanglement between subsystems since correlators between operators living in different regions factorize for unentangled states
\begin{equation}
\ket{\Psi} = \ket{\psi_A} \otimes \ket{\psi_{\bar{A}}} \hspace{0.5cm} \Leftrightarrow \hspace{0.5cm} \langle\CO_A\CO_{\bar{A}} \rangle-\langle\CO_A \rangle\langle\CO_{\bar{A}} \rangle=0\, ,
\end{equation}
but not in general.

Going back to the Von-Neumann entropy, we see that we can use it now as a diagnostic tool and quantifier for entanglement since
\begin{equation}
0 \leq S(\rho_A)\leq \log \left(\dim \CH_A \right) \, ,
\end{equation}
where the inequalities are saturated whenever $\ket{\Psi}$ is respectively a product state or maximally entangled under that bipartition. For this reason, $S(\rho_A)$ is often referred as \textit{entanglement entropy} in this context.

Before jumping back to black holes, there is a last natural question that will be relevant. As we just saw, given a pure state $\ket{\Psi}$ of a system, any subsystem $A$ can be described by an ensemble $\rho_A$, but, is it the opposite true? i.e. given some $\rho_A$ for a system, can we find the `parent' pure state in a larger system such that $\rho_A$ is its reduced density matrix on $A$? As $\rho_A$ has no information whatsoever about $\bar{A}$ it seems unlikely that we will be able to get \emph{the} original state $\ket{\Psi}$ but we may find \emph{a} state $\ket{\Phi}$ able to do the job for some $\bar{A}$. In general it is possible to come up with an infinite number of such \emph{purifications} but some of them might be more useful than others. A particularly interesting one can be defined for the thermal state in the canonical ensemble
\begin{equation}
\label{canonicalensemble}
\rho_A = \sum\limits_i e^{-\beta E_i } \ket{E_i} \bra{E_i}\, ,
\end{equation}
where $\beta$ is the inverse temperature of the system. Choosing the complementary system to be an identical copy of the original one, we may write the purification as
\begin{equation}
\label{TFD}
\ket{\Phi} = \sum\limits_i e^{-\beta E_i/2} \ket{E_i}_A \otimes \ket{E_i}_{\bar{A}}\, .
\end{equation}
This is what is known as a the \emph{thermofield double state} (TFD), a very particular state that possesses the property of yielding the the thermal state at the same temperature upon reduction to either of the two subsystems. We may think about this purification as a minimal choice among the possible `heat baths' to which to couple our original system in order to achieve thermal equilibrium\footnote{Actually, this picture as it stands might sound a little misleading since the regions $A$ and $\bar{A}$ are completely decoupled in the TFD state. If we want to physically prepare such state however it seems obvious that some Hamiltonian had to couple both systems somewhere in the past in order to thermalize/entangle both sides.}, and as a bonus, we get a new interpretation for the thermodynamical entropy of the thermal ensemble as measuring the entanglement of the system with its bath. 

The reason we are talking about thermal states is because it is time to go back to black hole physics. As we pointed out in the previous section, the equivalence of the thermal partition functions for the bulk and boundary descriptions suggests that the proper identification of thermal states in the CFT on a sphere is that of very massive black holes in the bulk since the former is certainly the preferred equilibrium state in AdS as long as the Hawking-Page phase transition is reached there. In this picture, the energy of the thermal state is just the ADM black hole mass and the entropy is given by the area of its horizon. 

An immediate question comes to mind: can we use the CFT to learn anything new about the black hole? in particular, does the thermal state contain information about the interior? As it turns out, the answer is: problably not, but a suitable purification might do. In \cite{MaldacenaBHinAdS}, Maldacena came up with a very suggestive picture to understand these purifications by looking at the simple and very special case of the TFD state. As he argued, such state should not be understood as describing a single black hole, but rather the complete Kruskal maximal extension comprising two asymptotically AdS regions connected by a wormhole. In this setup thus, there are two independent but identical CFTs on $\mathbf{S}^{d-1} \times \mathbf{R}$ with a very particular and symmetric entanglement pattern given by \eqref{TFD}. The proposal can also be formally motivated by a path integral construction and passes a number of checks that establish the TFD/Eternal-black-hole state as one of the most studied systems in the context of AdS/CFT.

Despite its apparent simplicity and naturalness, Maldacena's proposal has a somewhat radical and surprising consequence. The  terms $\ket{E_i}_A\otimes \ket{E_i}_{\bar{A}}$ in the TFD superposition are simple product states in the completely decoupled system of two CFTs, a fact that is not changed by time evolution since the Hamiltonian of the joint system is simply the sum of each CFT Hamiltonian and there is no interaction between the two. As the states on each side have nothing to do with each other, we would expect the TFD to correspond to a pair of identical but separate asymptotically AdS spacetimes and not a single connected geometry. The striking conclusion, which was later emphasized in \cite{EREPR, VanRaamsdonkBuilding} is that a suitable entanglement pattern between disconnected spacetimes can act as `spacetime glue', connecting both geometries into a single one. This idea has lead to the general slogan that \emph{entanglement builds space} or ``ER$=$EPR'', which tries to stress the importance of the concept of entanglement and its measures (and more generally, other quantum information tools) in the construction of a theory of quantum gravity, the unravelling of the emergence of the holographic picture and the resolution of long standing problems such as the black hole information paradox.

\begin{figure}[t]
$$\includegraphics[width=10cm]{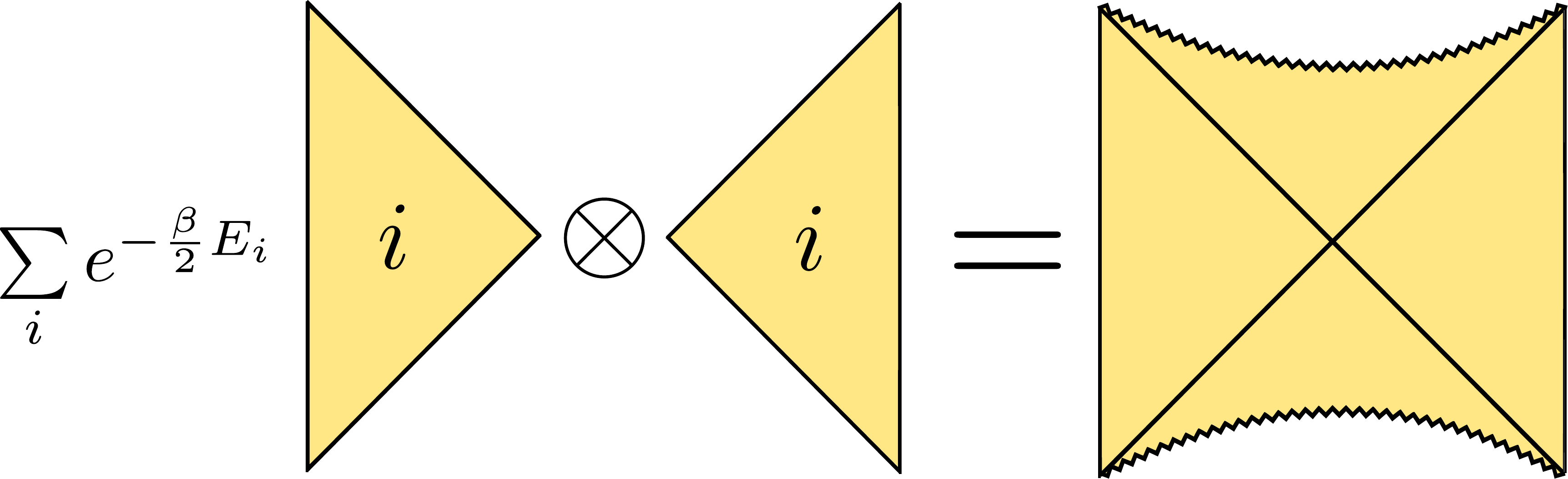} $$   
\begin{center}
\caption{\emph{Thermo Field Double state for two copies of a CFT on a sphere. The particular superposition above of completely disconnected spaces gives rise to a single connected eternal black hole geometry. }
\label{fig:TFD}}
 \end{center}
\end{figure} 

Before ending this chapter it is worth pointing out some findings that habe been able to put more meat into the slogans above, generalizing the connections between entanglement and geometry for a large class of states admitting gravity duals. For the last time, let us take the TFD exposed in the last paragraphs and think about the result we just got: the entanglement entropy of each of the CFT copies $A$ and $\bar{A}$ corresponds to the area of a codimension-two surface (the black hole horizon) which splits the bulk in two pieces. But, why the horizon? A black hole horizon is certainly a rater special surface from many points of view but it is one of their properties which will be the key one in the beatiful generalization proposed by Ryu and Takayanagi in \cite{RT}: it is the minimal area surface able to split the bulk in two.

When stated in this form, such generalization comes naturally and the proposal is the following: given any holographic state $\ket{\psi} \in \CH$ with a smooth gravity dual and admitting a local\footnote{With this we mean that $A$ is to be understood as a \emph{region} on the boundary manifold. Exotic splittings of the Hilbert space such as for example those that would be natural in momentum space will not satisfy the RT/HRT proposal.} bipartition $\CH = \CH_A \otimes \CH_{\bar{A}}$, it is possible to compute the entanglement entropy of $A$, $S(\rho_A)$ by finding the extremal codimension-two surface $\chi$ in the bulk dual geometry satisfying the condition $\partial \chi = \partial A $. Then, to leading order in $1/N$ the entropy is given by

\begin{equation}
S(\rho_A ) = \dfrac{\Area(\chi)}{4G}\, .
\end{equation}

The expression above, know as the Ryu-Takayanagi (RT) or Hubeny-Rangamani-Takayanagi (HRT) formula has passed a plethora of non-trivial checks in different scenarios were its agreement with independent calculations has shown to be remarkably accurate, specially for AdS$_3/$CFT$_2$ where both sides can be computed exactly down to the precise coefficients in a number of cases of interest. In \cite{LewkowyczMaldacena} a `folk proof' was given to the formula, which is considered nowadays as one of the most solid pieces of the AdS/CFT dictionary.

\begin{figure}[h]
$$\includegraphics[width=10cm]{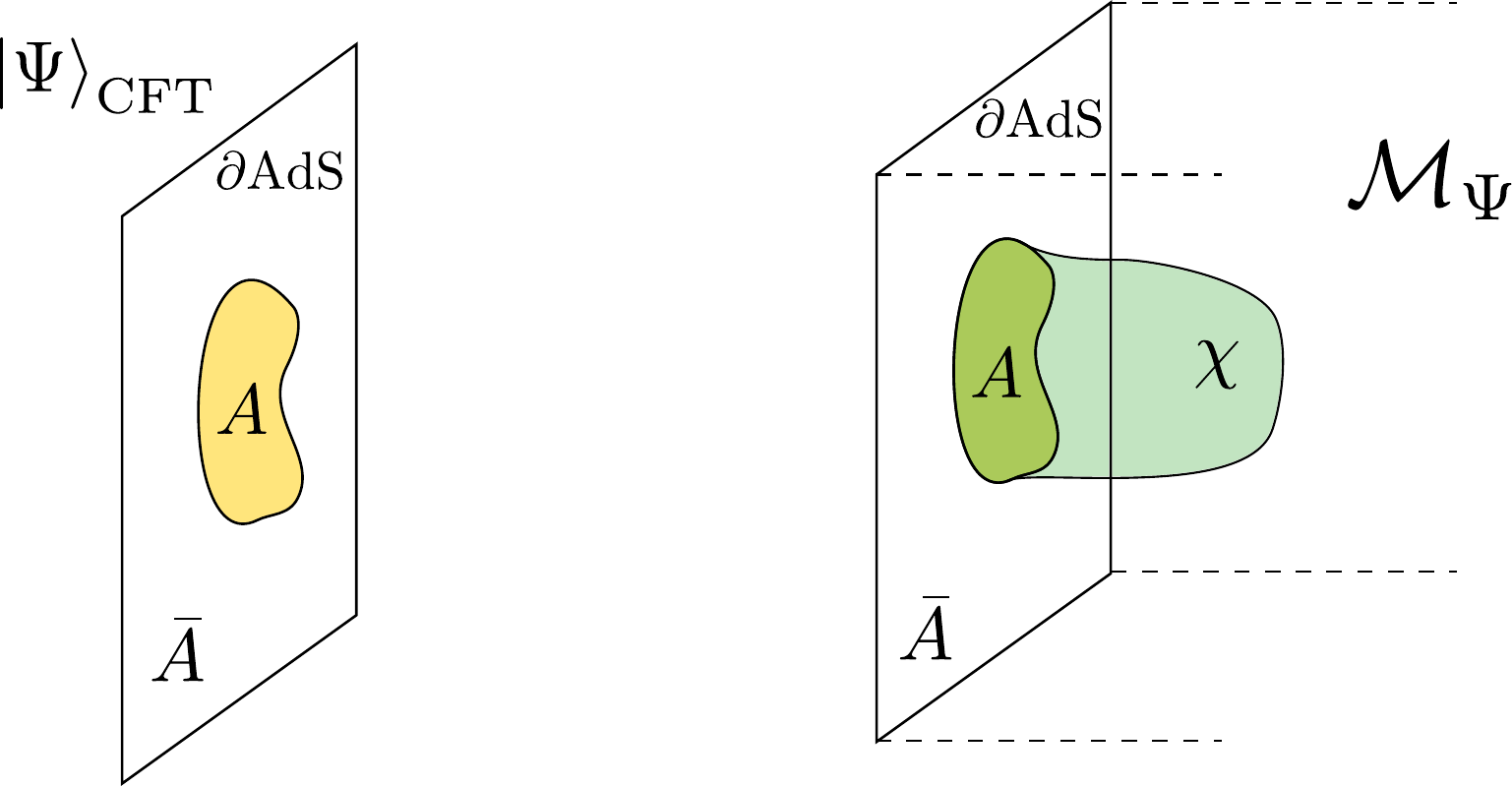} $$   
\begin{center}
\caption{\emph{Schematic picture of the RT proposal. The codimension-2 surface $\chi$ is anchored at the boundary of the region $A$ and it surface is extremalized over the bulk geometry $\CM_\Psi$.}
\label{fig:RT}}
 \end{center}
\end{figure}

\section{But entanglement entropy is not enough}

As we have seen, the developments of the last decade suggest a fascinating relation between quantum information and the physics of black holes, at least in the context of the AdS/CFT correspondence. Out of these developments many fruitful insights have been acquired from the careful study of entanglement entropy, from improving our understanding of black hole evaporation \cite{Penington, Almheiri:2019psf, Almheiri:2019hni, MaldacenaReview}, to that of the laws of thermodynamics \cite{WallTenproofs, WallSurvey} the role of energy conditions in gravity \cite{BoussoQFC,BoussoQNEC,FaulknerANEC, FaulknerQNEC }, bulk reconstruction of local operators \cite{FLM, JLMS} or even the very dynamics of gravity as a consequence the entanglement properties of the CFT \cite{VanRaamsdonkEE, VanRaamsdonkEE2}.

Nonetheless, it is also fair to say that many questions are still unanswered and clues seem to point that a deeper and deeper study of entanglement entropy (or any of its several cousins) by itself will not be enough to address those, specifically when it comes to the description of the black hole interior. Let us go back to the simplest model of the TFD state. By definition, the restriction \eqref{canonicalensemble} to either of the two coupled systems has no information whatsoever about the interior and it is in in fact time independent, implying that any measure of entanglement that is built from $\rho_A$ will also be static. The complete pure state \eqref{TFD} however does have a non trivial dynamics under the Hamiltonian $H= H_A + H_{\bar{A}}$, suggesting that we are still missing some physics in the interior of the black hole. One possible strategy is to choose a different bipartition of the TFD, for example taking a pair of mirror regions $B_L, B_R$ on each of the two CFTs and studying the entanglement entropy of $B_L \cup B_R$ with the rest of the system. This setup was analyzed in \cite{maldahartman} both from the bulk and boundary theory, finding an RT surface for this region that extends trough the wormhole, probing the interior at early times. For finite regions $B_{L,R}$ of size $L \ll 1$, however, the entanglement entropy encounters a topological phase transition at $t \sim \beta$, with the HRT surface switching to a pair of surfaces that do not cross the horizon. As it turns out, these sort of limitations are rather ubiquitous when trying to probe bulk geometries behind more general horizons, a phenomenon that has received the name of \emph{entanglement shadows} \cite{BalasubramanianShadows, FreivogelShadows}. As a result it seems that a complete characterization of the bulk geometry by merely sampling every possible geometric region in the boundary and obtaining their HRT surfaces may not be possible at all. The HRT surfaces are not always able to prove the complete bulk manifold and particularly can tell us little about the interior of the black hole.

The next question is obvious: is there then any other property of the CFT with a suitable geometric interpretation in the bulk able to do this job? and, if so, can it explain this mysterious evolution inside of a black hole? As it turns out, a more refined measure of  entanglement that we will refer to as \emph{quantum complexity} might.

As it happens for the case of the entropy, the concept of complexity already enjoys a number of inequivalent definitions in classical physics, their study defining a vast field in mathematics with hundreds of practical applications. When jumping into quantum mechanics, this zoo enlarges even more (see \cite{Aaronson} for a fantastic introduction to the subject) and merely scratching the surface of the research on the definitions of complexity in QFT\footnote{See for example \cite{nielsen2005geometric, Nielsen1133, Cesar, CesarTime, JeffersonMyers, HugoQFTComplexity, MyersFermions, MyersTFD, Myerscoherent, Myersmixed, BuenoMagan } for further discussion of {\it ab initio} approaches to holographic complexity.} would take us several chapters. Furthermore, as no consensus really exists about whether definition to use, we will bypass this awkward issue by taking a fairly simple and direct route to complexity by means of the language of \emph{tensor networks}, which will allow us to jump back quickly to a gravitational interpretation.

\subsection{Tensor Networks}

\begin{figure}[t]
\begin{center}
\includegraphics[width=8cm]{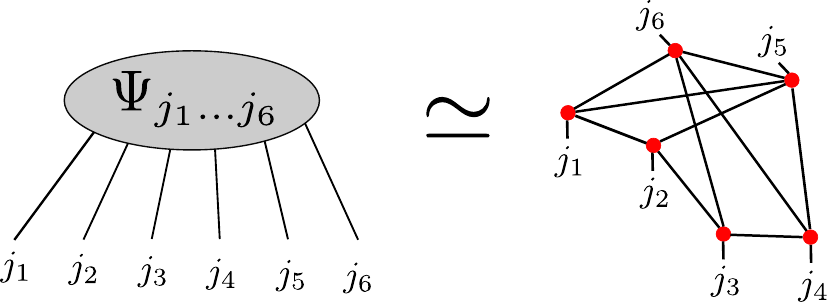} 
\end{center}
\caption{\textit{Tensor network representation of a state in a six-partite Hilbert space. Red dots represent tensor `building blocks' whereas links represent tensor indices. Free links correspond to the basis vectors whereas others represent contraction with other tensor indices. In this example, the dimensionality of the tensor network reduces from an original $d^6$ to $6\times d^4$.}}
\label{fig:TNexample}
\end{figure}

In order to define a tensor network, consider a multipartite  Hilbert space
\begin{equation}
\CH = \CH_1 \otimes \CH_2 \otimes ... \otimes \CH_n
\end{equation}
where, we typically may take each factor as corresponding to some local subsystem (say as in a spin chain, or an Ising model). Now, given the basis $\lbrace \ket{\psi_j}_i \in \CH_i \rbrace$ for each factor, we can write a general state of the system as
\begin{equation}
\ket{\Psi} = \sum\limits_{\lbrace j\rbrace} \Psi_{j_1...j_n} \ket{\psi_{j_1}}_1... \ket{\psi_{j_n}}_n.
\end{equation}

where $\Psi_{j_1...j_n}$ is a tensor of $\CO (\exp (n))$ components that codifies all information about the state. The key to the idea of tensor networks is the realization that most interesting states in physics are in a sense very simple and do not need a detailed specification of an exponential number of coefficients in $\Psi_{j_1...j_n}$. Instead, a fairly good approximation can be achieved with a decomposition into smaller tensor structures
\begin{equation}
\ket{\tilde{\Psi}} = \sum\limits_{\lbrace j, i\rbrace}  A_{j_1}^{i_1\, i_2}A_{j_2\, i_1}^{i_3} \hdots A_{j_n\, i_{m-1}}^{i_m}  \ket{\psi_{j_1}}_1... \ket{\psi_{j_n}}_n \, ,
\end{equation}
whose pattern of contractions can be graphically expressed in a suitable graph (see Figure \ref{fig:TNexample}). As a result, an only polynomially large number of tensor coefficients turns out to be enough to approximate within some $\epsilon$-accuracy\footnote{With this we mean that $|\braket{\Psi | \tilde{\Psi}}| \leq \epsilon$ in the usual metric. In other words, considering the coarse-grained set of total unit vectors on $\CH$, we will say that two states are approximately the same if they live within the same $\epsilon-$sized lattice cell. } states that are `not far from the vacuum' of a local Hamiltonian, in the sense that describe small perturbations around it or are reachable from those by unitary evolution in a short time (polynomial in units of the characteristic scale of the system).

\begin{figure}[t]
\begin{center}
\includegraphics[width=7cm]{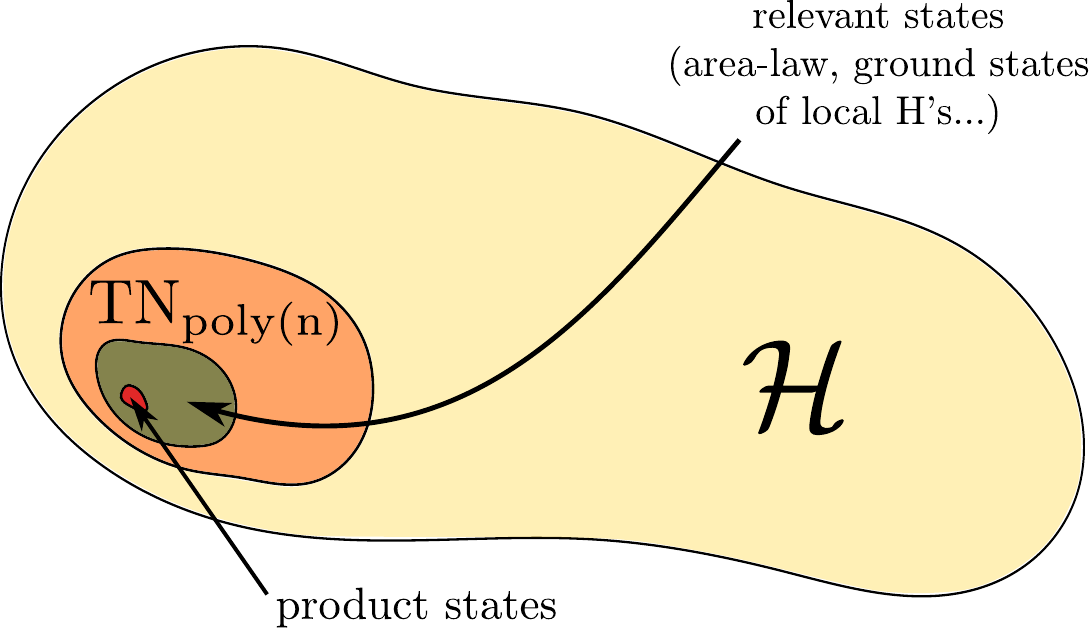} 
\end{center}
\caption{\textit{The set of states in the Hilbert space that is meaningful for quantum field theory, i.e. the ground states of local Hamiltonians and perturbations around those is a exponentially small corner of the total space. Small (polynomially large) Tensor Networks are generally able to cover such corner.}}
\label{fig:Hilbertspace}
\end{figure}

Different tensor networks ans\"atze  have proved to be extremely useful tools in a number of questions in quantum many body physics, from the problem of finding local Hamiltonian ground states to the study of quantum phase transitions or the accurate determination of CFT parameters (see \cite{Orus} for a nice introduction on the subject). In essence, a tensor network consist of two basic elements: a choice of elementary building blocks, or \emph{tensors}, and a pattern of connections among them, or \emph{graph}. As tensor networks are an approximation to a complete and much more detailed state, similar results might be obtained for completely different choices of such two elements, allowing the possibility to shift information about the state from one into another. For example, we might choose a very simple graph formed by a few high-dimensional tensors or a bigger (more \emph{complex}) graph made out of simple tensors with fewer independent coefficients. Once the list of building block tensors is chosen, there is certainly an optimal graph able to recover the state within a given accuracy with the minimal possible number of tensors. For such \emph{optimal tensor networks} we will refer to the `size' or number of basic tensors as the \emph{complexity} of the tensor network.

Within this definition there seems to be too much freedom in the choice of the set of elementary building blocks. Certainly, complexity does depend on such choice and different tensor elements can yield qualitatively different optimal graph structures. Rather than choosing these blocks at random, we may adapt the whole scheme to the symmetries of our system in such a way that both elements of the TN are able to realize explicitly some properties of the state without an excessive fine-tuning of parameters. As it turns out, when such adjustment can be achieved a remarkable split of the state properties takes place among the two tensor network elements: whereas detailed information about local operator correlations is usually encoded in the tensor coefficients, global properties of entanglement (such as different measures of entanglement entropy) become encoded in the graph structure. In other words, the structure of entanglement becomes `geometrized'. 

\begin{figure}[t]
\begin{center}
\includegraphics[width=7cm]{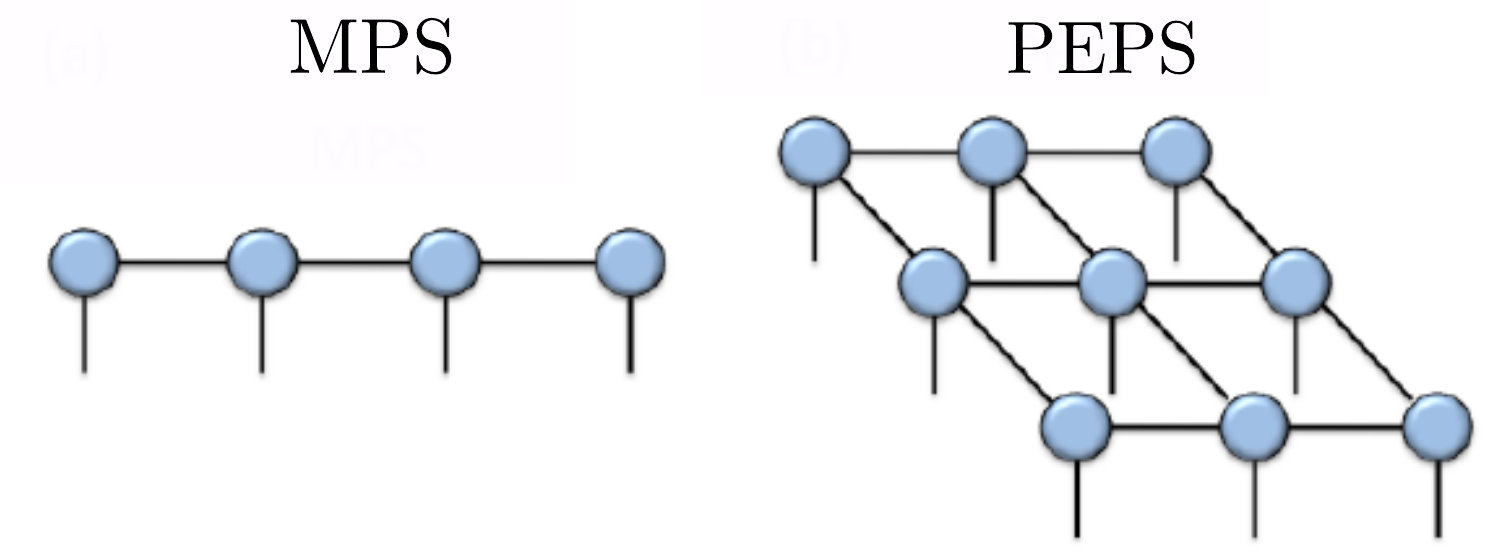} 
\end{center}
\caption{\textit{Examples of tensor network graphs. On the left, an MPS describing a spin-chain with four sites. On the right a PEPS describing a $3 \times 3$ lattice. }}
\label{fig:MPSPEPS}
\end{figure}

Going into the particulars, it is easy to guess appropriate graph geometries for systems that enjoy certain symmetries, such as those well described by translational invariant lattices in various dimensions. For those, TN receiving the names of Matrix Product States (MPS) or Projected Entangled Pair States (PEPS) have shown to be very efficient variational ans\"atze for a number of applications. As was pointed out in \cite{VidalMERA}, however, those simple tensor network graphs may not be the best option at hand if we use them to describe systems in equilibrium at a critical temperature that are well described by a CFT. Rather, as the symmetry group of the system enhances to the conformal group, a description in terms of a hyperbolic graph becomes more adequate, not only computationally but also naturally implementing scale invariance into a fractal-like graph and allowing for a simple real-space renormalization scheme. This proposal, known as Multiscale Entanglement Renormalization Ansatz (MERA) has received a lot of attention in the last decade, as it succeeds in obtaining an accurate description of a plethora of systems that are well described by a CFT.

\begin{figure}[t]
\begin{center}
\includegraphics[width=13cm]{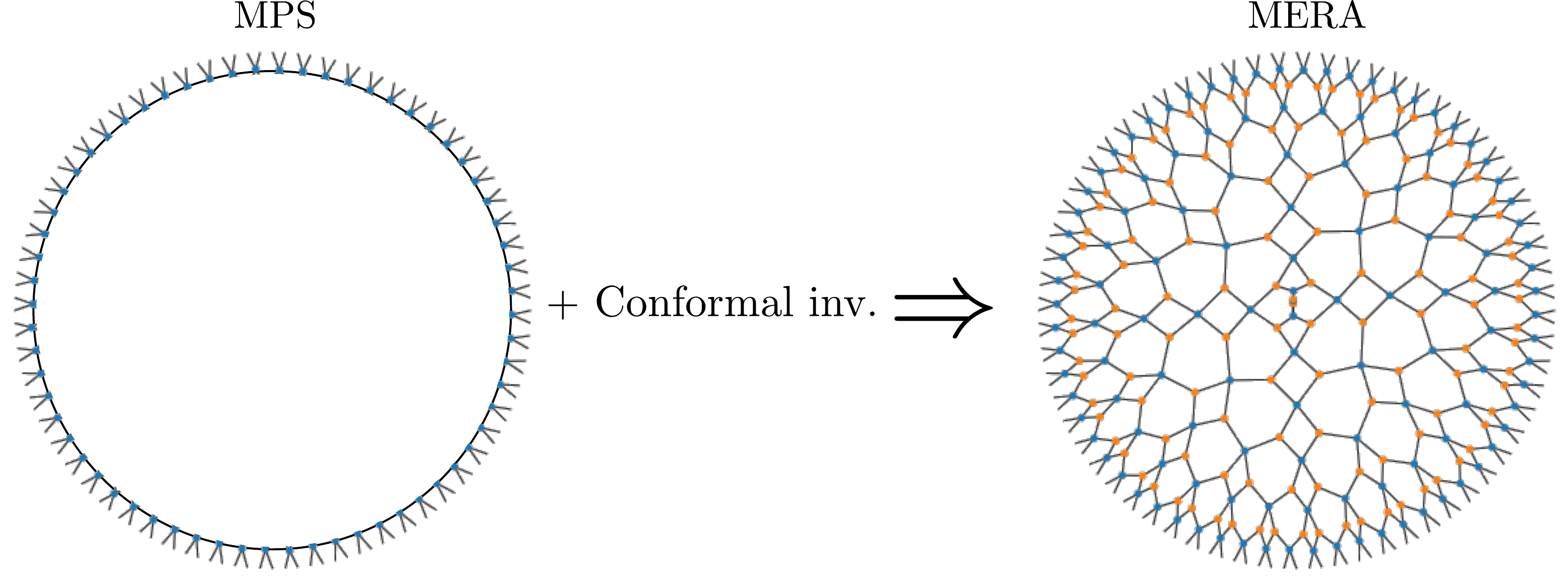} 
\end{center}
\caption{\textit{Two different tensor networks appropriate to describe the vacuum state of a $1+1$-dimensional system on a circle. When the critical temperature is reached, the MERA network (right) reproduces the state better than MPS (left).}}
\label{fig:MPS_MERA}
\end{figure}

Staring into the MERA network in figure \ref{fig:MPS_MERA} we may feel tempted to identify a number of features that suggest a `holographic' interpretation of this graph. First, in order to accommodate conformal symmetry on the system, an extra dimension has emerged in the graph. We can think of this dimension as implementing a sort of Kadanoff renomalization group, with the outermost layers corresponding to the physical UV degrees of freedom and the inner ones corresponding to coarse grained versions of them\footnote{The key point in \cite{VidalMERA} was to be able to do this while keeping the correct amount of entanglement at all scales that is required in local scale invariant Hamiltonians.}. Second, the geometry of the graph is no other than a discrete version of the Poincaré disk, suggesting a picture of the MERA as describing a constant time slice of AdS space, the would-be-dual geometry of a holographic CFT vacuum state described by the tensor network. Finally, the detailed construction of the MERA network allows its use as an isometric map from inner to outer layers, hinting towards a possible lattice implementation of the `bulk reconstruction' procedure of encoding bulk physics into the boundary.

\begin{figure}[h]
\begin{center}
\includegraphics[width=7cm]{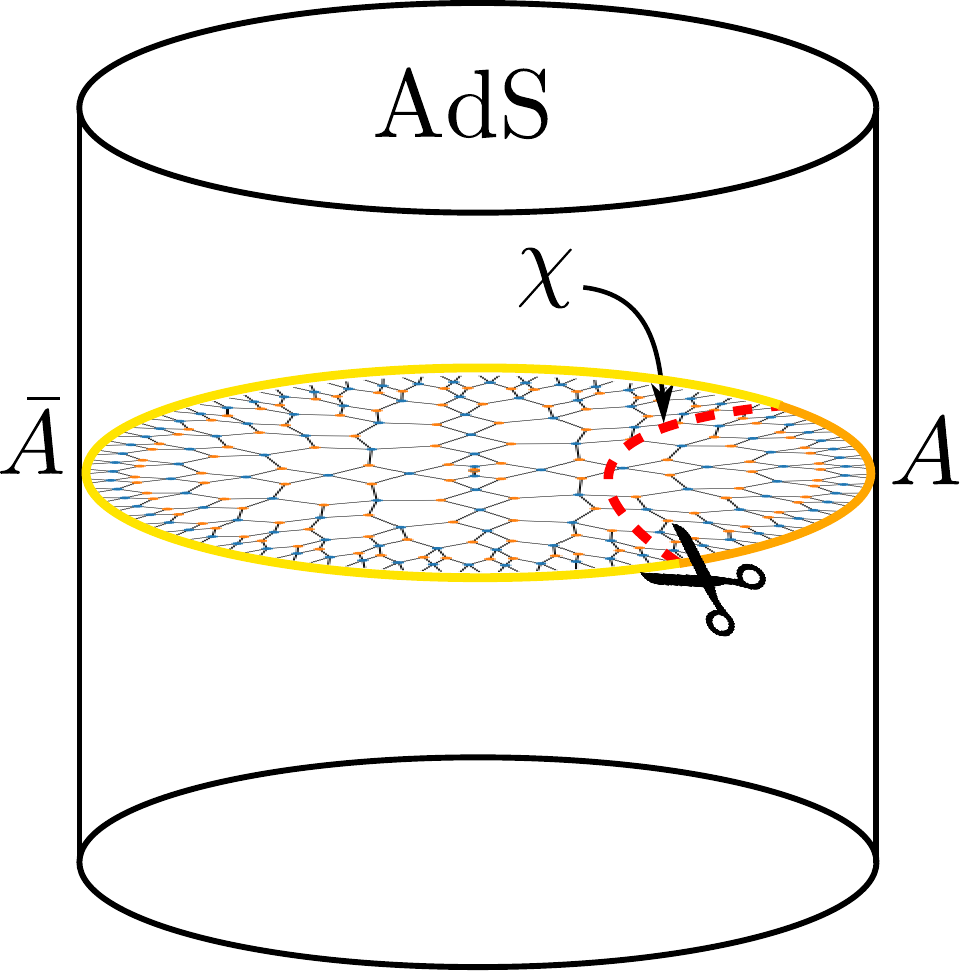} 
\end{center}
\caption{\textit{MERA tensor network interpreted as a discretization of a constant time slice of AdS spacetime. The RT formula enjoys a natural discrete variant in this language, where $S(A)$ can be compute by means of a minimal link cut along the tensor network.}}
\label{fig:RTTN}
\end{figure}

As we pointed out a few paragraphs above, truly optimal tensor networks enjoy a sort of geometrization of entanglement properties built in the network graph, and the MERA is no exception to this. In particular, entanglement entropy for subregions in the physical system is well approximated by an optimal process of `tensor network cutting out', i.e. $S(A)$ can be found to be proportional to the minimum number of links that one needs to cut in order to split the tensor network in two pieces. For ordinary systems that are well described by lattice tensor networks, like PEPS, this feature realizes the well known \emph{area-law} that is known to hold generically for a wide class of ground states of local Hamiltonians. In the case of MERA networks describing a CFT, it implements a discrete version of the Ryu-Takayanagi formula (see Figure \ref{fig:RTTN}).

These and other `holographic' features of the MERA network were emphasized in \cite{SwingleMERA}, were the picture of the graph geometry as giving a discrete description of the dual gravitational bulk geometry was proposed. Later on, many works  have built on this idea (cf. \cite{HaPPy, HaydenRandom, WallTN}), suggesting further generalizations to describe other holographic states and trying to come up with necessary and sufficient conditions for the tensor networks to properly reproduce the known properties of AdS/CFT.

As tensor networks are merely efficient tools to encode the physics and entanglement structure of quantum states, this is nothing but a reincarnation of the ER=EPR slogan, suggesting that the space of network graphs for holographic CFT states may not be entirely formal, but rather an approximation to the dual gravity theory.

\subsection{Holographic complexity}

For a tensor network definition of complexity as the one sketched in the previous section (number of `simple' tensor nodes in the graph for the optimal TN) and building on the intuitions for a holographic interpretation of the MERA, one can come up easily with a candidate for a new entry in the AdS/CFT dictionary. Indeed, if complexity is to be properly defined for CFT states, the volume of a suitable associated Cauchy slice may be a good candidate to compute this quantity in the bulk. In order to choose this slice in a covariant manner and inspired by the RT formula for entanglement entropy, Leonard Susskind conjectured in a series of works\cite{SusskindBridgetonowhere, StanfordShockWave, SusskindShocks, ShenkerStanford, SusskindEntnotenough} what is know as the Volume-Complexity (VC) conjecture, stating the following:

\begin{itemize}
\item \emph{Given a state $\ket{\Psi(t)}$ of a holographic CFT possessing a smooth gravity dual geometry $\CM_\Psi$, there exists a notion of computational complexity $ \CC_V (\ket{\Psi (t)})$ that at leading order in $1/N$ can be calculated  by
\begin{equation}
\label{CVansatz}
\CC_V (\ket{\Psi (t)}) = \dfrac{{\rm Vol}(\Sigma_t)}{\ell G}
\end{equation}
where $\Sigma_t$ is an spacelike extremal volume codimension-1 surface anchored at the constant $t$ surface on $\partial \CM_\Psi$, and $\ell$ is a yet-to-be-fixed length scale.}
\end{itemize}

As inspirational as the MERA construction might be in suggesting this proposal, there is certainly not much meat in the study of complexity for a vacuum state. The huge size of Hilbert space suggests an enormous variety of tensor networks to describe other states of interest. Particularly, states with non-trivial Hamiltonian evolution may force us to change our optimal tensor network at each instant of time, therefore rendering complexity a dynamical quantity. For a non-trivial test of the VC proposal thus we would like to test it in a dynamical system, studying the evolution of complexity in time.

If we view the tensor network construction as a quantum unitary circuit\footnote{Tensor networks and quantum circuits are certainly different concepts, the latter corresponding only to special instances of the former. Computational complexity and its properties, typically defined in the context of circuits,  will be useful however to characterize the growth of wormholes in black hole systems for which a tensor network representation will have a nice circuit interpretation.} (with the tensors playing the role of small quantum gates) the question is how and `how far' Hamiltonian evolution can take us from our initial state at each instant of time, i.e. how big is the smaller circuit able to perform such evolution.

\tcbset{breakable}
\begin{tcolorbox}[enhanced, breakable, title=Complexity: bounds and generic dynamics]
\small
In order to show the main features of complexity dynamics for typical systems we lay down here a few definitions and estimations. Let us take a quantum system with a Hilbert space $\CH$ of complex dimension $\sim e^S$ and consider the limit $S \gg 1$ in order to simplify the estimations. Such space is naturally equipped with the usual Fubini-Study (FS) metric, that we may visualize as an standard Euclidean metric on a real $e^S-$dimensional hyper-Bloch sphere $\mathbf{CP}(e^S)$. Such metric has of course many uses but it is better suited to measure closeness that remoteness. As an example, two similar systems sharing many properties can be made orthogonal by the presence of a simple control qubit, assigning a maximal FS distance between them and therefore yielding the same result as the one that we could obtain by measuring the distance to a completely different state
\begin{equation}
\braket{\text{\faCar }\, \otimes \uparrow | \text{\faCar }\, \otimes \downarrow}=0 = \braket{\text{\faCar}| \text{\faCoffee}}\, .
\end{equation}
The purpose of complexity is to provide a different metric able to distinguish between these two situations. For that, we would like to count the total number of unit vectors in $\CH$. It is of course infinite, but we may regulate it by means of a coarse graining of $\mathbf{CP}(e^S)$. If we do so by balls of radius $\epsilon$ it is easy to get the total number of states 
\begin{equation}
\#_{\rm{states}} \sim \exp(e^S \log(1/\epsilon)).
\end{equation}
In this setup, any two vectors belonging to the same cell will be at a FS distance of order $\epsilon$, meaning that all observables will match within that accuracy and therefore we will consider them equivalent. In fact, given some privileged $k-$local basis \footnote{The term $k-$locality formalizes the notion that the Hamiltonian can be approximated by quantum gates or tensors that are small in the sense that they act on few ($k$) qubits, or have few tensor `legs' as compared with the total degrees of freedom, i.e. $k \ll S$.} we may choose a TN representative of each cell defined as the simpler TN able to do the job. Accordingly, a complexity will be assigned to each cell.
\begin{figure}[H]
\begin{center}
$$\includegraphics[width=15cm]{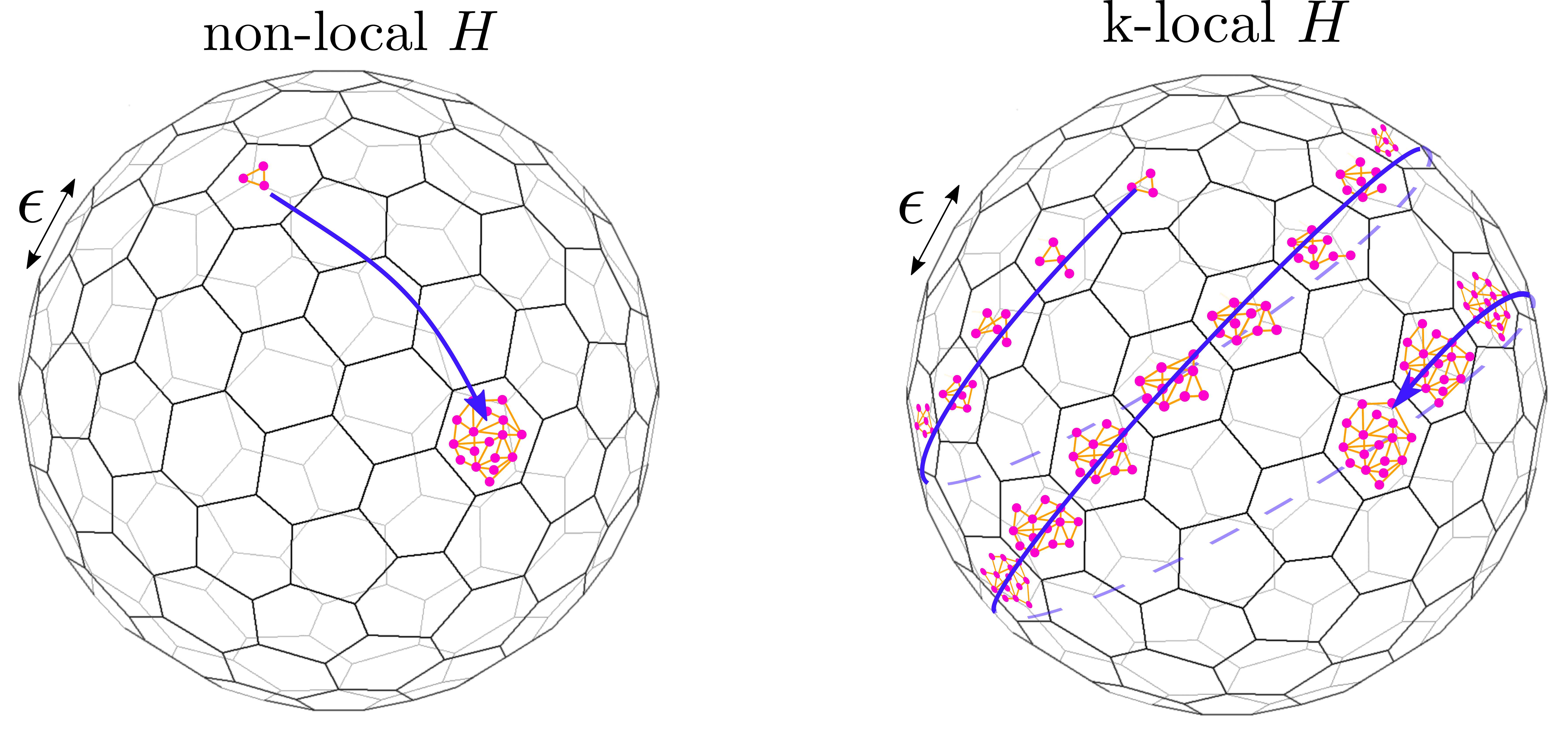} $$   
\caption{\small \emph{Tessellation of the Bloch sphere by $\epsilon-$balls. For each ball, we choose a TN representative. In general, shortest paths in Hilbert space can be achieved through evolution driven by non-local Hamiltonians. Restricting the Hamiltonian to be $k-$local forbids most directions in the sphere and gives rise to ergodically-long paths. }
\label{fig:twopaths}}
 \end{center}
\end{figure} 

Complexity dynamics in this picture has to do with the way states move around the hyper-Bloch Sphere under unitary evolution, where all possible Hamiltonians generate the $e^S$ directions of $\mathbf{CP}(e^S)$. Restricting the set of Hamiltonians to those that are $k-$local, however, kills most of such directions, allowing movement only on a subset of order $S$ directions. As we have discretized the Hilbert space, we may think of a typical unitary evolution as as quantum circuit

\begin{equation}
U(t_{\rm{comp}})= \prod\limits_{i}^{\mathcal{D}} U_i (t_{\rm{step}})\,.
\end{equation}

where the Hamiltonian acts with an order $S$ number of $k-$local gates at each step and $\CD$ is the \textit{depth} of the circuit. The physical time that is taken in each step will be given by some natural energy scale in the problem. In the case of a thermal system, $t_{\rm{step}} \sim T^{-1}$, which is the typical orthogonalization time for such systems.

\begin{figure}[H]
$$\includegraphics[width=3.5cm]{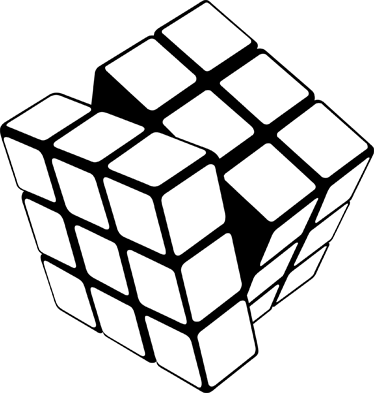} \hspace{2cm} \includegraphics[width=6cm]{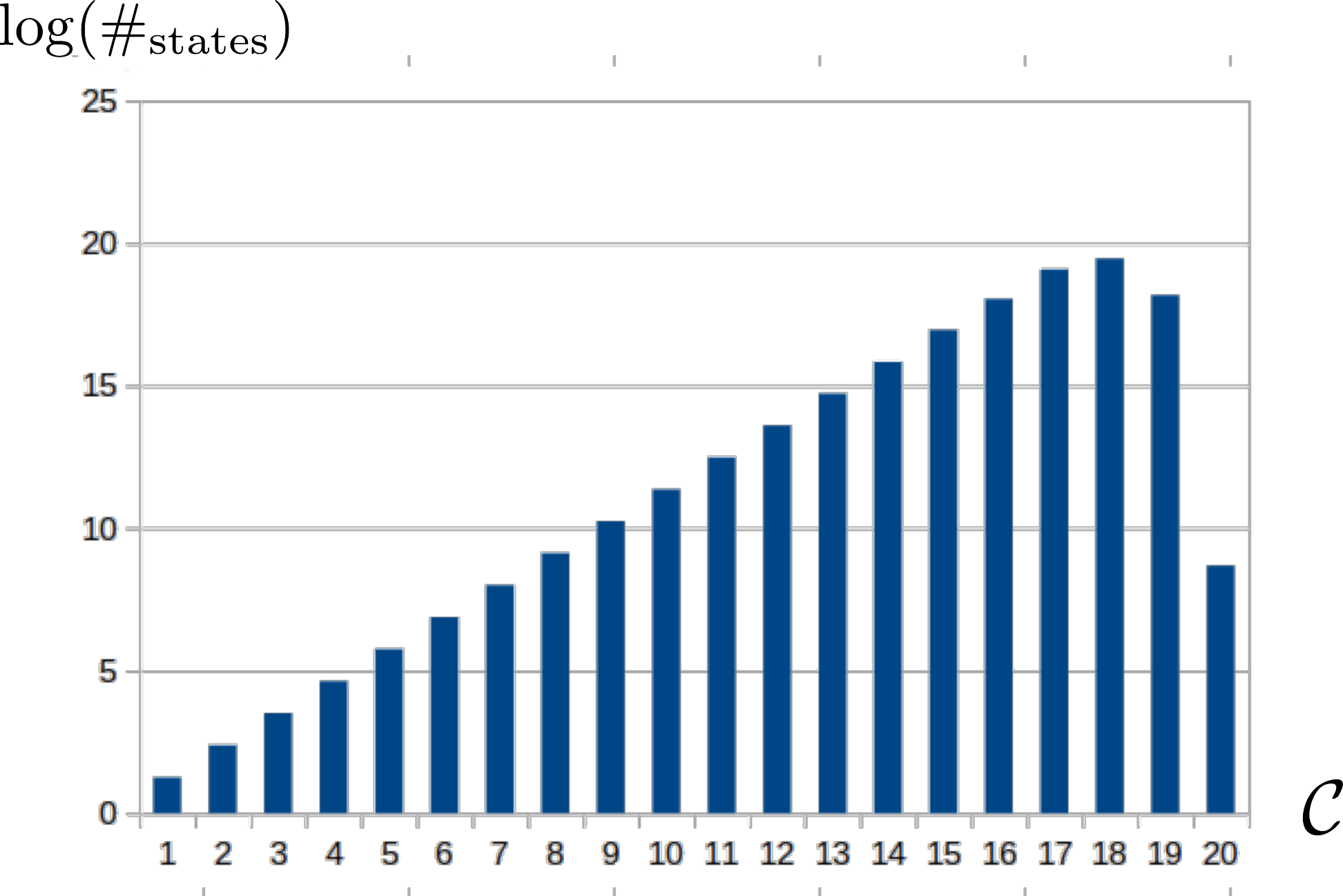} $$   
\begin{center}
\caption{\small \emph{Despite being a classical example, few problems are as well-studied and easy to visualize as the Rubik's Cube. The number of configurations in this system is about $\sim 4.3\times 10^{19}$, but it was recently proved in \cite{rubik} that all positions can be solved by a maximum of 20 turns. As we see in the right figure, there are are exponentially many more configurations closer to maximal complexity than to lower values, meaning that generic configurations are likely to have $\CC \sim \CC_{\rm{\max}}$. Since the number of configurations of a given complexity increases monotonically with $\CC$, we expect that starting from the solved cube, any generic algorithm that scrambles the cube reaches maximal complexity in about $\sim \CC_{\rm{\max}}$ steps.  }
\label{fig:rubik}}
 \end{center}
\end{figure}

Since complexity is defined by an optimization of such circuits, an important issue is whether real Hamiltonians build optimal circuits themselves or rather are inefficient and better ways exist to approximate $U$. In other words, is the complexity of $U$ just $\CC \sim S \times \CD$ or is there a shortcut which makes it smaller? Indeed, we do know shortcuts will eventually appear because the number of circuits is finite. To see this, consider all the programs of depth $\CD$ that you can construct with a finite $g = \CO(1)$ set of $k-$local gates. Each gate can be put down in about
\begin{equation}
\binom{S}{k}^g
\end{equation}
ways. This scales like $S^\alpha$ for $\alpha$ some $\CO(1)$ exponent which depends on how many basic gates we have and also the precise value of $k$. So, if we have about $S \CD$ gates, which we call the complexity of the program, the number of possible programs is of order
\begin{equation}
S^{\alpha S \CD}  \sim S^{\alpha \CC} \sim S^\CC\, ,
\end{equation}
but, with $\epsilon-$accuracy, we fill out all possible cells when
\begin{equation}
S^{\CC_{\rm{max}}} \sim \exp(e^S \log(1/\epsilon))\, ,
\end{equation}
which gives the complexity bound
\begin{equation}
\CC \leq \CC_{\rm{max}} \sim \dfrac{e^S}{\log S} \log(1/\epsilon) \simeq e^S \log(1/\epsilon)\,.
\end{equation}
\begin{figure}[H]
$$\includegraphics[width=13cm]{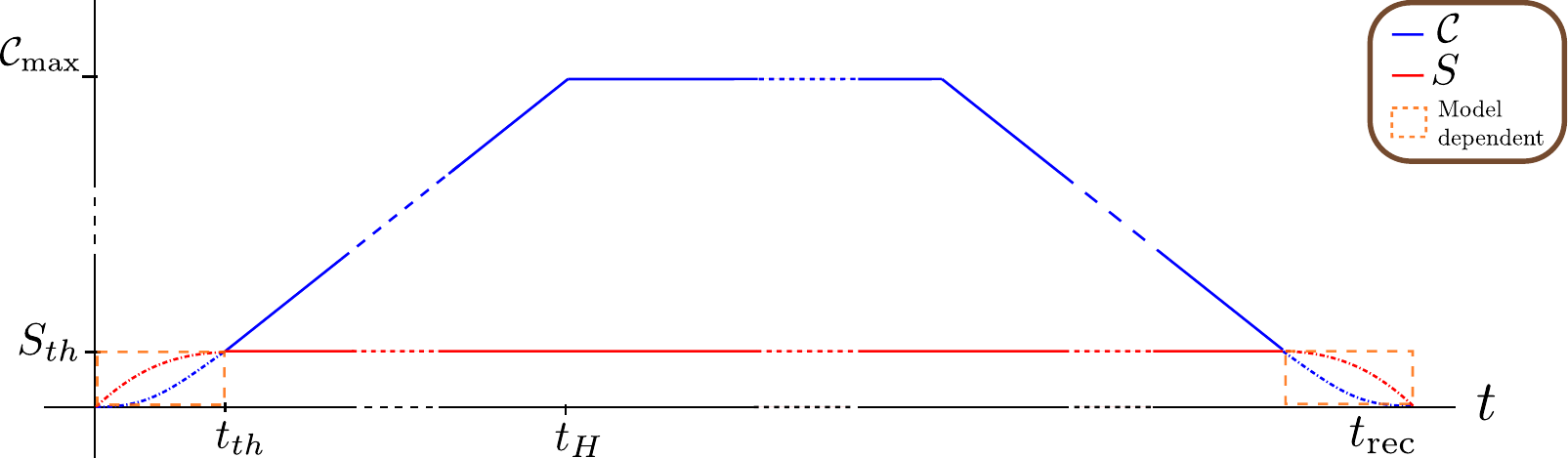} $$   
\begin{center}
\caption{\small \emph{Expected behaviour of complexity and entropy for a state undergoing thermalization. After the process of thermalization (inside orange box) the thermal entropy of the state stabilizes at a constant value. The complexity on the other hand keeps growing for exponentially larger times as the state evolves under a chaotic Hamiltonian.Complexity becomes bounded by $\CC_{\rm{max}}\sim e^S \log(1/\epsilon)$ at times of the order of the Heisenberg time $t_H \sim T^{-1} e^S \log(1/\epsilon)$. Furthermore, when all phases go back to their initial values (within an $\epsilon$ tolerance) after filling the torus, the state will recur to its original state. This happens at the quantum Poincaré recurrence time $t_{\rm{rec}}\sim T^{-1}\exp(e^S \log(1/\epsilon))$. We must think of the picture above as an average over a set of chaotic Hamiltonians since a single occurrence might describe some noise around this graph.}}. 
\label{fig:Cvst}
 \end{center}
\end{figure} 
As a result, we do expect long circuits to saturate eventually this bound, where shortcuts will inevitably appear. A remaining question is whether these shortcuts may appear earlier in the evolution, i.e. before saturating the bound. Generically we do not expect such phenomenon to happen if the Hamiltonian is generic (chaotic). Given a particular time-independent Hamiltonian $\hat{H}$, the time evolution operator 
\begin{equation}
U(t) = e^{-i \hat{H}t} = \sum\limits_{i=1}^{ N} e^{-i E_i t} \ket{i} \bra{i}
\end{equation}
explores an N-dimensional torus $\mathbf{T}^N$ defined by the set of $N\sim e^S$ phases. Since the Hamiltonian is fixed, this manifold represents an exponentially small corner of the full set of unitary operators $SU(e^S)$ which has dimension $\sim e^{2S}$, but has nevertheless a huge dimensionality and can contain circuits of exponentially big complexity. Whether a particular $k-$local Hamiltonian does generate such very complex circuits depends on its ability to densely cover the torus. Although there is not a rigorous proof, it is generally believed \cite{SusskindLectures, Susskind2law} that sufficiently chaotic \footnote{The very definition of quantum chaos could again take us an entire chapter. Incommensurability of the eigenvalues $E_i$ is enough to guarantee the ergodicity on $\mathbf{T}^N$ , but that might not be a general enough characterization of chaos. See \cite{StanfordLectures} for an introduction to the subject.} Hamiltonians are indeed ergodic in $\mathbf{T}^N$, with recurrences occurring for times that are doubly exponential in $S$ .

We do expect therefore complexity to grow for exponentially large times as the Hamiltonian evolution takes the state wandering around the torus (see Figure \ref{fig:Cvst} for some more details). For pure states that are approximately thermal as measured by some set of small operators (say, $n$-point functions with $n \ll S$) dimensional analysis suggests such growth to be behave as 
\begin{equation}
\label{CVrate}
\dfrac{\dd \CC (t)}{\dd t} \propto S\,T \sim M\,
\end{equation}
where $S$, $T$ and $M$ are the entropy, temperature and total energy of the system.

\end{tcolorbox}

Since we do expect a non-trivial evolution of complexity for thermal systems (see the box above), this gives us the opportunity to get a qualitative test of the VC conjecture for our favourite state: the TFD. As we mentioned in the beginning of this section, despite the fact that the TFD looks static as seen by any measure of $\rho_A$ (or $\rho_{\bar{A}}$), it is obvious that the complete pure state enjoys a non-trivial evolution under the joint Hamiltonian $H= H_A + H_{\bar{A}}$
\begin{equation}
\ket{{\rm TFD}(t)} = \sum\limits_i e^{- 2 i\,  E_i\, t} e^{-\beta E_i/2} \ket{E_i}_A \otimes \ket{E_i}_{\bar{A}}\, ,
\end{equation}
which, for a generic spectrum of energies, yields a complex structure of phases with no periodicities. If complexity is to be a meaningful quantity, it should be able to capture the structure of those phases, for which other entanglement measures, like entropies, are blind. In the gravitational side, hence, it should be able to probe the interior of the black hole.

\begin{figure}[h]
$$\includegraphics[width=7cm]{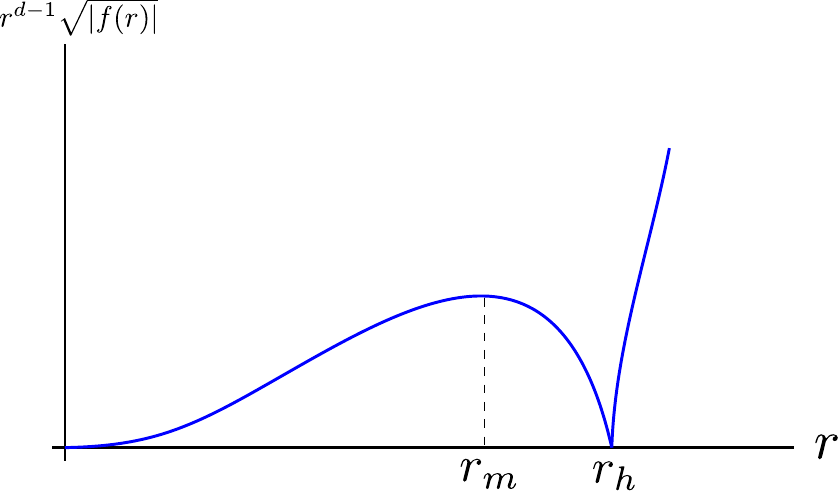} $$   
\begin{center}
\caption{\emph{Volume functional for constant $r$ surfaces at the black hole interior. A local maximum is found at the `limiting surface'  $r=r_m$.  }}
\label{fig:volumefunctional}
 \end{center}
\end{figure}
\begin{figure}[h]
$$\includegraphics[width=5cm]{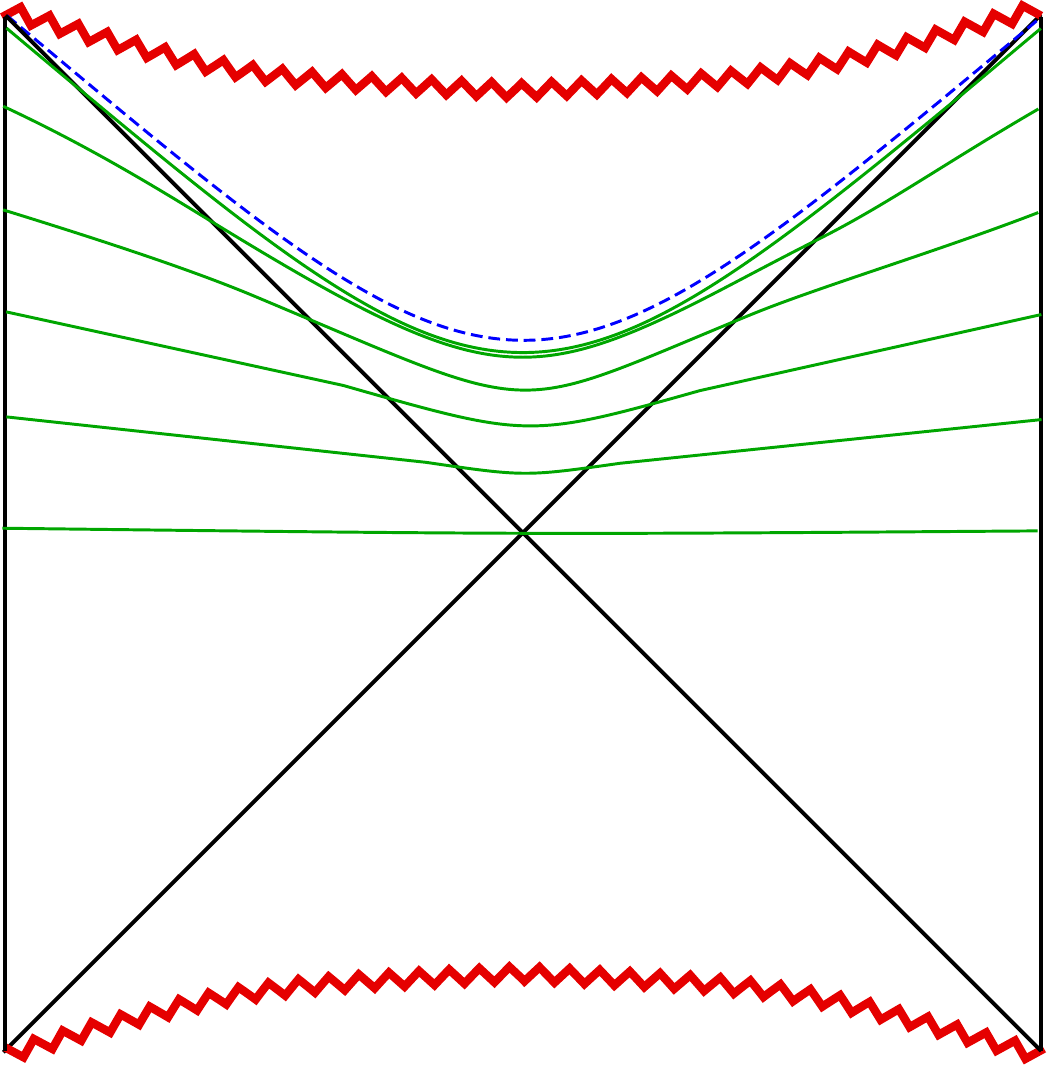} $$   
\begin{center}
\caption{\emph{Maximal slicing in the full Kruskal extension. In the interior, maximal surfaces lie along the limiting surface $r=r_m$, picking a larger portion of it as the boundary time moves upwards and changes the boundary conditions of the slice.   }
\label{fig:maximalPenrose}}
 \end{center}
\end{figure}

\begin{figure}[h]
$$\includegraphics[width=9cm]{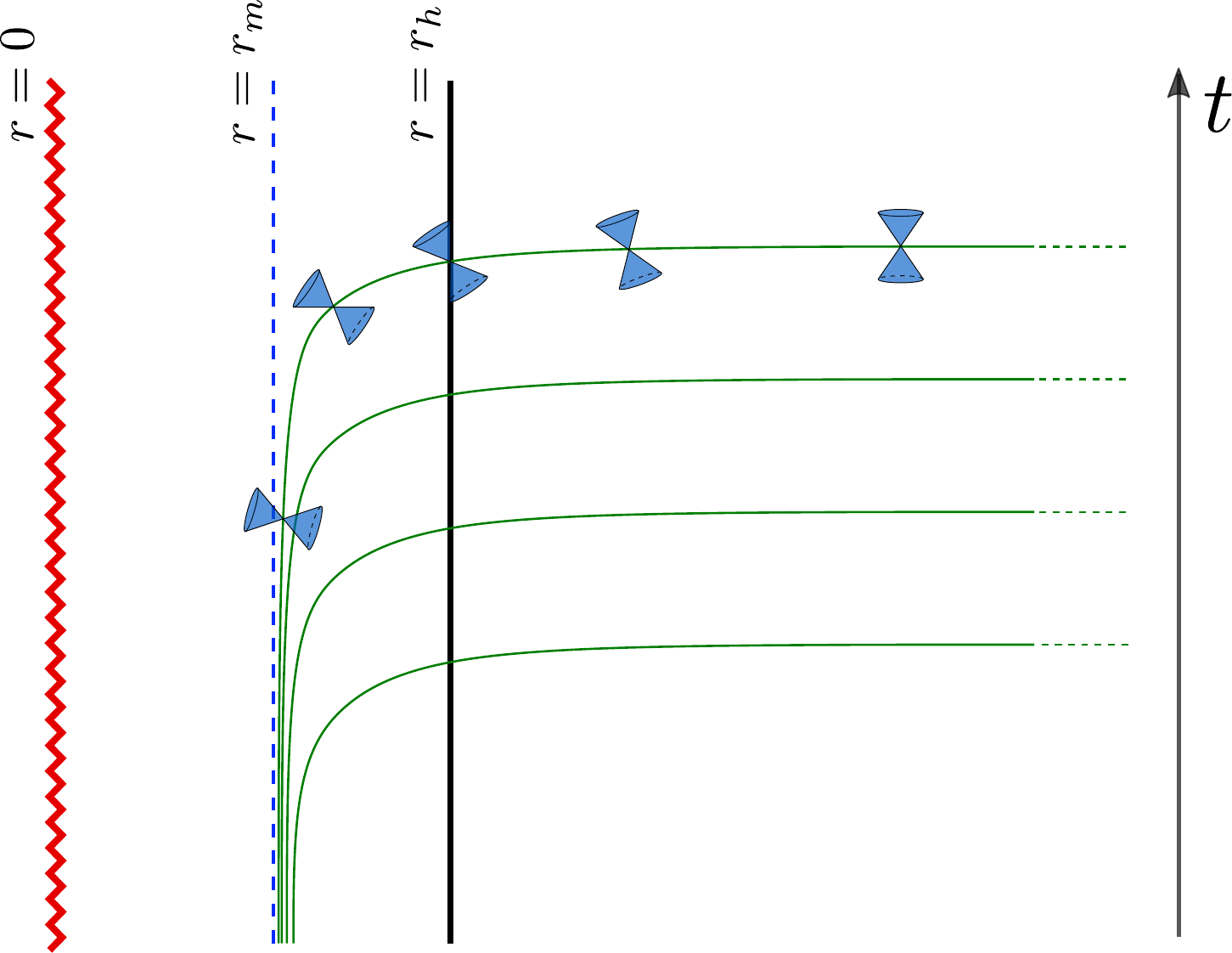} $$   
\begin{center}
\caption{\emph{Maximal slicing in Eddington-Finkelstein coordinates. This coordinates make manifest the fact that the volume outside the horizon remains constant, whereas the length of the cylinder $r=r_m$ that is picked by the maximal slice grows linearly with time.  }
\label{fig:maximalEF}}
 \end{center}
\end{figure}

 
As it turns out, this intuition is correct, and codimension-1 volumes in the eternal Schwarzschild solution not only explore the black hole interior, but also grow at the linear rate \eqref{CVrate}, confirming the expectations from the QFT side. Without the need of going into the details, we can readily confirm this result by looking at a couple of diagrams. First, by plotting the volume functional for the Schwarzschild solution  (see Figure \ref{fig:volumefunctional}), we see that there exists a maximal volume surface in the interior located at a constant $r$ value $r_m$, not far from the horizon. This surface is a cylinder of infinite length and base area $\sim S$.  Maximal surfaces that are anchored at the two asymptotic AdS boundaries lie at the natural constant time bulk slices close to the boundary, but enter across the horizon towards the interior of the black hole as they approach it, lying close to the cylinder before coming out through the other horizon (see Figure \ref{fig:maximalPenrose}). While the piece of the maximal surfaces remains almost constant due to the boost symmetry of the exterior regions, the length of the cylindrical piece that is picked by each extremal slice grows linearly with time. This is easy to see in Eddington-Finkelstein coordinates as in Figure \ref{fig:maximalEF}.

\begin{figure}[h]
$$\includegraphics[width=9cm]{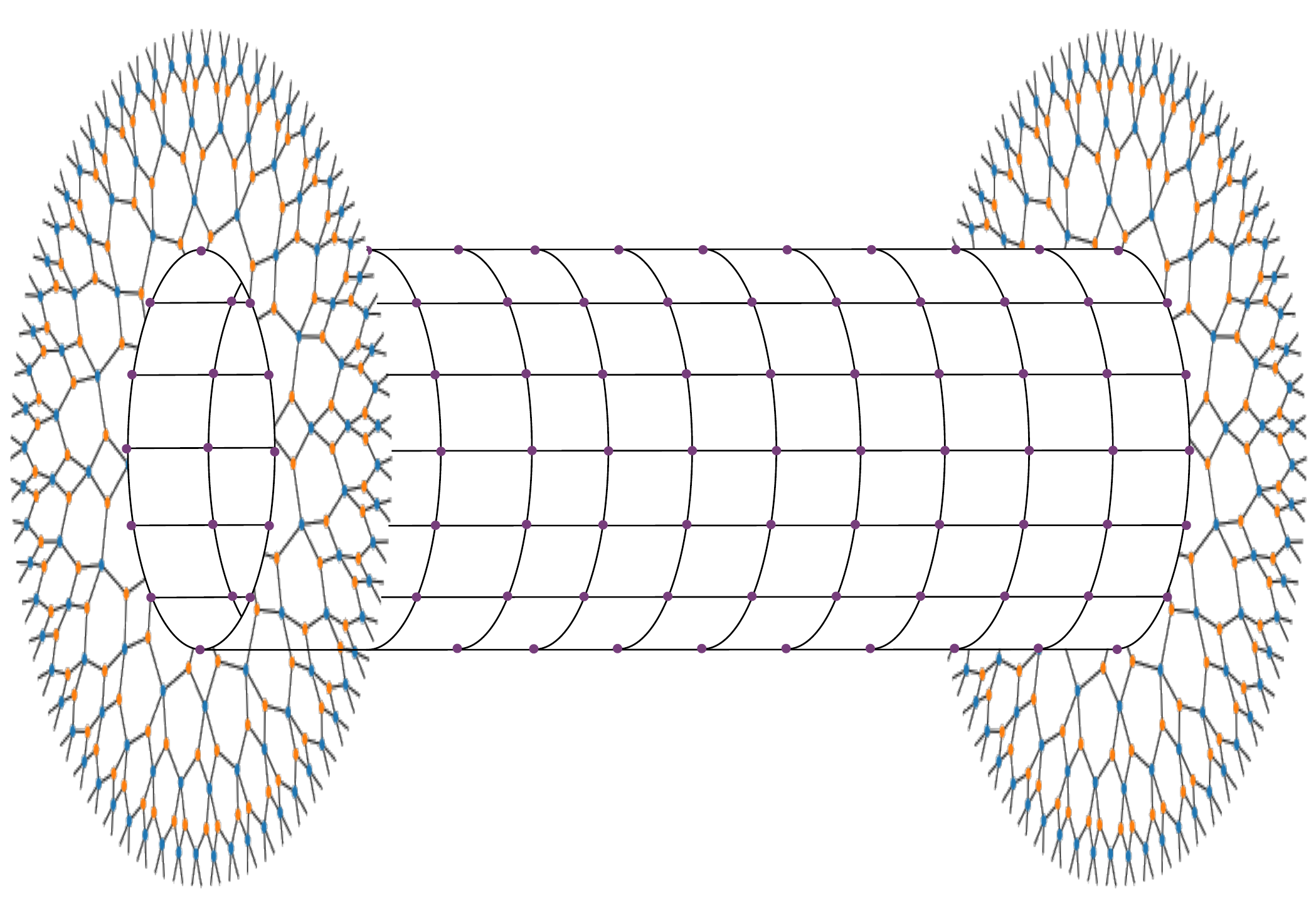} $$   
\begin{center}
\caption{\emph{Tensor Network representation of a TFD state at late times. At both sides, a truncated version of a MERA network represents the exterior of the black hole, whereas the interior is well approximated by a growing euclidean quantum circuit.}
\label{fig:TNwormhole}}
 \end{center}
\end{figure}

%

As we see, holographic complexity gives a meaning to the fact that black hole solutions possess growing structures in their interiors (the so called Einstein-Rosen bridges or wormholes), adds evidence to the ER=EPR proposal and promises to give us a new tool to study the nature of black holes from a pure boundary description. 

Despite the reasonable success in describing this (an other more refined) phenomenology, the VC conjecture suffers from a number of undesirable features

\renewcommand\labelitemi{\tiny$\bullet$}

\begin{itemize}
\item There seems to be no natural choice\footnote{See though \cite{JacobsonVolume}.} for the arbitrary length scale $\ell$ in \eqref{CVansatz}. In order to match the expectation \eqref{CVrate} for AdS-Schwarzschild solutions one has to choose this scale by hand at different values depending on the size of the black hole.
\item Quite disappointingly, maximal slices do not foliate the whole black hole interior, but merely a sort of `inner Rindler' region of low curvature. In this sense, complexity would still not allow us to learn about the greatest mystery of black holes: the singularity.
\end{itemize}
\renewcommand\labelitemi{$\bullet$}

\begin{figure}[h]
$$\includegraphics[width=5cm]{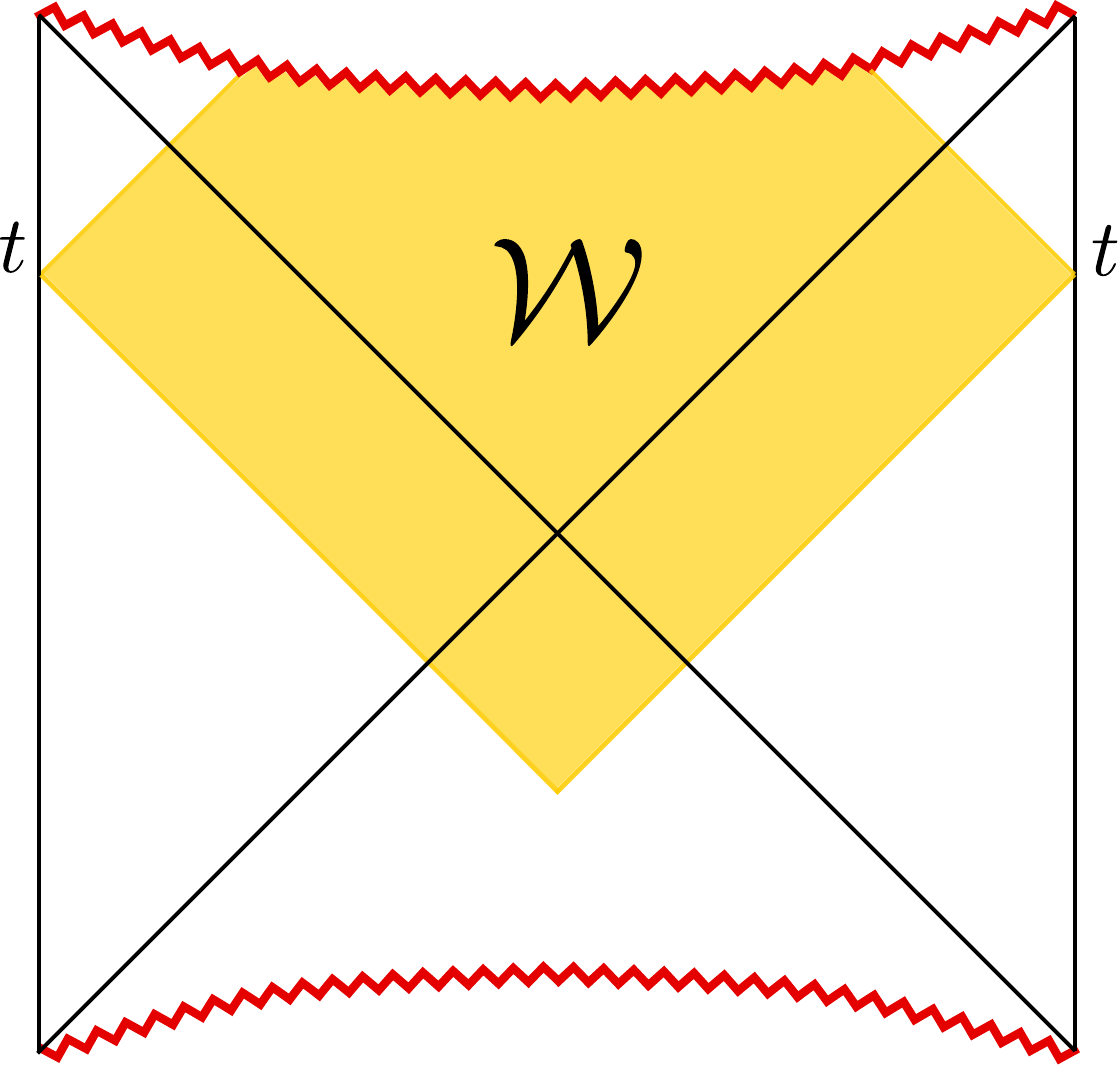} $$   
\begin{center}
\caption{\emph{WdW patch for boundary times $t_L=t_R=t$. This patch is the full causal development of Cauchy slices anchored at those boundary times and, as we see, penetrates into the black hole interior all the way up to the singularity. }
\label{fig:eternalbhaction}}
 \end{center}
\end{figure}

With the aim of curing these issues, it was pointed out in \cite{SusskindACShort,SusskindACLong} that a different quantity
in the bulk could be defined solving these problems at the same time as inheriting all the nice features of VC duality. In such proposal the full set of all possible Cauchy slices anchored at some boundary time $t$ (and not only the extremal one), known as de Wheeler-DeWitt (WDW) patch is now the element of interest, and holographic complexity is computed by evaluating the Einstein-Hilbert action on this patch. The ansatz \eqref{CVansatz} therefore becomes
\begin{equation}
\label{CAansatz}
\CC_A (\ket{\Psi(t)}) = I[\CW_t]
\end{equation}
where $I$ stands for the Einstein-Hilbert action with appropriate boundary terms, and $\CW_t$ is the WDW patch anchored at the boundary at time $t$, as shown in Figure \ref{fig:eternalbhaction}. This proposal is often known as the Action-Complexity (AC) duality, and exhibits similar phenomenology as the VC duality, at least for the benchmark model of eternal black holes. In fact, as claimed in \cite{SusskindACLong}, the late time complexity growth can be proven to yield the sharp result
\begin{equation}
\label{CArate}
\dfrac{\dd \CC_A}{\dd t} = 2M
\end{equation}
where $M$ stands for the ADM mass of the black hole and the overall coefficient appears as universal for all energies and spacetime dimensions, suggesting that \eqref{CArate} might be a strong model-independent prediction for complexity growth in thermal CFT states.

\subsubsection{Holographic complexity phenomenology}

Since the proposal of the AC and VC conjectures, many works have explored the details of both hypotheses trough the careful examination of a number of systems in different regimes. Let us quickly review some of them.

In a series of works \cite{SusskindBridgetonowhere, StanfordShockWave, SusskindShocks, ShenkerStanford}, the VC conjecture was tested to study its robustness against small perturbations, comparing the qualitative expected behaviour in the context of quantum circuits  with that of the perturbations of black hole solutions. The setup is roughly the following: consider a small perturbation by a local operator $\CO$ inserted at some time $t_0$ into the quantum thermal system. If the operator injects some energy $\delta M$ into the system, we expect on general grounds the complexity to modify its dynamics by $\delta \dot{\CC} \sim \delta M$. Following a simple epidemic model for the quantum circuit, however, we can readily see that such enhanced complexity rate, cannot be instantaneous, but must rather be delayed by a process of operator scrambling, during which $\CO$ grows up to the size $S$ of the original state following a logistic curve. As a result, the change in complexity growth is only appreciable after a the so called \emph{scrambling time} $t_* \sim \beta \log S $, a phenomenon that has been given the name of \emph{switchback effect}. In the gravitational side, the action of these small operators was modelled by perturbing the geometry with shockwaves thrown from the AdS boundary into the black hole. As a result, a remarkable agreement was found for a number of different scenarios, confirming the sensitivity of VC complexity to the switchback phenomenon. In \cite{SusskindACLong} this behaviour was also found for the AC proposal. Further results along these lines have been studied in \cite{VaidyaI, VaidyaII} for both prescriptions.

Another obvious test for holographic complexity is its application to richer black hole solutions, such as the charged or rotating ones. Reissner-N\"ordstrom black holes were studied in \cite{SusskindNotEnough, SusskindACLong} in both prescriptions, whereas an analytic treatment of the rotating ones has only been successfully achieved within AC in the simple three-dimensional case. As there is not much intuition about the influence of charges in the quantum mechanical definition of complexity, results within these gravitational setups are not particularly illuminating. In both cases, however, holographic complexity exhibits a slower computation rate, a fact that agrees with the naive intuition that conserved charges provide obstacles to fast complexification as a consequence of energy being tied up in non-computing degrees of freedom.

In \cite{MyersFormation, Myerstdep}, attention was paid to the early stages of holographic complexity before the stabiliziation of its late-time behaviour \eqref{CVrate} and \eqref{CArate}. In particular, the concept of \emph{complexity of formation} was outlined in \cite{MyersFormation} with the aim of quantifying the cost of preparing non trivially entangled states as compared with the vacuum. In \cite{Myerstdep} a thorough study of the early evolution of complexity was performed, signalling a number of discrepancies between the VC and AC proposals and  stressing the importance in these regimes of some formal choices in the definition of the WDW action.

Trying to deepen on the foundations of complexity, a series of works \cite{Susskind2law, MyersFirstLaw1,  MyersFirstLaw2} have speculated about possible relations between complexity and thermodynamics, suggesting  the existence of universal laws analogous to those of thermodynamics. Roughly, the idea is that the definition of complexity by means of a coarse-graining in Hilbert space resembles many aspects of the definition of entropy by a coarse-graining of phase space, a similarity that could point towards a duality between the two quantities. 

As a consequence of these ideas, it was proposed in \cite{SusskindThingsFall, SusskindNewton} that the natural tendency of complexity to increase (as suggested by a \emph{second law of complexity} \cite{Susskind2law}) might have a deep connection to the clumping nature of gravity in the context of AdS/CFT. More precisely, some notion of infall momentum can be defined in the bulk in such a way that the dynamics enjoyed by this quantity matches that of complexity growth. Remarkaby, the VC proposal provides a concrete framework in which this notion is realized in a very precise manner. In particular, as pointed out in \cite{PCfirst, PCZhao} the maximal foliation of spacetime that is built in VC defines a natural momentum component $\CP^\mu = -T^{\mu \nu}N^\Sigma_\nu$ with $T^{\mu \nu}$ the bulk energy-momentum tensor and $N^\Sigma_\nu$ standing for the normal vector at each point of the maximal slices $\Sigma$. After defining a suitable `infall vector' $C^\mu$, a general expression of the form 
\begin{equation}
\dfrac{\dd \CC }{\dd t} = - \int_\Sigma N_\Sigma^\mu \, T_{\mu \nu}\, C^\nu\, ,
\end{equation}
can be shown to hold for some sufficiently symmetric systems.

In \cite{PCProof} this statement was proven to hold exactly for asymptotically AdS$_3$ spacetimes and general spherically symmetric solutions in arbitrary dimensions. Curiously, this Momentum/Complexity duality appears within this framework as a consequence of the momentum constraint of general relativity.

\begin{figure}[h]
$$\includegraphics[width=7cm]{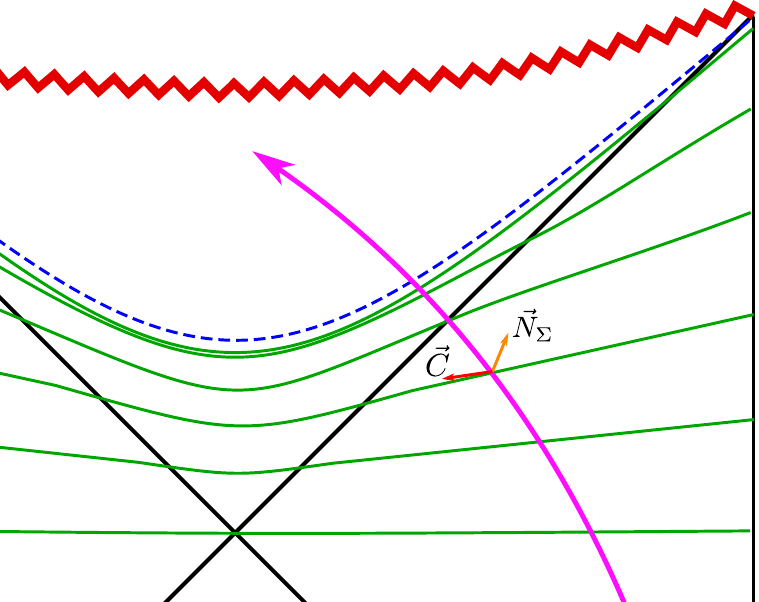} $$   
\begin{center}
\caption{\emph{The maximal volume foliation of spacetime defines naturally a notion of momentum for any probe particle moving on that spacetime. A correct choice of an infall vector $C^\mu$ guarantees a precise relation of this momentum and the dynamics of VC complexity.}}
\label{fig:PC}
 \end{center}
\end{figure} 
\part{\sc{Degenerate AC\faBolt VC}}
\chapter{Holographic complexity of exotic systems}
\label{ch:capitulo3}


\section{Introduction}

As presented in the previous chapter, the eternal black hole has conformed the main inspiration and principal testing ground for the AC and VC conjectures, showing a high level of agreement in some of the standard tests, both between the two prescriptions and with the expected insights from quantum many-body physics. In this chapter we aim to take  both holographic prescriptions out of the realm of the vanilla Schwarzschild solution in order to test the conjectures in richer scenarios while keeping the thermality condition. The target of the analysis is twofold: first, we seek to test the robustness of the standard features of holographic complexity when applied to exotic systems, as well as the limitations of some perturbative expansions to capture that feature. Second, we will search for states that present significant disagreement between the two prescriptions, arguing for those systems as conforming a key lab test to confront both conjectures and  settle down the correct one.

More specifically, we will study various situations possessing `exotic' thermodynamics, meaning that that we seek to account for the effect on complexity of a large ground state degeneracy in the spectrum. As a result, we find new IR divergent contributions to the complexity in the volume prescription, contrasting with the results of action complexity, which seems to ignore such effects. Furthermore, we find a common feature in the complexity dynamics of some of these systems, i.e. that the picture of a linear complexity growth is  modified by an strictly constant complexity, representing an instance of a `holographic non-computer'. This behaviour appears for some extremal systems both in the AC and VC prescriptions but it is found to hold for particular finite temperature states only for the latter, signalling a major difference between the two proposals. Finally, we explore more deeply the nature of other `non computers'  that can be obtained from ordinary thermal systems when studied within a mean-field approximation. 

In order to work within controlled AdS/CFT scenarios, we will avoid introducing $U(1)$ charges in the bulk setup. In principle, either electrically or magnetically charged solutions of the Reissner-Nordstrom type exhibit the kind of `exotic' thermodynamics that we seek in this chapter but they do so at the expense of showing perturbative instabilities. In particular, string theory embeddings of these finite-density systems show the that near extremal black holes are unstable to the condensation of clouds of classical charged hair. For this reason and with the aim of studying the simplest possible scenarios, we will work within pure gravity solutions all along.

In the section \ref{sec:hyperbolic} of this chapter we consider the example of hyperbolic black holes as an instance of a pure gravity solution enjoying an extremal regime. Within this solution, we observe the appearance of a IR divergent contribution to the volume, coming from the development of a long AdS$_2$ throat as extremality is approached. Regarding the dynamics we find again discrepancies between the two holographic prescriptions signaled by the occurrence of a \textit{non-computer} behavior for the finite range of temperatures that span its near-extremal regime \cite{MyersFormation}.

Next, we devote section \ref{sec:noncomp} to a different class of non-computing behavior,  exhibited by higher dimensional black holes. As showed in \cite{BrownSusskindAction}, black holes in four or higher spacetime dimensions enjoy a period of constant Action-Complexity at early times, postponing the usual linear growth after some delay lapse which depends on physical properties of the black hole. As we will show, this behavior gets enhanced as the dimension grows and can lead to an eternal non-computer system at leading order in a large $d$ expansion. The large-$d$ expansion of General Relativity has illuminated a number of classical dynamical regimes in various black-hole systems (cf. \cite{EmparanlargeD, Minwalla}). While its status at the quantum level is rather unclear, we find it interesting that a non-trivial statement can be made for such highly quantum properties as the computational complexity of black holes.

\section{Hyperbolic black holes: a degenerate system in pure gravity}
\label{sec:hyperbolic}
\noindent

The standard test offered by the analysis of eternal AdS black holes gives the usual qualitatively similar results for the late time growth of AC and VC complexity. i.e.

\begin{equation}
\label{rate}
 {\dd \CC_V \over \dd t} \sim T\,S\;, \hspace{2.5cm}  {\dd \CC_A \over \dd t} = 2M\, ,
\end{equation}

for $t  \gg T^{-1}$, where $T$ is the temperature and $S$ the entropy of the eternal black hole. The calculation of \eqref{rate} within the eternal black hole geometry assumes implicitly that $T$ is sufficiently large to neglect finite-size effects. For standard AdS black holes, this means that $T\ell \gg 1$ where $\ell$  has the dual interpretation as the AdS$_{d+1}$ radius of curvature in the bulk and also the radius of the $(d-1)$-sphere where the $d$-dimensional CFTs are defined.   In this particular case the $T\,\ell \ll 1$ limit is on the other side of the Hawking--Page transition, and the $\CO(1/G)$ contribution to the complexity must be calculated in the vacuum AdS manifold, giving no contribution at this order to \eqref{rate}.

Alternatively, we can remove finite-size effects by working with black branes of non-compact horizon, where all integrated quantities, such as entropy and complexity, are extensive  in the CFT volume. In this case  we  implicitly refer to a  `complexity density'. Black-brane metrics have the general form 
\begin{equation}
\label{bbrane}
\dd s^2 = -f(r) \,\dd t^2 + {\dd r^2 \over f(r)} + {r^2} \,\dd \Sigma_{d-1}^2 \;,
\end{equation}
where $\dd  \Sigma_{d-1}$ stands for the spatial $(d-1)$-dimensional boundary metric and we measure length in units of the AdS curvature radius $\ell =1$. Solutions of this ansatz satisfying vacuum AdS asymptotics can be found in the form
\begin{equation}
\label{efes}
f(r) = k + {r^2 \over \ell^2} -{\mu \over r^{d-2}}
\;.
\end{equation}
with $k=0,-1$ respectively for flat and hyperbolic boundary metrics. For large enough values of the parameter $\mu$, these solutions exhibit a a non-degenerate horizon at $r=r_h$, and the thermodynamic quantities can be readily calculated to be
\begin{equation}
\label{physicalquantities}
T= \dfrac{dr_h^2+k(d-2)}{4\pi r_h }, \hspace{1.5cm} S = \dfrac{V}{4G} \,r_h^{d-1}, \hspace{1.5cm} M = \dfrac{V (d-1)}{16 \pi G} \mu,
\end{equation}
with $V$ the volume factor in $d-1$ dimensions. For the flat case, the usual UV/IR relation $r_h \sim T$ holds down to zero temperature, with the entropy vanishing as $T^{d-1}$. In the second case, the solution represents a thermal state of a CFT which lives on a hyperboloid of curvature radius $1$. Alternatively, the maximally extended geometry can be interpreted, following \cite{Maldacenaeternal},  as dual to a thermofield double state on the direct product of two copies of the CFT on respective hyperboloids. This system, known as hyperbolic (or topological) black hole, has exotic properties at low temperatures \cite{roberto}, in particular a large ground state degeneracy, signalling a gross violation of the third law of thermodynamics 
\begin{equation}
 \label{zen}
  \lim_{T\to 0} S \longrightarrow S_0 = N_* \,V\,\ell^{1-d}\;,
 \end{equation}
 where  $N_* \sim \ell^{d-1} /G \gg 1$ is the effective number of `species' in the strongly-coupled CFT.

In the following we study some properties of  the holographic complexity, as defined by the AC and VC {\it ansätze} \eqref{CVansatz} and \eqref{CAansatz},  in such degenerate systems. In particular, we shall consider the concrete case of thermofield double states for pairs of CFTs on hyperboloids, as defined by eternal AdS hyperbolic black holes \cite{Lemos:1994xp,Lemos:1994fn,Aminneborg:1996iz,Mann:1996gj,Cai:1996eg,Brill:1997mf,Vanzo:1997gw,Birmingham:1998nr,Birmingham:2001dq,Birmingham:2007yv}. We begin in section \ref{sec:highT} with the analysis at high temperatures and continue in section \ref{sec:lowT} with the study of the low-temperature near extremal regimes.

As the calculation of VC complexity is not fully tractable within analytic methods, we include an appendix in which we develop a piece-wise approximation to the extremalization problem. Along this section, we take the  scale factor of \eqref{CVansatz}   equal to the AdS curvature radius $\ell =  \ell_{AdS}$, keeping in mind that the correct overall normalization factor will only be fixed by means of a precise calculation on the CFT side.
  
\subsection{High temperatures}
\label{sec:highT}
\noindent 

In the high temperature (  $T\gg 1$  ) phase of topological black holes we have  $f(r) \approx r^2 - r_h^d /r^{d-2}$ with $r_h\approx 4\pi T /d$, resembling the same functional form of the thermodynamics of large spherical black holes and implying therefore a very similar complexity phenomenology both in AC and VC prescriptions. Precisely, in evaluating the volume of $\Sigma_t$, we find the standard result \eqref{rate} for the long-time growth rate, with $S\sim N_* V T^{d-1}$ the high temperature entropy of the large-$N_*$ CFT on the hyperboloid. 

As detailed in appendix \ref{sec:TBHappendix}, a splitting of the interior, Rindler and exterior region of the extremal slice $\Sigma_t =\Sigma_{\rm WH} \cup \Sigma_{\rm R} \cup \Sigma_{\rm UV}$ allows us to approximate all contributions to the VC complexity. At $t=0$, the wormhole has vanishing length  on the bifurcation sphere and the total complexity is given only the two qualitatively different contributions. First we have the  UV contribution of $\Sigma_{\rm UV}$, 
\begin{equation}
\label{uvc}
{ \CC_V}[\Sigma_{\rm UV}] = 2  \dfrac{V}{  G} \int_{r_R}^{r_\Lambda} {\dd r\,r^{d-1} \over \sqrt{f(r)}} \sim N_* \,V \, \left(\Lambda^{d-1} - T^{d-1} \right)\;,
\end{equation}

where we can as well neglect the $T$-dependent term coming from the lower limit of the integral, since we are assuming $\Lambda \gg T$.
Second, we have a threshold contribution coming form $\Sigma_{\rm R}$:
\begin{equation}
\label{ritz}
{ \CC_V}[\Sigma_{\rm R}]\Big |_{t=0} = 2  {V \over   G} \int_{r_0}^{r_R} {\dd r\,r^{d-1} \over \sqrt{f(r)}} \sim N_* \, V\,T^{d-1} \sim S\;,
\end{equation}

where we have used the Rindler approximation to the metric to estimate the integral in order of magnitude. In this expression, as well as others that follow, the matching ambiguity coming from the precise location of $r_R$ and the various errors from the piecewise matchings  of $\Sigma_t$ can be estimated by shifting $r_R$ an amount of $\CO(1)$, resulting in  an additive ambiguity of order $S$  for $\CC_{\rm R}$.  

Finally, the wormhole contribution of $\Sigma_{\rm WH}$ can be approximated at large times by the volume of a cylinder of constant radius $r_m$ (see details on appendix \ref{sec:TBHappendix}), yielding the usual contribution

\begin{equation}
{ \CC_V}[\Sigma_{\rm WH}] \simeq ST\, t\, .
\end{equation}

The UV contribution to the complexity is constant in time, which allows us to define a subtracted complexity  $\Delta \CC_V (t) = \CC_V[\Sigma_t] - \CC_{V}[\Sigma_{\rm UV}]$ which is finite and takes only into account the IR degrees of freedom of the black hole interior. The behaviour at high temperatures for such quantity is 
\begin{equation}
\label{cit}
\Delta \CC_V(t) =\CO(S) \;\;{\rm for}\;\;t < T^{-1}\;, \qquad \Delta \CC_V(t) \sim S\,T\,t\;\;{\rm for} \;\;t\gg T^{-1}
\;.
\end{equation}
in a very similar fashion as the benchmark spherical black hole. Similarly, the analysis of this system in the AC prescription can be performed in detail \cite{MyersFormation, Myerstdep}, showing that the standard calculation goes trough every step, and finding the usual late time asymptotics for the regularized complexity \footnote{The structure of UV divergencies in AC and VC conjectures present some discrepancies as they are highly sensitive to the precise prescriptions for the null boundary contributions. See \cite{Ross} for a careful analysis. In the following, we will ignore those constant UV contributions. } 
\begin{equation}
\dfrac{\dd I}{\dd t} = 2M \approx 2\,\dfrac{d-1}{d-2}\, TS \, .
\end{equation}

\subsection{Cold and Frozen Hyperbolic Horizons: a first instance of non-computers} 
\label{sec:lowT}

As seen from the previous section, nothing particularly interesting happens for hot hyperbolic horizons, and if we consider physical solutions to lie within $M>0$ this is all there is. Careful analysis of the thermodynamics, however, shows that finite temperature states exist with have negative mass respect to this vacuum for the parametric region
\begin{equation}
-\dfrac{2 (d-2)^{\frac{d}{2}-1}}{d^{\frac{d}{2}} } \leq \mu  <0 \, .
\end{equation}
 In the  $T\rightarrow 0$ limit, a pure gravitational extremal black hole state is reached, with the horizon dropping to a minimum radius
\begin{equation}
\label{minrad}
r_{\rm ext} = \sqrt{d-2 \over d}
\;,
\end{equation}
and corresponding to the negative mass parameter
\begin{equation}
\label{muc}
M_{\rm ext} = -{2 \over d} \left({d-2 \over d}\right)^{d-2 \over 2}
\;.
\end{equation}
In this extremal case the function $f(r)$ develops a double zero at the horizon, so that we may write in the vicinity of $r=r_{\rm ext}$ 
\begin{equation}
f(r)_{T=0} = d\cdot (r-r_{\rm ext})^2 + \dots
\end{equation}

where the dots stand for terms of order $(r-r_{\rm ext})^3$ or higher. This suggests that we can parametrize the low-temperature geometries in terms of the radial variable $\rho= r-r_{\rm ext}$. Then, to first non-trivial order in $\rho$ and $\rho_h = r_h - r_{\rm ext}$ we have 
\begin{equation}
f(r) \approx d\cdot (\rho^2 - \rho_h^2) + \dots\;, 
\end{equation}

an approximation good for $\rho_h \leq \rho \ll r_{\rm ext}$. The low-temperature horizon sits at $\rho=\rho_h \approx 2\pi T /d  +\CO(T^2)$.

\begin{figure}[h]
\begin{center}
\includegraphics[height=10cm]{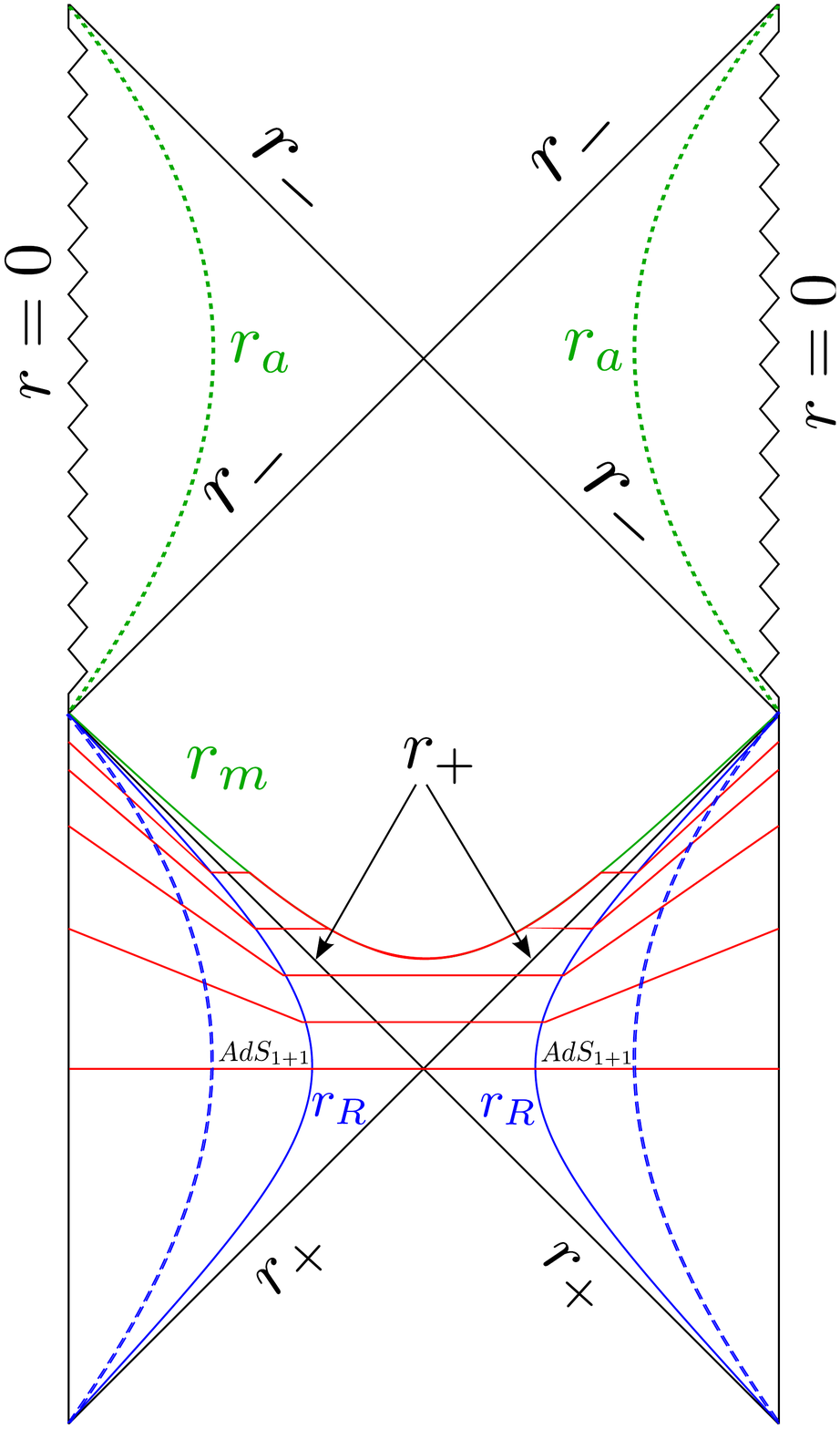}
\includegraphics[height=8cm]{HyperbolicPenrose.pdf}
\end{center}
\caption{\emph{The global structure of the hyperbolic black hole interior is similar to that of charged black holes. The inner horizon $r_-$ separates the space-like and time-like character of fixed-$r$ surfaces in the interior. The intermediate AdS$_2$ region (between the dashed and the continuos blue lines) develops at low temperatures.  }  }
\label{fig:nearextremal}
\end{figure}

As a consequence of this structure in the blackening factor, two usual aspects of near-extremal solutions become patent. The fist one is the fact that as the temperature crosses the threshold $T=1/2\pi$, a transition occurs in the structure of the conformal diagram, whose topology resembles that of a Reissner-N\"ordstrom black hole (see Figure \ref{fig:nearextremal}). As a consequence of this topology change, the complexity of these states, characterized by a temperature $0<T<1/2\pi$, is not given by the standard \eqref{CArate} but will require a separate calculation. The second feature, also known in near-extremal charged black holes is the modification of the geometry in the vicinity of the horizon trough the development of a large AdS$_2$ `throat'. More specifically the region $\rho_h \ll \rho \ll r_{\rm ext}$, arising at very low temperatures,  is approximately described by ${\rm AdS}_{1+1} \times {\rm H}_{d-1}$, i.e. effectively decoupling the hyperbolic `space' from an asymptotic AdS$_2$ factor.  The corresponding curvature radii are given by 
\begin{equation}
\ell_{{\rm AdS}_2} = {1\over \sqrt{d}}\;, \qquad \ell_{{\rm H}_{\rm IR}} = \sqrt{d-2 \over d}\;,
\end{equation}

measured in units $\ell =1$. We will refer to this factorized geometry as the CQM region, to signify the formal AdS$_2$/CFT$_1$ duality to some hypothetical  Conformal Quantum Mechanical (CQM) system that would describe the deep infrared regime.

\begin{figure}[h]
\begin{center}
\hspace{2cm}
\includegraphics[height=8cm]{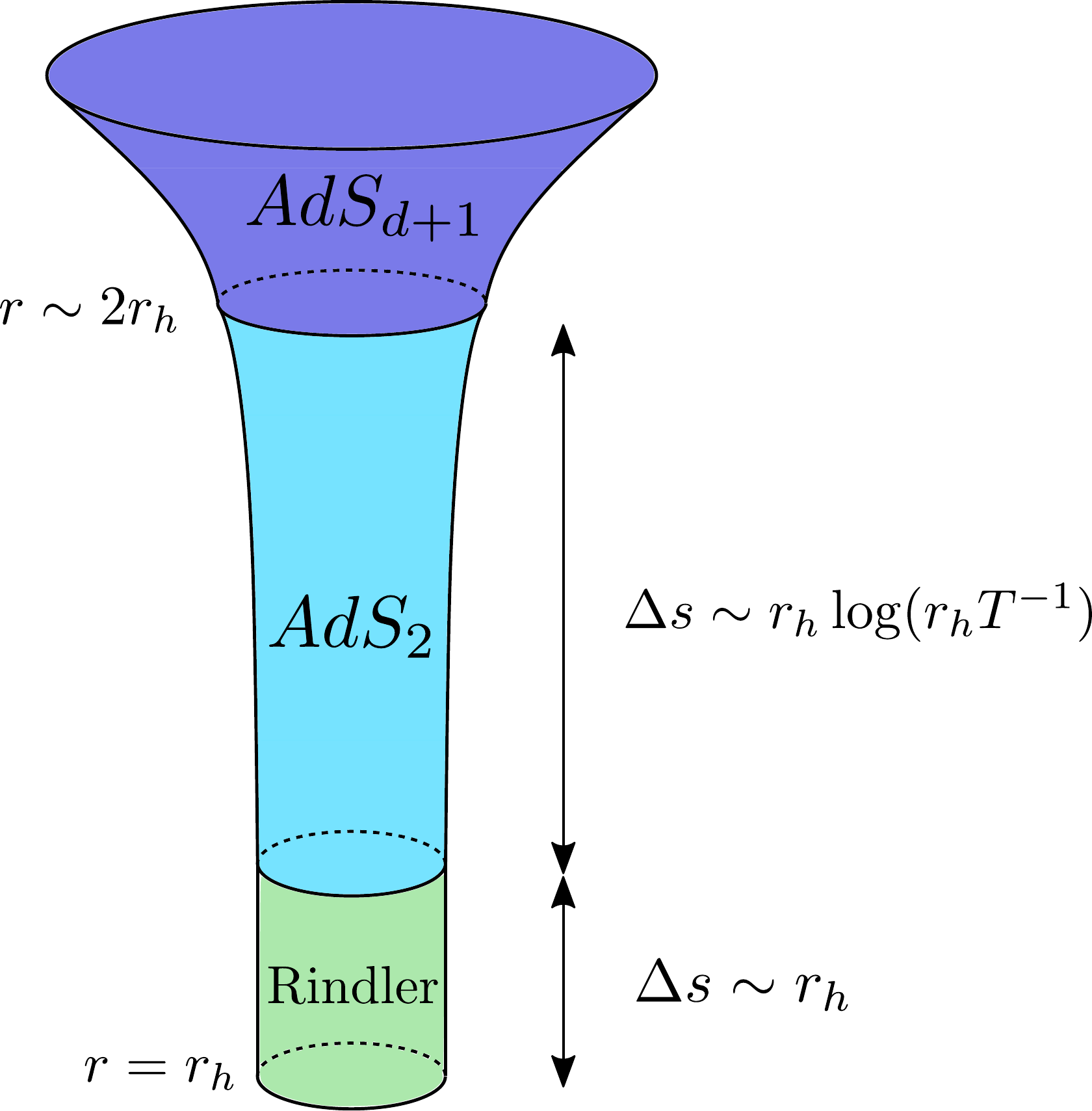}
\end{center}
\caption{\emph{General structure of the near-horizon geometry of a near-extremal black hole. As the temperature approaches the $T \rightarrow 0$ limit, an AdS$_2$ throat develops at finite coordinate distance from the horizon. The proper length of the throat however diverges logarithmically.}  }
\label{fig:throat}
\end{figure}
 
In light of these remarks we will proceed now to the calculation of AC and VC quantities in this cold phase of the hyperbolic black hole.

In evaluating the volume of extremal surfaces, we must distinguish the qualitatively different regions of the bulk geometry, namely for $r\gg r_{\rm ext}$ we have an approximately AdS$_{d+1}$ geometry with a time slicing adapted to the ${\bf R} \times {\rm H}_{d-1}$ CFT frame, and for $\rho_h \ll \rho\ll r_{\rm ext}$ 
we have a ${\rm AdS}_{1+1} \times {\rm H}_{d-1}$ geometry. Accordingly, the codimension-one  surfaces split as  (see Figure \ref{fig:nearextremal}) 
\begin{equation}
\Sigma_t \sim \Sigma_{\rm WH} \cup \Sigma_{\rm R} \cup \Sigma_{\rm CQM} \cup \Sigma_{\rm UV}
\;.
\end{equation}

Here $\Sigma_{\rm UV}$ extends for $r\gg r_{\rm ext}$. The new portion extending   along the AdS$_{1+1}$ radial slice $\rho_h \ll \rho\ll r_{\rm ext}$
will be denoted  $\Sigma_{\rm CQM}$.  Finally, in the deep infrared region we have the Rindler portion $\Sigma_{\rm R}$ given by the interval $\rho_h < \rho < \rho_R$, with $\rho_R$ an $\CO(1)$ multiple of $\rho_h$.  In the interior we find  the wormhole portion $\Sigma_{\rm WH}$ along $r=r_m$. For all partitions except $\Sigma_{\rm UV}$ we can regard the hyperbolic H$_{d-1}$ factor as an spectator. 

We first discuss the situation at $t=0$, where $\Sigma_{\rm WH}$ is absent. The contribution from $\Sigma_{\rm UV}$ is the standard 
$N_* V \Lambda^{d-1}$. The contribution from $\Sigma_{\rm CQM}$ is interesting because the complexity picks equal contributions for every region of the CQM region,
\begin{equation}
\label{cqm}
C_{\rm CQM} (0) \approx 2 {r_c^{d-1} V \over G} \int_{\rho_R}^{r_{\rm ext}} {1\over \sqrt{d}}{d\rho \over  \rho}  = {8 \over \sqrt{d}} \,S_0 \, \log(1/T)\;,
\end{equation}

leading to a logarithm with a characteristic coefficient controlled by the zero-temperature entropy of the system. The $\CO(1)$ ambiguities at the endpoints of the integral amount to an additive error of order $S_0$. Notice however that the coefficient of the logarithm, given by $8 S_0 /\sqrt{d}$,  is robust in the low $T$ limit.

Finally, the Rindler contribution coming from $\Sigma_{\rm R}$ is of order
\begin{equation}
\label{rtzero}
C_{\rm R} (0) \approx 2 {r_c^{d-1} V \over G} \int_{\rho_0}^{\rho_R} {1 \over \sqrt{d}} {d\rho \over \sqrt{\rho^2 - \rho_0^2} }\sim S_0\;,
\end{equation}

where the matching errors are also of order $S_0$.

As before, the exterior surfaces  in both the CQM and UV regions have a time-independent volume. Hence the time development of the complexity proceeds by the gradual deformation of $\Sigma_{\rm R}$ into $\Sigma_{\rm WH} \cup \Sigma_{\rm R}$. As can be seen from figure 2, the volume of $\Sigma_{\rm R}$ is negligible at large times, whereas that of $\Sigma_{\rm WH}$ is controlled by the local maximum of $r^{d-1} \sqrt{|f(r)|}$. Since we are working at very low temperatures,  it is tempting to pick the $\CO(1)$   radius $r=r_a$ which maximizes  the $T=0$ function
\begin{equation}
r^{d-1} \sqrt{\left|1-r^2 + {M_{\rm ext} \over r^{d-2}}\right|}\;
.
\end{equation}

However, there is a subtlety. This $\CO(1)$ maximum at $r=r_a$ survives for small but non-zero $T$, but in fact we have $f(r_a) >0$, implying that $r=r_a$ is a time-like surface (shown in figure 2). It turns out that there is a small $T$-dependent local maximum of $r^{d-1} \sqrt{|f(r)|}$, with height of $\CO(T)$,  within the interior Rindler region (see figure 3). The corresponding $r=r_m$ surface is space-like, since  $f(r_m) <0$. In this regime the function to be maximized  is approximately given by
\begin{equation}
(r_{\rm ext} + \rho)^{d-1} \sqrt{d|\rho_h^2 - \rho^2|}\;,
\end{equation}

which is maximized close to $\rho_m =0$, so that the WH surface is given by $\rho \approx 0$. Again, it is roughly the symmetrical  of  the $\rho=\rho_R$ surface by a reflection with respect to the horizon, implying that $t_{\rm exit} \sim  t$ and thus  a 'wormhole length' of order $\Delta t \approx 2t$. The resulting   large $t$ complexity is
\begin{equation}
\label{whiii}
C_{\rm WH} (t) =2\,t\, \, {V r_c^{d-1} \over G} \sqrt{d} \,\rho_0 \approx  {16\pi T \over \sqrt{d}} \,S_0 \, t\;.
\end{equation}

 \begin{figure}[h]
\begin{center}
\includegraphics[height=6cm]{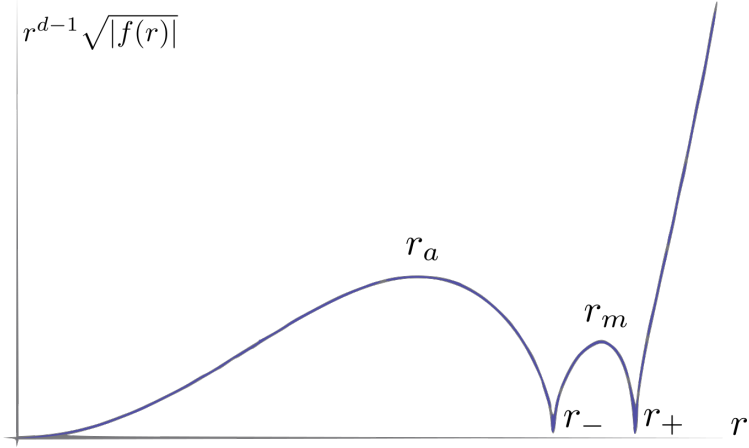}
\end{center}
\caption{\emph{The function $r^{d-1} \sqrt{|f(r)|}$ at low temperatures, showing the small maximum of $\CO(T)$ at $r=r_m$ in the near-horizon (Rindler) region, and the $\CO(1)$ maximum at $r=r_a$. In the high-$T$ regime the Rindler bump grows larger than the local maximum at $r=r_a$, which becomes a small detail near the singularity.     }  }
\label{fig:fofrnearextremal}
\end{figure}

Grouping together these results and restoring the curvature radius $\ell$, we find a total low-temperature subtracted complexity  given by
\begin{equation}
\label{fins}
\Delta C(t) \approx {8\over \sqrt{d}} \,S_0 \, \log (1/\ell \,T) \;,\;{\rm for} \;\;t< T^{-1}\;,
\end{equation}

at small times and 
\begin{equation}
\label{lod}
\Delta C(t) \approx {8 \over \sqrt{d}} \,S_0 \, \log (1/\ell \,T)
+ {16\pi  \over \sqrt{d}} \,S_0 \, T\,t \;\;{\rm for}\;\;t\gg T^{-1}\;.
\end{equation}

at long times. It should be noted that, while we have kept the coefficient found in \eqref{whiii}, it must be understood as an estimate with 
$\CO(S_0)$ additive ambiguities, unlike the coefficient of the time-independent logarithmic term, which is a robust prediction for the strongly coupled CFT.

Turning now our attention to the AC complexity, we find two major differences with VC. The first one is the fact that the action is not sensible to the infinitely long AdS$_2$ throat, leading a finite substracted complexity $\Delta \CC(0) \sim S_0$ in the $T \rightarrow 0 $ limit (cf. \cite{MyersFormation}). The second one, noticeable from the conformal diagram in Fig \ref{fig:nearextremal}, is the absence of the boundary YGH term, a fact that in turn entails a rather exotic behaviour, i.e. a vanishing late-time complexity growth.

\begin{figure}[h]
\begin{center}
\includegraphics[height=7cm]{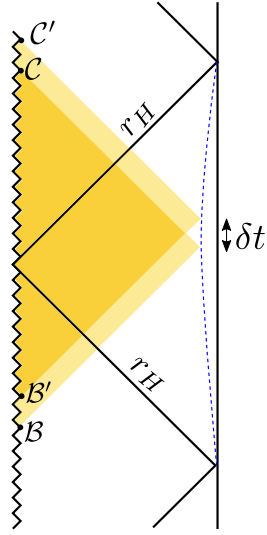}
\end{center}
\caption{\emph{WdW patch for an extremal black hole. The bulk volume and boundary contributions are conserved by the isometries of the spacetime, whereas the joint piece vanishes due to the sphere shrinking at the singularity. }}
\label{fig:extremal}
\end{figure}

In particular, the computation rate of near-extremal hyperbolic black holes only gets contributions from the bulk and null boundary contributions to the action. Computing the difference between $\CW_1$ and $\CW_2$ in figure \ref{fig:nearextremal} we can get the contribution from the Einstein-Hilbert term
\begin{equation}
\label{coldbulk}
I_{\CW_1}-I_{\CW_2} = -\dfrac{V}{8 \pi G}(r_\CB^d-r_\CC^d) \delta t\, .
\end{equation}
The contribution of the joints is given by
\begin{eqnarray}
\label{coldjoints}
 I_{\CB \CB'} +  I_{\CC \CC'}= \delta t \dfrac{V}{8 \pi G} \left[ 2r^{d}+ (d-1)r^{d-2}f(r)\log\left(f(r) \right) \right]^{r_{\CB}}_{r_{\CC}}\, ,
\end{eqnarray}
and the null boundary counterterms
\begin{eqnarray}
\dfrac{\dd I_{\Theta}}{\dd t}&=& -\dfrac{V (d-1)}{8\pi G} \left[r^{d-2} f(r) \log \left( \dfrac{r}{(d-1) \LL} \right) \right]^{r_{\CB}}_{r_{\CC}} \, .
\end{eqnarray}
Adding up the three contributions we get that the total rate is given by
\begin{equation}
\label{coldi}
\dfrac{\dd I}{\dd t}=\dfrac{V(d-1)}{8 \pi G} \left[r^{d-2} f(r)\log\left(\dfrac{(d-1)f(r) \LL}{r} \right) \right]^{r_{\CB}}_{r_{\CC}}
\;,\end{equation}
which indeed clearly vanishes in the late time limit due to the vanishing of $f(r)$ as $r_{\CB, \CC}  \rightarrow r_{\pm}$. As we see, the above cancellation holds independently of the temperature of the black hole, as long as it lies within the near-extremal regime $0\leq T \leq 1/2\pi$ smoothly recovering the strictly extremal result, in which the steadiness of complexity is exactly enforced by the symmetries of the WdW patch (see Figure \ref{fig:extremal}) in the same fashion as in Reissner-N\"ordstrom extremal black holes. In this very last instance such result is not very surprising from the field theory point of view. After all, extremal black holes have zero temperature and we expect every property to be static in such states. Nonetheless, extremal black holes provide the first non-trivial state with vanishing computation rate, and might constitute a very  relevant example in the elucidation  of holographic quantum complexity as a microscopic quantity in the CFT side.

More surprisingly, cold hyperbolic black holes provide an example of a finite temperature state enjoying a non-computer behaviour in AC. This contrasts with the result obtained within the VC proposal in which the linear growth behaviour held also in this regime.

Despite the finite-temperature nature of these solutions, it must be said  that such states are unlikely to be  stable,  but rather should be interpreted as highly degenerate unstable or perhaps metastable systems. Evidence in this direction comes from the embedding of these solutions into fully fledged string theory systems, such as stacks of type IIB D3-branes, yielding a canonical example of AdS$_5 \times {\bf S}^5$ duality with maximally supersymmetric Yang--Mills theory on an spatial hyperboloid ${\bf H}^{3}$. A marginally tachyonic scalar, saturating the AdS$_5$  BF bound with $m^2 = -4$ and corresponding to BPS-protected scalar mass operators with $\Delta =2$, will have zero modes that actually violate the BF bound in the emerging AdS$_{1+1} \times {\bf H}^3 \times {\bf S}^5$ 
geometry of the near-horizon region of cold hyperbolic black holes. In this case the AdS$_{1+1}$ radius of curvature is $\ell' = 1/2$ and the corresponding BF bound $m'^{\,2} \geq -1/4\ell'^{\,2} = -1$. Such systems are therefore expected to undergo tachyonic instabilities. Even if the perturbative instabilities are somehow checked out,  \cite{BarbonTopBH} shows that the cold branch of hyperbolic black holes is likely unstable to non-perturbative D3-brane fragmentation processes.  

It is interesting to notice that the difficulties associated to the emergence of an approximate AdS$_{1+1}$ geometry in the near-horizon region are also responsible for the mismatch between the AC and VC ans\"atze observed for these solutions, perhaps hinting towards a failure of a correct definition of holographic complexity for the effective CQM description of the throat \cite{HugoJTcomplexity, SusskindJTcomplexity}. On general grounds, the precise nature of AdS$_2$/CQM duality has recently been put into question \cite{MaldacenaStanfordSYK, StanfordJT, StanfordWittenJT}, suggesting perhaps a missing piece in the puzzle.

\section{Black holes in a large number of dimensions: a hot non-computer}
\label{sec:noncomp}

Away from the late time approximation, higher dimensional black holes are known to exhibit a delay in their AC computation rate for any $d\geq3$ \cite{BrownSusskindAction}. This phenomenon, absent in the VC proposal, arises as a consequence of an extra symmetry of the WdW patch at early times which postpones the start of the complexity growth to a later time $t_{\CC}$. In particular, as the spacetime dimension gets bigger, the past and future singularities bow into the Penrose diagram, effectively splitting the time symmetry in two separate left and right time-shift symmetries,  $t_{L,R} \rightarrow  t_{L,R} + c_{L,R}$, for every WdW patch touching both past and future singularities. As the past boundaries leave the singularity, this symmetry breaks down to the smaller boost symmetry $c_L + c_R =0$ and the black hole starts computing. In the following we will calculate the value of these delays and explore its behaviour respect to the spacetime dimension.

\subsection{Computation delays for $d \geq 3$ black holes}
\label{sectiondelays}

We begin by recalling the form of the metric for spherical AdS$_{d+1}$  black holes
\begin{equation}
\label{otram}
 \dd s^2  = -f(r) \dd t^2 + \dfrac{1}{f(r)}\dd r^2 + r^2\,\dd \Omega_{d-1}^2\, ,
\end{equation}
where $\dd \Omega_{d-1}^2$ is the volume form of the unit ${\bf S}^{d-1}$ sphere with volume
\begin{equation}
\label{sphevol}
V_\Omega = {2 \pi^{d\over 2} \over \Gamma(d/2)}\;,
\end{equation}
and the warping function is given by
\begin{equation}
\label{warp}
f(r) = 1 + {r^2 \over \ell^2 } - \left({r_h\over r}\right)^{d-2}\, \left(1+ {r_h^2 \over \ell^2} \right)\;,
\end{equation}
after we have restored the  dependence on $\ell$, the curvature radius of AdS. 
The basic thermodynamic quantities are given by 
\begin{equation}
\label{physicalquantitie}
T= \dfrac{d\,r_h^2+(d-2) \ell^2 }{4\pi r_h \, \ell^2}, \hspace{1.3cm} S = \dfrac{V_\Omega}{4G} \,r_h^{d-1}, \hspace{1.3cm} M = \dfrac{V_\Omega (d-1)}{16 \pi G} \left(r_h^{d-2} + {r_h^d \over \ell^2}\right)\;.
\end{equation}

In order to study the delay, it is necessary to construct the Kruskal extension for general dimensions. 
The first step will be to define the tortoise coordinate, given by
\begin{equation}
\label{deftortoise}
r_*(r) = \int\limits^{r}_0 \dfrac{\dd r}{f(r)} +C\, ,
\end{equation}
where the constant $C$ is chosen so that the coordinate is real in the exterior region. Analytic expressions for this integral cannot be found in general. For our purposes however, it will suffice to find the asymptotic limit $r_*(\infty)$, whose value will be crucial in the construction of the conformal diagram. In terms of this coordinate, the Kruskal extension is defined in the lightcone coordinates as follows

\begin{eqnarray}
uv &=& -e^{4 \pi T {r}_*(r)}\, , \\
\label{uoverv}
\dfrac{u}{v}&=&-e^{4 \pi T t}\, .
\end{eqnarray}
With this choice, the singularity will be located at $uv=1$ for any dimension, whereas the AdS boundary is located at

\begin{equation}
uv = e^{4 \pi T {r}_*(\infty)}\,.
\end{equation}
The value of ${r}_*(\infty)$ will in general depend both on the dimension and physical parameters of the solution, with qualitatively different behaviours depending on the relative size of the black hole respect to the curvature radius.

\subsubsection{Large AdS black holes}

In the large black hole limit\footnote{For flat ($k=0$) AdS Black holes, this condition is not needed and the result holds for any hierarchy of $r_h$ and $\ell$.} ($r_h \gg \ell$) we might approximate

\begin{equation}
f(r) = 1 + \dfrac{r^2}{\ell^2} - \left( \dfrac{r_h}{r} \right)^{d-2} \left(1+ \dfrac{r_h^2}{\ell^2} \right) \simeq \dfrac{r^2}{\ell^2} - \dfrac{r_h^d}{\ell^2 r^{d-2}}\, ,
\end{equation}
and we can calculate the integral in \eqref{deftortoise} analytically 

\begin{equation}
\int\limits^{r}_{0} \dfrac{\dd r}{f(r)} = \frac{\ell^2}{r} \left[_2F_1\left(1,-\frac{1}{d};1-\frac{1}{d};\left(\frac{r}{r_h } \right)^d\right)-1\right]\,,
\end{equation}
which forces us to choose $C= -i \frac{\pi}{d}$. Using the asymptotic expansions for the hypergeometric function at large $r/r_h$ we get (cf. \cite{Kim, Myerstdep})

\begin{equation}
\lim\limits_{r \rightarrow \infty} r_{*}(r) = \dfrac{1}{4 T} \cot \dfrac{\pi}{d}\, .
\end{equation}
As we see, the value of $uv$ at the boundary depends only on the dimension for large black holes
\begin{equation}
uv = e^{\alpha(d)}\, ,
\end{equation}
with $\alpha(d)=\pi \cot \frac{\pi}{d}$, an approximately linear function of $d$. This means that as $d$ grows, the corresponding hyperbola in the Kruskal diagram is further apart from the origin. In order to construct now the Penrose diagram, we might choose to flatten one of the two pairs of hyperbolas. If we choose (as usual) to flatten the AdS asymptotic boundary, we may perform the change of coordinates
\begin{eqnarray}
\label{Penrosecoordinates}
v &=& e^{\frac{\alpha(d) }{2}} \tan \dfrac{\tilde{v}}{2}\, , \\
u &=& e^{\frac{\alpha(d) }{2}} \tan \dfrac{\tilde{u}}{2}\, ,
\end{eqnarray}
in which the full spacetime is now compactified in a finite region. Undoing the lightcone coordinates
\begin{eqnarray}
\tilde{u} &=& \tau + \rho\, , \\
\tilde{v} &=& \tau-\rho\, ,
\end{eqnarray}
it is easy to see that the AdS boundary at $uv=e^{\alpha(d)}$ is now given by the straight lines $\rho= \pm \frac{\pi}{2}$. The singularity, on the other hand, becomes bowed in \footnote{Had we chosen to flatten the singularity in the Penrose diagram, the result would have been that the AdS boundary becomes bowed out. One might wonder if there exist a suitable conformal transformation that could flatten out both boundaries at the same time. Symmetries guarantee that this is not possible in this case \cite{Hubeny}.} with a form given implicitly by

\begin{equation}
\tan\left(\dfrac{\tau+\rho}{2} \right) \tan\left(\dfrac{\tau-\rho}{2} \right) = e^{-\alpha(d)}.
\end{equation}
Assuming a symmetric evolution for the action growth (i.e. the WdW patch starts at the same asymptotic time in both sides $t_L=t_R$ ), it is possible to calculate the time at which the `south tip' of the WdW patch leaves the singularity for the first time. This will correspond to the time at which the black hole starts computing. Finding the intersection of the past singularity with $\rho=0$ and inverting back to the asymptotic time $t$ we get that the delay is given by the simple expression

\begin{equation}
\label{bigdelay}
t_\mathcal{C} = \dfrac{\alpha(d)}{4 \pi T} \simeq \dfrac{d}{4 \pi T} + \mathcal{O}(1/d).
\end{equation}
As we could have intuitively expected, the computation delay increases as the singularity bows further into the diagram for higher and higher dimensions, suggesting an strict non-computing behaviour in the large $d$ limit. As the spacetime dimension changes, however, physical properties of the black hole might become trivial unless the scales in the problem are made $d$-dependent. The latter interpretation, thereby, can depend on such possible scalings. In section \ref{sectionscalings}, we will discuss such scalings and their implications in the study of complexity for large $d$ black holes.

\begin{figure}[h]
\begin{center}
\includegraphics[height=7cm]{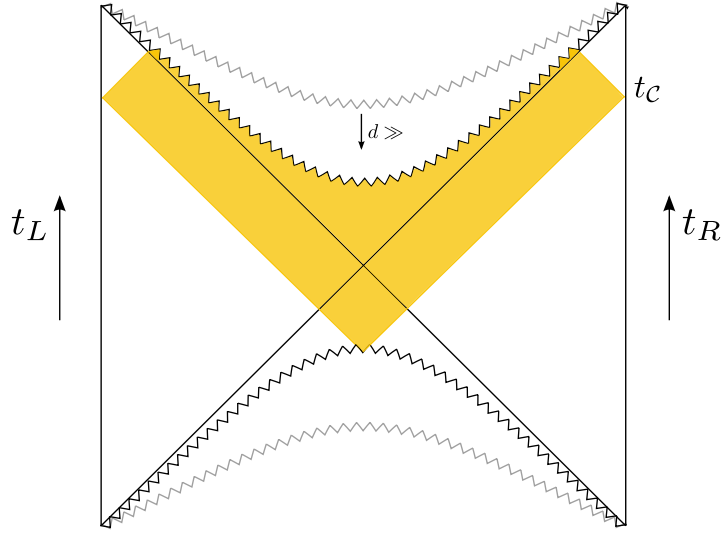}
\end{center}
\caption{\emph{Conformal diagram for higher dimensional black holes and WdW patch at the moment of computation starting.}}
\label{penroselargeD}
\end{figure}

\subsubsection{Small AdS black holes. Schwarzschild black holes.}

As the shape of the large-$d$ conformal diagram shows some significant differences for the case of small AdS black holes, we treat them here separately. In particular, as the dimension grows the singularity does not bow arbitrarily into the conformal diagram, but rather saturates at a finite distance from the horizon. Furthermore the value $r_*(\infty)$ becomes $d-$independent in this limit, changing slightly the complexity phenomenology. Specifically, we may approximate $f(r) \simeq 1+\frac{r^2}{\ell^2},$ for $r_h \ll \ell$  and use the definition in \eqref{deftortoise} to get
\begin{equation}
r_*(r) = \ell \arctan \frac{r}{\ell}\, ,
\end{equation}
whose $r\rightarrow\infty$ limit gives us the corresponding delay
\begin{equation}
\label{smalldelay}
t_\mathcal{C} \simeq \dfrac{\pi}{2} \ell\, .
\end{equation}
Equivalently, asymptotically flat Schwarzschild black holes in a box give a similar solution, i.e. a computational delay that is only controlled by the size of the box. Indeed, for $f(r) = 1- (r_h/r)^{d-2}$ we get
\begin{equation}
r_*(r) = r \, _2F_1\left(1,\frac{1}{2-d};1+\frac{1}{2-d};\left(\frac{r_h}{r}\right)^{d-2} \right)\;.
\end{equation}
If we regard the WdW patch as anchored at the walls of the box, we must evaluate the tortoise coordinate at the location of the box in order to find the corresponding delay. For a well-contained black hole, $L \gg r_h$, we have
the asymptotic behaviour $r_* (L) \sim L$ and we obtain 
$$
t_\mathcal{C} = r_* (L) \simeq L\;.
$$
We see that well-contained black holes have computation delays controlled by the  size of box rather than the black hole itself. In other words, the non-computing feature is a property of the combined system, including both the black hole and its `container'. 

\begin{figure}[h]
\begin{center}
\includegraphics[height=6cm]{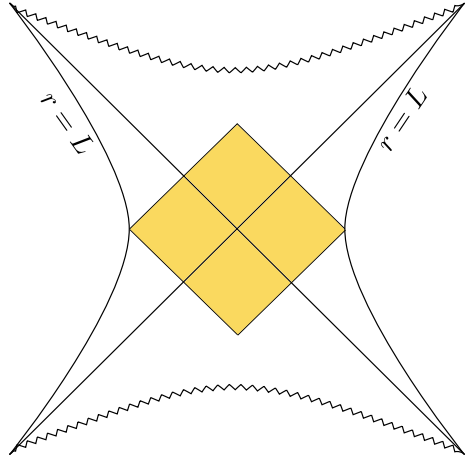}
\end{center}
\caption{\emph{If a box is too small ($L< L_c$) the non-computing features can disappear.}}
\label{fig:smallbox}
\end{figure}

It is then interesting to ask what happens when we shrink the box down to the size of the black hole. For AdS black holes, there is
a smooth transition from small to large black holes. At the transition region we have $T \sim 1/\ell$, so that the small black-hole behaviour 
(\ref{smalldelay}) smoothly morphs into the large black-hole behaviour (\ref{bigdelay}). On the other hand, for asymptotically flat  Schwarzschild black holes with a WdW patch anchored at the location of walls, there is always a limiting value of the box size, $L_c$,  below which the combined system of black hole and box cease to present non-computer behaviour, since the WdW patch eventually becomes too small to simultaneously touch both past and future singularities (cf. Figure \ref{fig:smallbox}). The smallest WdW patch with a non-computer delay is anchored at a zero of the tortoise coordinate, i.e.
$$
r_* (L_c) = 0
$$

\subsection{Holographic non-computers in the large $d$ limit}
\label{sectionscalings}

In the previous section we have seen that large AdS black holes feature a computational delay which becomes parametrically large at large dimensions. This suggests the analysis of holographic complexity in the $1/d$ expansion of general relativity \cite{EmparanlargeD, Minwalla}. These large-$d$ approximations are a kind of mean-field expansion which reveal interesting structure in many classical gravitational phenomena.  A non-trivial question is whether there exist a set of large-$d$ scalings which preserve the standard `phenomenology' of complexity, namely the existence of a linear growth and a large-complexity saturation at very long times (cf. \cite{Susskindtypical}). 

\begin{figure}[h]
\begin{center}
\includegraphics[height=4.5cm]{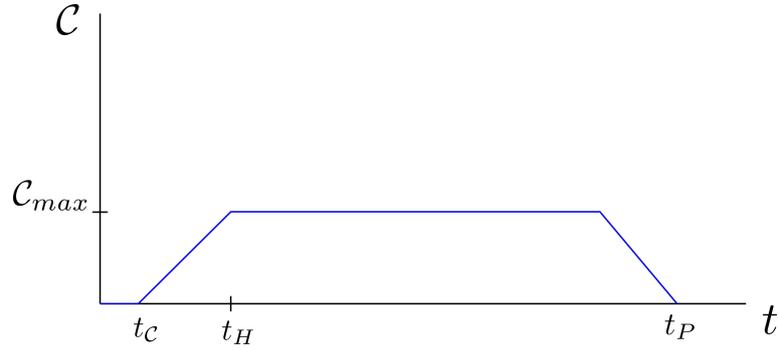}
\end{center}
\caption{\emph{Scheme for complexity pattern of a finite $d$ black hole.}}
\label{plateau}
\end{figure}

The holographic prescription captures the growth of complexity at a rate of order $ST \sim M$, up until we reach complexities of order
\begin{equation}
C_{\rm max} \sim \log(1/\epsilon) \,e^S\;,
\end{equation}
where $\epsilon$ is a coarse-graining parameter in Hilbert space, controlling the degree of approximation we require to `stop the computation'. It is unclear to what extent $\epsilon$ could have an interpretation in the bulk geometry. Assuming $\log (1/\epsilon)$ of order unity, the time of complexity saturation is thus of the order of the Heisenberg time of the system, $t_H \sim T^{-1} e^{S}$ up to subleading terms in the exponent. Over periods of the order of the quantum Poincar\'e recurrence time, $t_P \sim T^{-1} \exp\left(e^S \log(1/\epsilon) \right)$, one expects the system to undo its evolution and decrease its complexity.   A caricature of this behaviour is shown in figure \ref{plateau}. Notice that any large-$d$ scaling preserving the plateau shape must keep finite both the mass and the entropy of the black hole.  

Making this choice however implies that we should not forget about the Hawking process, which actually becomes rather violent in the large $d$ limit. Indeed, typical frequencies for Hawking quanta scale as $\omega \sim d^2/r_h$ , yielding a radiation power that grows factorially in $d$ \cite{Hod} and implying thus an almost immediate evaporation in the large $d$ limit. Of course, a suitable scaling of quantities could be made in order to keep the evaporation time finite, but would limit our possibilities to do so with other quantities of interest. Instead, we will require our `computers' to remain   in thermodynamic equilibrium for exponentially long times, as is the case for large AdS black holes or Schwarzschild black holes inside a suitable box,  in order to avoid the evaporation process. 

In the following sections we show that the requirement of thermodynamic stability is actually non-trivial for small Schwarzschild black holes. On the other hand, no  obstructions are found for large AdS black holes, which are always thermodynamically stable. 

\subsubsection{$1/d$ scaling for large AdS black holes}

In the large AdS black hole regime, $r_h \gg \ell$, the relevant thermodynamic  quantities behave as follows
\begin{equation}
T= \dfrac{dr_h}{4\pi \ell^2}, \hspace{1.5cm} S = \dfrac{V_\Omega}{4G} r_h^{d-1}, \hspace{1.5cm} M = \dfrac{V_\Omega (d-1)}{16 \pi G \ell ^2} r_h^d,
\end{equation}
satisfying the Smarr relation
\begin{equation}
\label{smarr1}
TS = \dfrac{d}{d-1} M,
\end{equation}
which stabilizes in the large $d$ limit and forces us to fix the temperature as well the entropy and mass if we are to maintain the plateau-shape of the  complexity function. Let us rewrite the temperature as
\begin{equation}
T= \dfrac{d}{4\pi} \left( \dfrac{r_h}{\ell} \right) \dfrac{1}{\ell}\, ,
\end{equation}
where the large black hole regime implies that the ratio $r_h/\ell$ is always above unity. In order to keep this hierarchy, we might choose a general set of scalings
\begin{equation}
\label{fded}
\dfrac{r_h}{\ell} = f(d)\, ,
\end{equation}
with $f(d)$ either a constant larger than unity or a growing function of $d$. Once this function is chosen, we must rescale the AdS radius in such a way that the temperature remains finite, i.e.
\begin{equation}
\ell \,T \sim  d\, f(d) \, ,
\end{equation}
To make the entropy finite we can now exploit our freedom to rescale the Planck length $G= (\ell_P)^{d-1}$ as 
\begin{equation}
\left( \dfrac{\ell}{\ell_P} \right)^{d-1} \sim \, \dfrac{S}{f(d)^{d-1} \,V_\Omega}\, ,
\end{equation}
with fixed $S$. 
In order to ensure consistency of the geometrical  description, $\ell_P \ll \ell$, we must limit the growth of $f(d)$ to remain below $\CO(\sqrt{d})$, since then the strong vanishing of the unit volume $V_\Omega \rightarrow \left({1/\sqrt{d}}\right)^d$ is enough to maintain $\ell_P$ as the hierarchically smaller length scale in the problem. 
 
Once we stabilize the scalings of $S$ and $T$, the mass $M$ is kept stable by the Smarr relation \eqref{smarr1}, thus keeping  the qualitative shape of the plateau as $d$ becomes large. Looking now at the computational delay, we can see that the finiteness of the temperature ensures that $t_{\mathcal{C}}$ blows up as $d$ becomes large, meaning that the complexity plateau becomes postponed away in the future
\begin{equation}
t_{\mathcal{C}} \sim \dfrac{d}{T}\, .
\end{equation}
At leading order in the $1/d$ expansion, we have thus  a parametric example of a holographic non-computer, i.e. a finite temperature state for which complexity seems to remain always at a constant value.

\begin{figure}[h]
\begin{center}
\includegraphics[height=4cm]{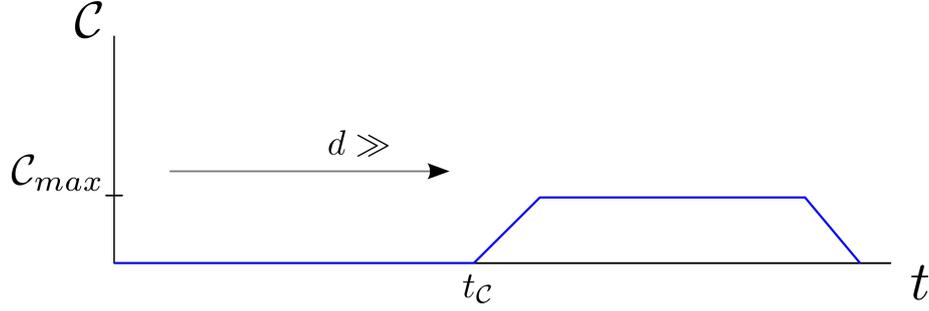}
\end{center}
\caption{\emph{Plateau shifting of large $d$ AdS black holes.}}
\label{plateau2}
\end{figure}

\subsubsection{Infinite-volume scaling}

It is interesting that the successful large-$d$ scaling of the complexity plateau involves a large-$d$ scaling of the  AdS `containment box' .   Since the radius of AdS becomes a physical box size in the CFT dual, it is interesting to reformulate the problem in terms of a `complexity density' which becomes stable in the infinite-volume limit of the CFT.  To this end we consider   black-brane solutions dual to thermal states on flat space.     Now we are free from any restrictions regarding the $r_h/\ell$ ratio, and  this additional freedom allows us to preserve the complexity plateau without scaling with $d$ every physical scale in the problem. 

The temperature formula 
\begin{equation}
T = {d \,r_h \over 4\pi \ell^2}
\end{equation}
remains the same as before. However, the horizon entropy  {\it density} is now given by
\begin{equation}
{S \over V}= {1 \over 4G} \, \left({r_h \over \ell}\right)^{d-1} = N_* \, \left({4\pi T \over d}\right)^{d-1} \;,
\end{equation}
where we have denoted $N_* = \ell^{d-1} /4G$ the effective number of `species' in the CFT (proportional to the central charge). 

At fixed $T$, the power-like behaviour proportional to $T^{d-1}$ implies that any notion of entropy which remains stable in the large-$d$
limit must factor out this term. A natural way of achieving this is to focus on the entropy per thermal cell, namely
\begin{equation}
S_{\rm cell} \equiv {S \over VT^{d-1}}\;, 
\end{equation}
and a similar definition for the thermal-cell energy:
\begin{equation}
M_{\rm cell} \equiv {M \over VT^{d-1}}\;.
\end{equation}
Scaling now $N_* \rightarrow \infty $ according to 
\begin{equation}
N_* = S_{\rm cell}   \left( \dfrac{d}{4 \pi}\right)^{d-1},
\end{equation}
as $d\rightarrow \infty$ with  fixed $S_{\rm cell}$, we make stable the `thermal cell complexity' given by 
\begin{equation}
C_{\rm cell} \equiv {C \over V T^{d-1}}\;.
\end{equation}
Then we find that $C_{\rm cell}$ should reproduce a plateau shape with parameters $S_{\rm cell}, T$ and $M_{\rm cell}$. 
As before, the delay time remains given by $t_{\cal C} \sim d/T$, which diverges linearly in the large $d$ limit.

\subsubsection{$1/d$ scaling for small  AdS black holes}

If we now consider small black holes in AdS, $r_h \ll \ell$, the thermodynamic quantities will behave as those of the the usual Schwarzschild black holes
\begin{equation}
T= \dfrac{d-2}{4\pi r_h}\,, \hspace{1.5cm} S = \dfrac{V_\Omega}{4G} r_h^{d-1}\,, \hspace{1.5cm}  M = \dfrac{V_\Omega (d-1)}{16 \pi G } r_h^{d-2}\,.
\end{equation}
And again, we can find a simple expression relating the three of them which stabilizes in the large $d$ limit
\begin{equation}
\label{smarr}
TS = \dfrac{d-2}{d-1} M\,.
\end{equation}
Keeping now a finite temperature requires that we scale up the horizon radius as 
\begin{equation}
r_h  \,T \sim  d\,, 
\end{equation}
whereas the entropy $S$  is fixed if we scale the Planck length as 
\begin{equation}
\label{planckscaling}
\left( \dfrac{r_h }{\ell_P} \right)^{d-1} \sim \dfrac{{S}}{V_\Omega}\,. 
\end{equation}

Up to this point, the AdS radius did not make an appearance. However, it will be the relevant scale for the computation delay \eqref{smalldelay} and it is constrained by the requirement that the black hole actually `fits the box', i.e. $r_h < \ell$. In general we can allow 
\begin{equation}
\left( \dfrac{r_h}{\ell} \right) = g(d),
\end{equation}
with $g(d)$ either a small constant or a decreasing function of $d$. Feeding these scalings into \eqref{smalldelay} we get a computation delay
\begin{equation}
t_\mathcal{C} \sim \dfrac{d}{g(d)} T^{-1},
\end{equation}
which again diverges in the large $d$ limit for any of the allowed behaviours of $g(d)$. The case of Schwarzschild black holes
well-contained in a flat box follows along similar lines, with the size of the box playing the role of the AdS radius, $\ell$. 

Our analysis shows that a blow-up of the `containment box' is essential to manufacture a large-$d$ Schwarzschild non-computer. Since
thermodynamic equilibrium of ordinary black holes in finite boxes requires certain ratios between the relative sizes of the black hole and the box, we must check the compatibility of stability  with the required  large-$d$ scaling. 

\paragraph{Stability analysis at large $d$}

 Having isolated large-$d$ scalings with parametric computational delay for both large and small black holes, we come now to the discussion of their thermodynamical stability. Since the discussion of action-complexity is formally tied to the two-sided eternal black hole geometries, we shall focus mostly on the canonical ensemble at fixed temperature, which is the effective one-sided description of the associated thermofield-double states. 
 
 The canonical thermodynamics for AdS black holes is well known (cf. \cite{page}). Large and small black holes form a continuous family of solutions labelled by the horizon radius $r_h$. For $r_h \ll \ell$ all small black holes have negative specific heat and their thermodynamics is  locally unstable. The associated temperature is large, and the dominant phase in this regime is a large AdS black hole with $r_h \gg \ell$ and positive specific heat. There is a critical temperature, the so-called Hawking--Page (HP) temperature, $T_{\rm HP} = (d-1)/2\pi \ell$,  below which the large AdS black hole has larger free energy than a gas of gravitons in AdS. Below the HP temperature there is a narrow window down to $T_l = \sqrt{d(d-2)} /2\pi \ell$ in which black holes are locally stable but globally unstable. In this narrow window the size of the black holes is of order $\ell$ and all of them have computational delays of order $\ell$. 
 
Locally stable but globally unstable  entangled black holes should behave as ordinary holographic computers for large periods of time, exponential
in $1/G$, where $G$ is Newton's constant, after which they are likely to fluctuate into a state of two entangled boxes filled with radiation, with a complexity of order $G^0$. It would be very interesting to study how this time scale compares to the Heisenberg time scale, controlling the saturation of complexity. At any rate, black holes whose thermodynamic state is both locally and globally stable, i.e. those with $r_h >\ell$,  are guaranteed to last beyond the saturation plateau and furnish the pattern of large-$d$ computational delay indicated in the previous section. 

The situation is different for asymptotically-flat Schwarzschild black holes contained inside entangled cages of size $L$. If each black hole is much
smaller than its cage, it is guaranteed to be locally unstable, so that it will decay very fast into a graviton-gas state (cf. \cite{gibbons-perry}). In the present interpretation, we say that the thermofield double state will look like an entangled pair of boxes full of radiation for almost all the time. Such states should have growing complexity of order $G^0$. On the other hand, for black holes which almost touch the cage, there are windows of local and global stability for growing complexity of order $1/G$. Following   \cite{York}, we can
determine these regimes by evaluating the Euclidean action of the black hole solution with two boundary conditions: the temperature is physically fixed at the walls of the box for both the black hole and the graviton gas states, and of course the metric is smooth at the horizon. 

Writing  the Euclidean black hole metric as 
$$
ds^2_{\rm bh} = \left(1-\left(\frac{r_h}{r} \right)^{d-2} \right) \,d\tau'^{\,2} + \dfrac{dr^2}{\left(1-\left( \frac{r_h}{r} \right)^{d-2} \right)}   + r^2 d\Omega_{d-1}^2\;,
$$
with $\tau' \equiv \tau' + \beta'$, we require that the ${\bf S}^1$ parametrized by $\tau'$ be smoothly contractible, which fixes 
$$
\beta' = {4\pi r_h \over d-2}\;.
$$
On the other hand, the physical temperature is measured as the inverse proper length of the ${\bf S}^1$ at the walls of the box, i.e.
\begin{equation}
\label{smo}
\beta = {1\over T} = \beta' \sqrt{ 1- (r_h /L)^{d-2}}\;. 
\end{equation}
The vacuum metric which is used for normalization is given by
$$
ds^2_{\rm vac} = d\tau^2 + dr^2 + r^2 d\Omega_{d-1}^2\;,
$$
with $\tau \equiv \tau + \beta$. The canonical free energy is computed in the saddle-point approximation by subtracting the corresponding  Euclidean actions. Ricci flatness of both solutions implies that only the YGH term contributes in both cases:
$$
-\log Z(\beta) \approx -{1\over 8\pi G} \int_{\partial {X_{\rm bh}}} K + {1\over 8\pi G} \int_{\partial {X_{\rm vac}}} K = \beta M_{\rm eff} - S\;,
$$
 where $S$ is the entropy of the black hole and $M_{\rm eff}$ is the quasilocal Brown-York mass (cf. \cite{brown-york}) given by
 \begin{equation}
 \label{meff}
 M_{\rm eff} = 2L^{d-2} {(d-1) \Omega_{d-1} \over 16\pi G} \left(1-\sqrt{1-(r_h /L)^{d-2}}\right)\;.
 \end{equation}
 Notice that this effective mass approaches the standard ADM mass of the black hole as we push the cage to infinity, $L\rightarrow \infty$. The form of  $M_{\rm eff}$ is completely fixed by the Bekenstein--Hawking formula
 $$
 S= {\Omega_{d-1} \over 4G} r_h^{d-1}\;,
 $$
 together with the smoothness condition (\ref{smo}). To see this, notice that we can rewrite the first law as
 $$
 \beta = {\partial S \over \partial E} = {\dd S \over \dd r_h} {\partial r_h \over \partial E}\;,
 $$
 where $E$ is the internal energy. 
 Since we know the functional dependence of both $\beta$ and $S$ on $r_h$, the previous relation is a simple differential equation for
 $E(r_h)$. This equation is easily solved with the condition that $E(r_h =0) =0$ to yield exactly the expression (\ref{meff}): 
 $$
 E(r_h) = M_{\rm eff} (r_h)\;,
 $$
 and the free energy follows then from the standard thermodynamic relation
 $$
 \log Z(\beta) = -\beta E + S\;.
 $$
 At any rate, our expression for $\log Z(\beta)$ as a function of $r_h$ determines a window of local stability for black holes which
 are sufficiently close to the walls of box. In terms of the parameter 
 $$
 x\equiv \left({r_h \over L}\right)^{d-2}\;,
 $$
 locally stable black holes exist inside the cage for $x_l < x < 1$ with 
 $$
 x_l = {2\over d}\;.
 $$
 Globally stable black holes are determined by a negative free energy, which requires  that $x_s < x< 1$ with
 $$
 x_s = 4{d-1 \over d^2}\;.
 $$
 Notice that, as $d\rightarrow \infty$, the stable black holes lie arbitrarily close to the walls of the box. 
 
\begin{figure}[h]
\begin{center}
\includegraphics[height=5cm]{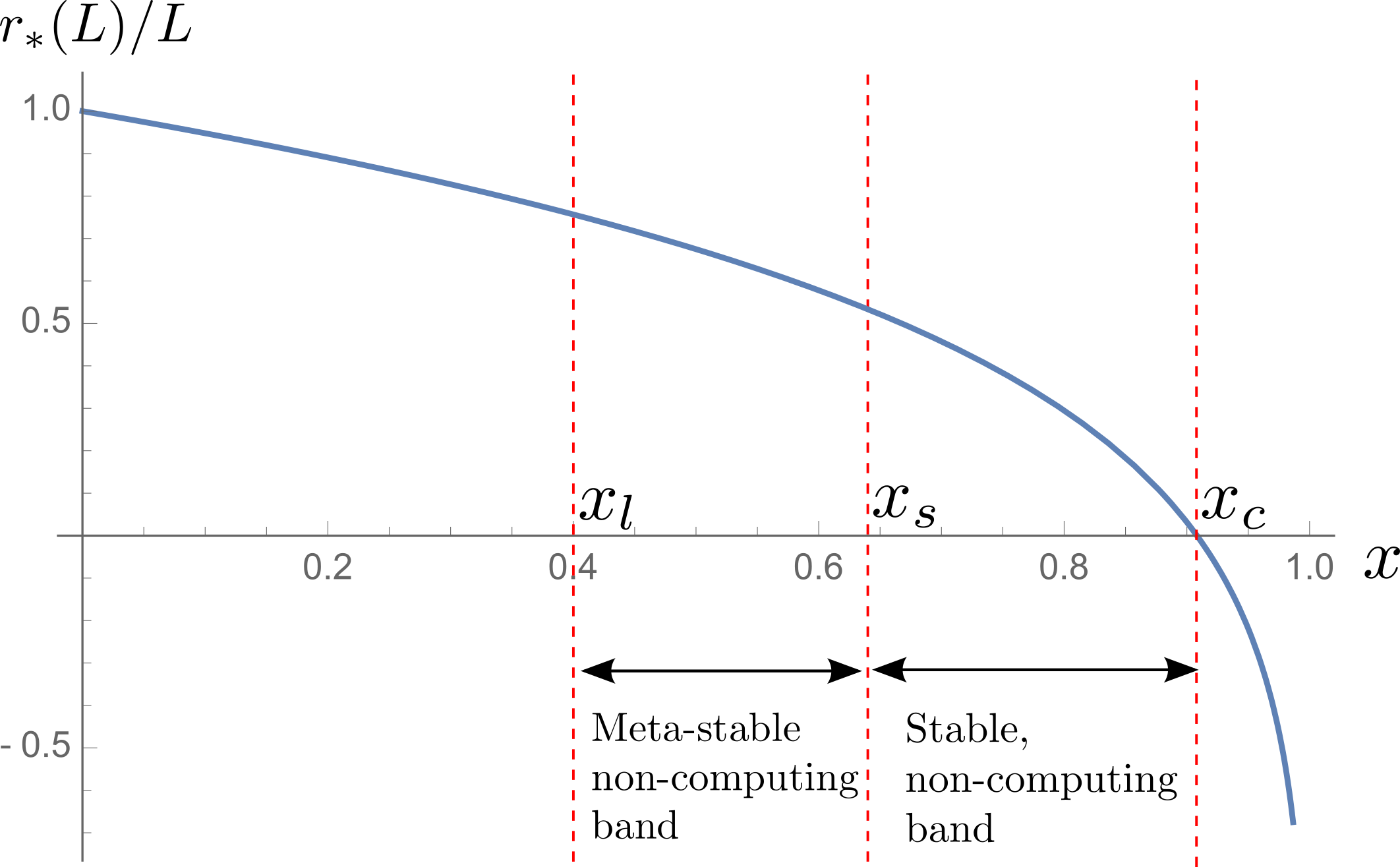}
\end{center}
\caption{\emph{Tortoise coordinate at the wall as function of $x=\left({r_h \over L}\right)^{d-2} $. The band compatible with global stability as well as non-computing features lies within $x_c> x > x_s$. }}
\label{fig:hyper}
\end{figure}

 These windows of stability combine in a non-trivial fashion with the requirement that they behave as holographic non-computers. As indicated in the previous section, the condition for the black hole to possess a computational delay is that the cage is not too small. In particular, the critical value for non-computing, determined by $r_*(L_c) =0$ must be such that $x_c = (r_h /L_c)^{d-2}$ be {\it smaller} than $x_s$. Only then we can find stable black holes with a non-computing WdW patch. Alternatively, we require that the tortoise coordinate at the wall be {\it positive} for the critically stable black hole at $x=x_s$. We show in Figure \ref{fig:hyper} that this is indeed the case, so that a  band of large-$d$ non-computers exist among the narrowly caged Schwarzschild black holes. 
 
\subsubsection{Firewalls as natural non-computers?}
\label{sectionfirewalls}

As we learned in section \ref{sectiondelays}, singularities of some black hole solutions become arbitrarily close to the horizon in the large $d$ limit, suggesting the fact that large $d$ black holes could provide a classical model of firewalls \cite{AMPS}. The exotic complexity dynamics of such solutions raises the question of whether firewalls might actually provide a natural candidate of non-computer systems. 

In order to check if large-$d$ black holes are really classical models of firewalls, we must check if the physical `thickness' of the black-hole interior is Planckian.  We can phrase this question  by  calculating the proper free-fall time through the interior geometry,  towards the singularity. For big (flat and spherical) AdS black holes, this is given by

\begin{equation}
\label{bigcrash}
\tau_{sing} = \int\limits^{r_h}_0 \dfrac{dr}{\sqrt{-f(r)}} \simeq \dfrac{\pi}{d} \ell,
\end{equation}
whereas this quantity is controlled by the size of the horizon for small AdS and Schwarzschild black holes

\begin{equation}
\label{smallcrash}
\tau_{sing} \simeq \dfrac{r_h}{d}.
\end{equation}

A classical model for a firewall would presumably correspond to a Planckian infalling time towards the singularity. The particular scalings defined in this paper, which are fixed by the requirement of keeping a qualitative plateau-shape for the complexity growth, imply an effective shrinking of the Planck length, so that the falling time is always large compared to the Planck length in the case of finite-entropy black holes (large or small). For the case of large AdS black holes we have
\begin{equation}
\left({\tau_{sing} \over \ell_P}\right)^{d-1}  \sim {S \over (d\, f(d))^{d-1} V_\Omega} \sim \left({\sqrt{d} \over f(d)}\right)^{d-1},
\end{equation}
which diverges at large $d$, under the condition $f(d) < \sqrt{d}$, which was imposed to ensure that the Planck length is indeed smaller than the AdS radius. In the case of small AdS black holes, a similar estimate yields a scaling proportional to $(\sqrt{d}\,)^{d-1}$, which again diverges as $d\rightarrow \infty$. Hence, we conclude that the large-$d$ `shrinking' of the interior geometry is not felt by an infalling observer as a Planckian wall. 

On the other hand, it is interesting to point out that for flat branes we do not need to scale $\ell$ in order to achieve stable `thermal-cell complexity'. In this case we can actually bring a `firewall' physically close to the horizon while maintaining the shape of the plateau. It would be interesting to study if these considerations have any significance for the  meaning of `firewall' states.

\section{Discussion}

We have studied the phenomenology of Holographic Complexity for a class of black hole solutions presenting non-standard thermodynamics, showing that some of such systems present not only singular complexity dynamics but also qualitatively different features when studied with either the AC or VC holographic prescriptions.

First, we have studied the structure of low-temperature thermofield double states in strongly coupled CFTs defined on hyperboloids. In particular, we have focused on properties characterized by the gravitational description in terms of AdS hyperbolic black holes, finding that in addition to the known zero-temperature entropy of order $N_*$, these states have a large holographic complexity as measured by extremal bulk volumes, logarithmically diverging in the $T\rightarrow 0$ limit. 

\begin{equation}
\Delta \CC_V (0)\Big|_{T \ll 1} \sim \dfrac{ S_0}{\sqrt{d}} \log(T^{-1}\ell) + \CO(S_0)\,, \hspace{2cm}\Delta \CC_A (0)\Big|_{T \ll 1} \sim \CO(S_0)\,.
\end{equation}

The coefficient of the logarithm is a  reliable strong-coupling prediction in the low-$T$ limit, once we define the absolute normalization of the complexity. Since this behavior is controlled by the emergence of the AdS$_2$ throat, it is tempting to take it at face value, as a general property of any near-extremal geometry of Reissner--Nordstrom type. This includes the benchmark model of AdS/CMT, the near extremal charged AdS$_4$ black brane, with either chemical potential $\mu$ or magnetic field $B$ (cf. \cite{rcmt} for a review). In those systems, the same expression follows, with the substitution  of the curvature scale $1/\ell$ by an effective mass of the order of $\mu$ and/or $\sqrt{B}$. For such systems, this structure was indeed analyzed an confirmed in both AC and VC complexity (cf. \cite{Myerstdep}), showing the large IR diverging contribution as a common feature of near-extremal RN solutions. Curiously enough, such IR divergence is not present when computing the action of the cold hyperbolic black hole, signalling a major difference between the two prescriptions.

Most strikingly,the `cold' regime ($0<T<1/2\pi$) of this system provides the first example of a finite-temperature system with constant AC complexity, contradicting the general expectations for quantum systems as well as the result in the VC prescription, which predicts the usual late-time linear growth.

\begin{equation}
\dfrac{\dd \CC_V}{\dd t}\Bigg|_{T< 1/2\pi} \sim S_0 T, \hspace{2cm} \dfrac{\dd \CC_A}{\dd t}\Bigg|_{T< 1/2\pi} = 0\,.
\end{equation}

The origin of this behavior, as well as the discrepancy between the two proposals, are not well understood and suggest that hyperbolic black holes might conform a suitable testing ground for the settlement of the correct complexity prescription. Furthermore, as the throat developed by these systems is expected to be described by an effective AdS$_2/$CQM, the problem begs the question of its relation with similar studies in the context of CQM models such as the SYK model (cf. \cite{Kitaev1, MaldacenaStanfordSYK, SarosiSYK}), recently argued to be dual to the two-dimensional theory of Jackiw-Teitelboim (JT) gravity. In \cite{SusskindJTcomplexity, HugoJTcomplexity} the AC complexity of JT gravity was studied in detail, finding again a \textit{non-computer} behaviour for that theory within the inclusion of the naive boundary terms. Nevertheless, instabilities appearing in consistent string theory embeddings of these systems could have a decisive impact on the prediction, and its understanding could lead to a clarification of the exotic properties of these degenerate systems.

Next, we have shown that a formal application of the large-$d$ expansion of GR to large AdS black holes produces parametric examples of holographic non-computers  with computational delays scaling linearly with $d$. From the gravitational point of view, the origin of this phenomenon can be traced back to the existence of a larger set of   independent symmetries acting on the WdW patches  for $t< t_{\mathcal{C}}$.  We find that small Schwarzschild black holes are somewhat puzzling. First of all, their computational delay does not appear to be intrinsic, but rather depends on  the infrared regulator, i.e. the containment box. Despite the apparent existence of a parametric delay of ${\cal O}(d)$ in the large-$d$ limit, one ultimately finds this incompatible with the requirement that the black hole be stable  unless we fine tune the walls of the box to approach the horizon as $d\rightarrow \infty$. Otherwise we are left with a trivial realization of the `non-computer' in this case, namely two entangled boxes full of radiation.  

 It is interesting to notice that this large$-d$ complexity-phenomenology seems certainly particular to the AC conjecture, and does not appear (at least in an obvious manner) in the VC proposal.  In this sense, it joins the properties of cold hyperbolic black  holes in the list of identifiable discrepancies between the two proposals, a question which deserves further scrutiny.

A major open problem is the understanding of the various non-computing systems described here in the language of the CFT. On general grounds, we expect the large $d$ limit of gravity to correspond to the mean field theory approximation of QFT. In this context, it might be not so surprising that some fine grained properties of the field theory, such as complexity, are not captured by this approximation, yielding a completely trivial dynamics for the leading order in the $1/d$ expansion. On the other hand, given the scarcity of CFTs in higher dimensions, the very existence of a parametric $1/d$ expansion in the AdS/CFT correspondence is a rather intriguing, albeit remote possibility. 

\newpage
\section{Appendix. Piecewise estimation of extremal volumes}
\label{sec:TBHappendix}

Following the formula   \eqref{CVansatz}, we are asked to compute the holographic complexity as the volume of extremal codimension-one surfaces in the given geometry, parametrized by the static asymptotic time variable.  

Due to the non-linearity of the the corresponding Euler-Lagrange equation, the exact variational problem is complicated, restricting our ability to find analytic solutions. However, a useful order-of-magnitude estimate can be obtained by an approximate description of the full metric \eqref{bbrane}, according to a piece-wise approximation for the function $f(r)$. For $r\gg r_h$ we can approximate the metric by the vacuum AdS$_{d+1}$ solution. In the near-horizon region $r_0 < r< r_R$, with $r_R$ an $\CO(1)$ multiple of $r_0$, we can take the Rindler approximation, whereby the metric is expressed as a product of two-dimensional flat space and the horizon:
\begin{equation}
\label{rme}
\dd s^2_{\rm Rindler} \approx -(\dd X^0)^2 + (\dd X^1)^2 + r_h^2  \;\dd {\rm H}_{d-1}^2
\;,
\end{equation}

where
\begin{equation}
\label{cha}
X^0 = \sqrt{r-r_h \over \pi T} \,\sinh(2\pi T t)\;, \qquad X^1 = \sqrt{r-r_h \over \pi T} \,\cosh(2\pi T t)\;,
\end{equation}

a change of variables valid for $r>r_h$ on one of the asymptotic regions. 
Finally, the interior geometry is parametrized in Schwarzschild coordinates $(r,t)$, formally continued to $r<r_h$, with $r$ now denoting a time-like coordinate and $t$ a space-like one. There is an analogous extension of the Rindler patch to the interior, with the analogous  change of variables 
\begin{equation}
\label{chain}
 X^0 = \sqrt{r_h-r \over \pi T} \,\cosh(2\pi T t)\;, \qquad X^1 = \sqrt{r_h-r \over \pi T} \,\sinh(2\pi T t)\;.
\end{equation}

Within this prescription we view the portion of the extremal surface lying outside the horizon as composed of two pieces: an asymptotic component $\Sigma_{\rm UV}$ which is well approximated by a constant $t$ surface in AdS$_{d+2}$, and  a `Rindler piece' $\Sigma_{\rm R}$, parametrized by
a curve on the $(X^0, X^1) $ plane of  \eqref{rme}. Within the Rindler patch, local volume for fixed $X^1$ interval is maximized by the $X^0 = {\rm constant}$ surfaces, and thus we take this ansatz for $\Sigma_{\rm R}$. 
For $t=0$, this is all there is, since the extremal surface is just the $t=0$ section of the extended geometry, with the two exterior geometries   glued by the horizon. However, as $t$ grows, the surface enters the horizon at higher values of $X^0$ and tends to extend through the interior patch of the black brane geometry. The $X^0 = {\rm constant}$ ansatz continues to be reasonable as long as the  complete surface stays inside the  interior Rindler region. Since the only length scale controlling the width of the Rindler region is $T^{-1}$, the approximation $X^0 = {\rm constant}$  must break down for large times, $t \gg T^{-1}$.

\begin{figure}[h]
\begin{center}
\includegraphics[height=6cm]{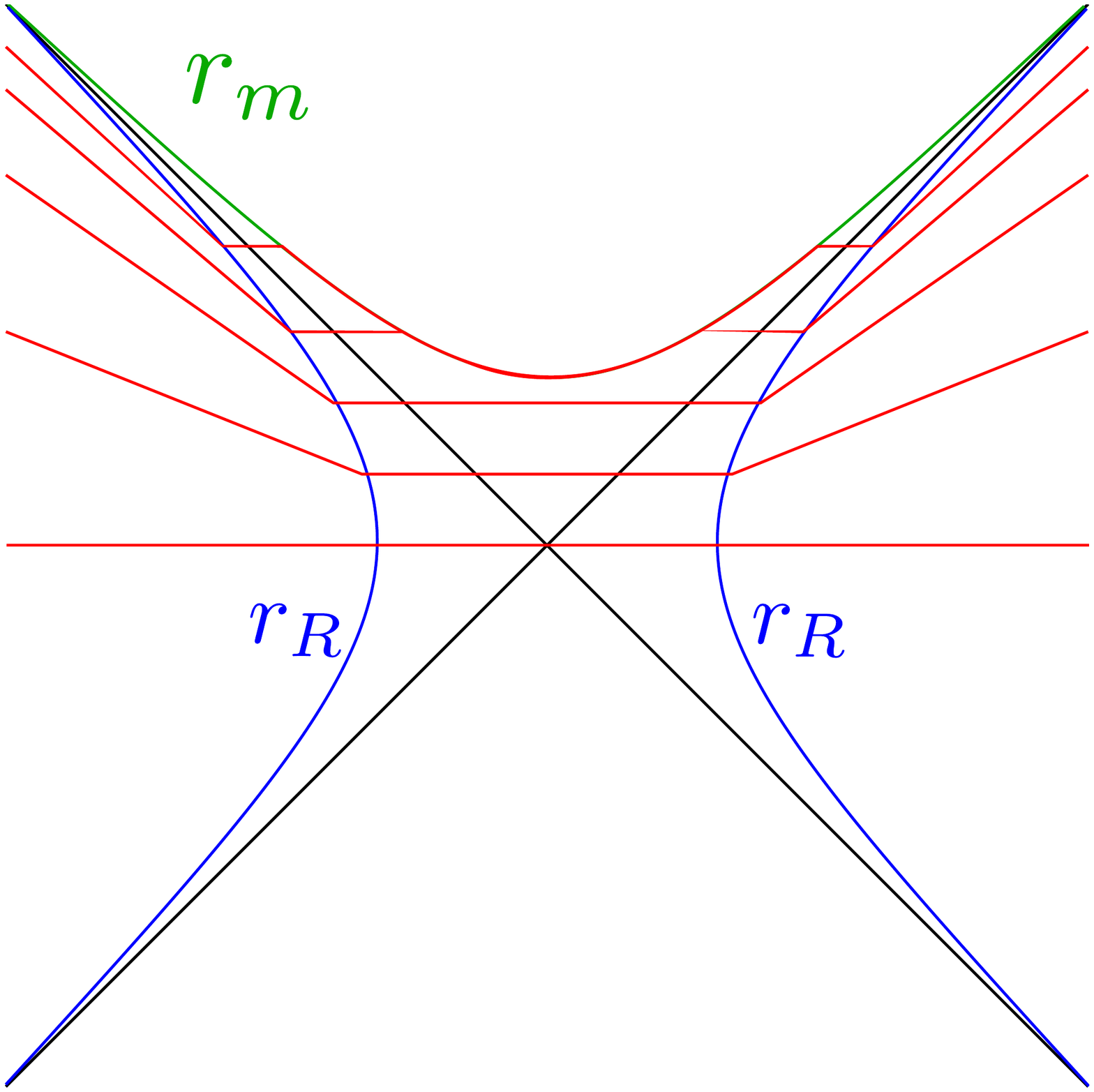}
\end{center}
\caption{\emph{Piecewise decomposition of $\Sigma_t$, represented by space like sections at different times. For generic times, straight tilted segments correspond to $\Sigma_{\rm UV}$, horizontal segments give  $\Sigma_{\rm R} $ and hyperbolic segments on the $r=r_m$ surface correspond to $\Sigma_{\rm WH}$. The $t=0$ surface lacks an interior component. As $t$ increases from zero, an interior component  begins to develop gradually, as $\Sigma_{\rm R}$ eventually transmutes into $\Sigma_{R} \cup \Sigma_{\rm WH}$. }  }
\label{fig:ninerm}
\end{figure}

At very long times, there is a natural answer for the variational problem in the interior, since the surfaces $r={\rm constant}$ are invariant under the $t$-translation isometry. The volume of a $\Delta t$ portion of such $r={\rm constant}$ surfaces is proportional to  
$$
\Delta t\, r^{d-1} \sqrt{|f(r)|}\;,
$$
so that stationary points $r_m$  of this function determine extremal surfaces far from the `exit point', i.e. for  large $\Delta t$. In all cases studied in this paper, one finds $|r_h - r_m | \sim |r_R - r_h |$, implying that $r_m$ is always close to the inner edge of the Rindler region and, in particular, it is roughly symmetrical of the $r=r_R$ surface by a reflection through the horizon (see figure 1). As a consequence, the `exit point' from the $r = r_m$ surface is approximately given by $t_{\rm exit} \approx t$, where $t$ is the time label of the exterior asymptotic surface $\Sigma_{\rm UV}$.

 The approximate ansatz for the extremal surface is thus  $\Sigma_{\rm WH} \cup \Sigma_{\rm R} \cup \Sigma_{\rm UV}$, where $\Sigma_{\rm WH}$ is the $r=r_m$ surface along the `wormhole' in the interior, cut off at $t_{\rm exit} \sim t$, with total $t$-length  of order $\Delta t \sim 2t$.

Within this construction, the volume of $\Sigma_{\rm UV}$ is independent of $t$, whereas the volume of $\Sigma_{\rm R}$ vanishes at large $t$, being delimited by two curves (interior and exterior) asymptotic to the same horizon. Therefore, the rate of growth of the complexity is controlled by $\Sigma_{\rm WH}$ at large times. A graphical representation of the piecewise decomposition of $\Sigma_t$ is shown in figure \ref{fig:ninerm}.

\part{\sc{Terminal Complexity and Singularities}}

\section*{Introduction}
\noindent

\addcontentsline{toc}{chapter}{Introduction}

From the origin of the universe to the interior of black holes, spacetime singularities seem to be present in some of the less understood phenomena within the field of fundamental physics. Not only they constitute one of the most misterious features of the theory of General Relativity, but also signal the very breakdown of it, calling for a UV completion able to resolve such singularities at distances below the Planck scale. Despite the success of perturbative string theory in the description of some timelike singularities, \comm{(through the expedient of exhibiting extra light degrees of freedom localized at the  singular locus)} much less is understood about the most realistic case of the spacelike ones, for which strong gravitational dynamics is believed to be needed.

Very broadly, there are two traditions regarding the interpretation of spacelike singularities: either they must be `resolved' so as to restore some type of  evolution across the singularity, or they must be accepted as  true `spacetime terminals'. To the extent that the black hole singularity is a general guide, the second option is preferred in modern discussions based on holography as the fist one putatively violates the entropic bounds\comm{. On the other hand, the straightforward application of holographic ideas requires an identification of appropriate AdS/CFT boundaries or at least some notion of holographic screen}. On the other hand, even within the realm of the well known benchmark examples of AdS/CFT, the search for a detailed mapping of the singularity in terms of the boundary degrees of freedom has been unsuccessfull. The ostensible obstructions to extend the so called `bulk reconstruction' procedure behind spacetime horizons \cite{Papadodimas, Harlowbulk} seems to render this problem both technically and conceptually non-trivial, hindering the search for useful CFT quantities that could describe or even diagnose the presence of bulk singularities.

A remarkable exception to this, however, is provided by AC complexity, which enjoys a quantitatively important contribution coming directly from the singularity through the evaluation of the YGH term. On the one hand, we might be suspicious of the validity of such contribution, as this is a term in the low-energy effective action that shoud be corrected in the gravitational strongly coupled regime (cf. \cite{Nally}). On the other hand, we are instructed to take this contribution seriously down to its precise dependence on coefficients, as this is crucial for the claimed uniformity of the growth law (\ref{CArate}) for AdS black holes in various dimensions, large and small. In a similar vein, the contribution (or lack of it) of the YGH term at the singularities is crucial for the `non-computing' behavior in various  systems, such as those studied in section \ref{sec:noncomp}.
  
These considerations suggest that holographic complexity is actually the piece of the holographic dictionary which most efficiently `sees' the properties of the singularities, a fact that encourages us to explore a broader phenomenology, seeking to associate holographic measures of quantum complexity to states which are linked to spacelike singularities by time evolution. This program was initiated in previous work \cite{BarbonRabinoComplexity, rabinoac} by the analysis of certain cosmological singularities with controlled AdS/CFT embedding. Here we seek to provide quasilocal notions of complexity which may be abstracted from particular AdS/CFT constructions, and  therefore   having a larger degree of generality. \comm{While we use the volume-complexity (VC) proposal \cite{SusskindEntnotEnough} as a heuristic guide, our discussion is tailored to the more covariant action-complexity (AC) proposal \cite{SusskindCAcorto, BrownSusskindAction}. }

The connection between spacetime singularities and complexity has a long history, going back to the occurrence of classical chaos in generic cosmological singularities   \cite{misner, BKL,BKL2,BKL3} (see \cite{libro} for  a recent review.) and other efforts to classify the different nature of singularities in the quest to seek for a gravitational definition of an `arrow o time' \cite{Penrose}. As we will see, holographic complexity provides a language in which some of these ideas can be realized in a precise way.

This part is organized as follows. In Chapter \ref{ch:terminaldef} we lay down our setup and motivate the definition of the notion of `terminal holographic complexity', a new variant of the AC prescription which is well adapted to isolate the complexity of generic spacetime singularities. In chapter \ref{ch:locking} we investigate the universality properties of this quantity in some  examples of singularities admitting a completely analytic treatment. We end with an epilogue in Chapter \ref{ch:arrow} where we speculate about the resemblance of terminal complexity with older proposals for local measures of gravitational complexity.


\chapter{Holographic complexity of cosmological singularities}
\label{ch:terminaldef}

\section{A quasilocal AC ansatz for terminals}
\label{sec:quasilocal}
\noindent

\begin{figure}[h]
$$\includegraphics[width=10cm]{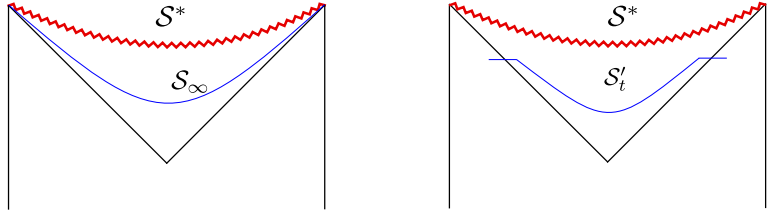} $$   
\begin{center}
\caption{\emph{  On the left, the codimension-one asymptotic surface ${\cal S}_\infty$, accounting for the total complexity `flowing' into the black-hole singularity $\CS^*$. On the right, the subtracted codimension-one surface ${\CS}'_t$ which accounts for the time-dependence of VC complexity in the eternal black hole geometry. }\label{fig:sigmainf}}
\end{center}
\end{figure}

In the benchmark model provided by the eternal black hole spacetime, the central object of interest for the VC ansatz is the extremal codimension-one surface $\CS_\infty$ shown in Figure \ref{fig:sigmainf}. This surface maximizes the volume locally and it lies entirely within the black hole interior, i.e. the past causal domain of the singularity.

The growth of complexity within the VC ansatz can be seen as the result of gradually accessing an increasing portion of $\CS_\infty$. More precisely, the portion of the extremal surface ${\CS}_t$ which has a significant contribution to time dependence  can be analyzed approximately as composed of two parts: a subset of $\CS_\infty$ with volume proportional to $t$, and a transition surface at the horizon, whose contribution to the complexity is of order $S$, the entropy of the black hole. Let us denote by ${\CS}'_t$ this, loosely defined,  `subtracted' surface as indicated in Figure \ref{fig:sigmainf}. 

\begin{figure}[h]
$$\includegraphics[width=10cm]{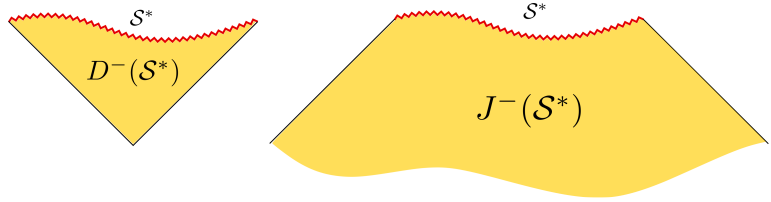} $$   
\begin{center}
\caption{\emph{Generic terminal set $\CS^*$ (in red) and its past domain of dependence $D^-(\CS^*)$ and causal past $J^-(\CS^*)$ .  }\label{fig:DS}}
 \end{center}
\end{figure}

Once we decide to focus on ${\cal S}'_t$ and its asymptotic limit $ {\cal S}_\infty$, we may consider  versions of these quantities for any terminal set $\CS^*$ (which may in particular be a proper  subset of a wider one.) The reason is that the analogue of ${\cal S}_\infty$ always exists given any spacelike terminal set $\CS^*$ and its associated past domain of dependence $D^-(\CS^*)$  (see Figure \ref{fig:DS}.) Since the volume is positive and the past boundary of $D^- (\CS^*)$ is null, the extremal surface is either a local maximum of volume or it coincides with $\CS^*$ in a degenerate case. The first situation occurs when $\CS^*$ is a standard singularity of the kind we encounter at black holes and cosmological crunches in General Relativity, since the volume of spatial slices vanishes at such singularities.

\begin{figure}[h]
$$\includegraphics[width=5cm]{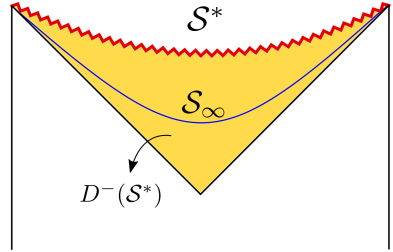} $$   
\begin{center}
\caption{\emph{  The total VC complexity flowing into the singular set ${\cal S}^*$ is the volume of the asymptotic surface ${\cal S}_\infty$. Its AC analog is the on-shell action integrated over the past domain of dependence $D^- ({\cal S}^*)$.  }\label{fig:c1}}
 \end{center}
\end{figure}

A more covariant version of $\CS_\infty$ and ${\CS}'_t $ could be obtained by adapting the AC ansatz to this situation (cf. Figure \ref{fig:c1}.) Since ${\cal S}_\infty$ is the extremal surface on $D^-({\cal S}^*)$, the natural AC version of the full  terminal complexity of the set ${\cal S}^*$ is the on-shell action 
\begin{equation}\label{fullc}
\CC[{\cal S}^*]  =  I\left[D^-({\cal S}^*)\right]
\;,
\end{equation}
evaluated over the set $D^- ({\cal S}^*)$. Since this definition  only makes reference to the terminal set ${\cal S}^*$ we regard this notion of complexity as `quasilocal' and will often denote it as such.

 \begin{figure}[h]
$$\includegraphics[width=5cm]{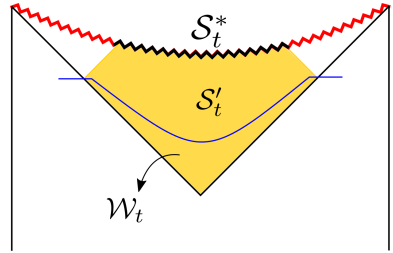} $$   
\begin{center}
\caption{\emph{  The WdW patch ${\cal W}_t$, associated to the cut-off surface ${\cal S}'_t$, intersects the singularity at ${\cal S}^*_t$. }\label{fig:c2}}
 \end{center}
\end{figure}

Next, a  notion of `time-dependence'  can be defined by considering a WdW patch anchored roughly at the exit points of the cut-off surface ${\cal S}'_t$, as indicated in  Figure \ref{fig:c2}. A more invariant definition can be obtained by noticing that these WdW patches are
nested into one another as time increases. For sufficiently `late' WdW patches, this `nesting' extends to the intersections of the WdW patches with the singular set. This suggests that  we may use the nested singular subsets as a starting point in the definition of the WdW nested family. To be more precise, let us pick  a sequence $ \lbrace {\CS}^*_u  \rbrace $ of terminal subsets labelled by $u$ , ordered by inclusion in the sense that
\begin{equation}
{\CS}^*_u \subset {\CS}^*_{u'}  \;, \; {\rm for}\;\; u<u'\;,
\end{equation}
and  converging to 
the full terminal set $\CS^*$ as $u\rightarrow u_*$, we can consider a set of WdW patches ${\cal W}_u$, defined as the intersection between $D^- (\CS^*)$ and 
the causal past of ${\CS}^*_u$, 
\begin{equation}\label{wdwu}
{\cal W}_u = J^- ({\CS}^*_u) \cap D^- ({\CS}^*)\;.
\end{equation}

\begin{figure}[h]
$$\includegraphics[width=8cm]{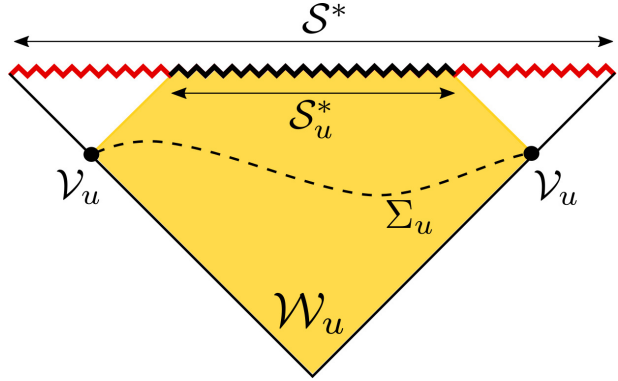} $$   
\begin{center}
\caption{\emph{ The WdW patch ${\CW}_u$ (in yellow), associated to a given ${\CS}^*_u$ subset (in black) of the full terminal set $\CS^*$ (in red).  The codimension-two set ${\CV}_u$ is the (possibly disconnected) boundary of Cauchy surfaces $\Sigma_u$ for ${\CW}_u$. }\label{fig:wdwu}}
 \end{center}
\end{figure}

For any given ${\cal W}_u$, its Cauchy surfaces $\Sigma_u$  have a common codimension-two boundary ${\cal V}_u = \partial \Sigma_u$ (cf. Figure \ref{fig:wdwu},) which would hold the `holographic data' for ${\cal W}_u$. For example, ${\cal V}_u$ is a spatial section of the event horizon when $\CS^*$ is a black-hole singularity.   Therefore, we would like to interpret the `area' of ${\CV}_u$ in Planck units \footnote{ We henceforth refer to  codimension-two volumes as `areas'.} as a measure of the effective number of holographic degrees of freedom `flowing' into the terminal subset ${\CS}^*_u$, assigning therefore the entropy

\begin{equation}
S= \dfrac{{\rm Area}[{\CV}_u]}{4 G}\, ,
\end{equation}

to the singular subset.

In defining the WdW patches ${\cal W}_u$ we may give privilege to the `anchors', namely the codimension-two sets ${\cal V}_u$, or alternatively we may consider  the nested family ${\cal S}^*_u$, as more fundamental. These two constructions  are not completely equivalent, since the WdW patch anchored at ${\cal V}_u$ may fail to intersect ${\cal S}^*$ at sufficiently `early times'. In this chapter we are more interested in the asymptotic `late-time' behaviour in which ${\cal W}_u$ does have a non-trivial boundary component at the singularity. Therefore, we tacitly adopt in what follows the nesting construction of the WdW patches and we will often refer to the
associated complexity measures as `nesting complexity'. 

\begin{figure}[h]
$$\includegraphics[width=6cm]{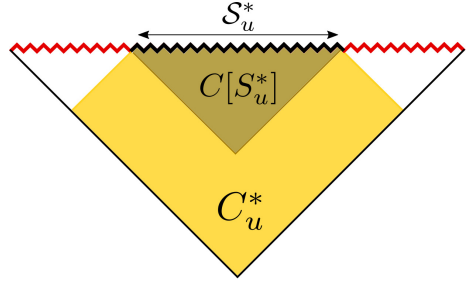} $$   
\begin{center}
\caption{\emph{ The difference between $\CC^*(u)$ and $\CC[{\cal S}^*_u] $ as determined by the different domains of integration. }\label{fig:nest}}
 \end{center}
\end{figure}

With the previous definitions, we are led to the following definition of nesting complexity associated to the given family of WdW patches ${\cal W}_u$, 
\begin{equation}
\label{terminaldef}
\CC^*(u)=  \,I[{\CW}_u]  \;,
\end{equation}
where $I[{\CW}_u]$ denotes the on-shell gravitational action, now integrated over the WdW patch ${\cal W}_u$. Once this nesting complexity is defined, we can now recover the notion of `total complexity flow' into the singularity, which was loosely defined in (\ref{fullc}), as the asymptotic limit of the nesting procedure. More precisely, we have
\begin{equation}\label{toal}
\CC[{\cal S}^*] =  \lim_{u\to u_*} \CC^*(u) \;,
\end{equation}
It is important to notice that, when considering singular subsets ${\cal S}_u^*$, the nesting complexity $\CC^*(u)$ is {\it different} from the `total complexity'   $\CC[{\CS}^*_u]$ flowing into ${\cal S}^*_u$, as shown in Figure \ref{fig:nest}. In other words,  
we regard $\CC[\CS^*]$ as the AC-analog of ${\CS}_\infty$, (cf. Figure \ref{fig:c1},)  and $\CC^*(u)$ as the AC-analog of ${\CS}'_t$, (cf. Figure \ref{fig:c2}.) 

\subsection{Technical aside: action prescription}
\label{sec:actionprescription}

Before entering into any details of terminal complexity, we must clarify some technical points. In particular, the gravitational action for the WdW patch is not unique, and some prescriptions are to be setlled down. As shown in \cite{Poisson}, the gravitational action for some patch $\CW$ with arbitrary boundaries can be defined with the following prescription for codimension-one and codimension-two boundary terms

\begin{eqnarray}
\label{masterformula}
16 \pi G \, \tilde{I}[\CW] &=& \int\limits_{ \CW} \text{d}^{d+1}x \,\sqrt{-g}\,(R-2\Lambda + \mathcal{L}_m) \\ \nonumber &+& 2 \sum\limits_{\CT_i} \int\limits_{ \CT_i} \text{d}^dx \,K   + 2 \sum\limits_{\CS_i} \sgn(\CS_i)\int\limits_{ \CS_i} \text{d}^d x \,K  \\ &-&  2 \sum\limits_{\CN_i} \sgn(\CN_i)\int\limits_{ \CN_i} \text{d}^dx \,\text{d}\lambda \,\hspace{0.1cm}\kappa  +2 \sum\limits_{j_i} \sgn(j_i)\oint\limits \text{d}^{d-1}x \,\alpha_{j_i} \, , \nonumber
\end{eqnarray}
where in this expression

\begin{itemize}
\item $\CT_i$, $\CS_i$ and $\CN_i$ are respectively the timelike, spacelike and null boundaries of the WdW patch and $K$ are the traces of the corresponding extrinsic curvatures for the first two cases. For the null boundaries, $\lambda$ represents an arbitrary parameter on null generators of $\CN_i$, with $\kappa$ the surface gravity associated to $\CN_i$ in this parametrization. The signs for the null and spacelike boundary contributions are defined depending on the relative location of the boundary to the WdW patch as follows $$ \text{sgn} \left( \raisebox{0pt}{\includegraphics[height=0.26cm]{spacelikesign.pdf}} \right)=\text{sgn} \left( \raisebox{-5pt}{\includegraphics[height=0.5cm]{null2.png}} \right)=\text{sgn} \left( \raisebox{-5pt}{\includegraphics[height=0.5cm]{null3.png}} \right)=-\text{sgn} \left( \raisebox{-2pt}{\includegraphics[height=0.26cm]{spacelikesign2.pdf}} \right)=-\text{sgn} \left( \raisebox{-5pt}{\includegraphics[height=0.5cm]{null4.png}}\right)=-\text{sgn} \left( \raisebox{-5pt}{\includegraphics[height=0.5cm]{null1.png}} \right)   =1.$$

\item $j_i$ are the codimension-two junctions between boundary components. For those joints that are formed by at least one null boundary, the form of the integrand in \eqref{masterformula} is given by

\begin{equation}
\alpha_{j_i} = 
\begin{cases}
\log |k_\mu n^\mu| \\
\log |k_\mu s^\mu|\\
\log |\frac{1}{2}k_\mu \bar{k}^\mu|,
\end{cases}
\end{equation}

where $k^\mu$ and $\bar{k}^\mu$ are taken to be the future directed vectors tangent to the null surfaces and $n^\mu$ ($s^\mu$) is the future-directed (outward-directed) unit normal to the spacelike (timelike) surface. For such terms the signs $\sgn(j_i)$ are simply given by the product of the surfaces signs. Joints that are formed only by spacelike and timelike boundaries have a different set of rules that we will not cover here as they do not appear on WdW patches (see \cite{Poisson} for a full discussion of all possible joint actions).
\end{itemize}

The boundary terms in \eqref{masterformula} require special consideration. As we see, the boundary piece of the action is given by a sum of contributions from  codimension-one and codimension-two components of the boundary $\partial {\cal W}$. The non-null codimension-one pieces and their joints  are given by the standard York--Gibbons--Hawking (YGH) term and a set of well understood joint contributions.  On the other hand, some formal choices are necessary in the  presence of null codimension-one pieces as the freedom to choose the parametrization of null generators renders the on-shell action a gauge-dependent quantity. The physics behind these choices remains somewhat unclear (see for example the considerations in \cite{Poisson, Ross, MyersFormation, WhichAction, SusskindJT},) and different prescriptions can lead to qualitatively different behaviours for the action and its dynamics. Accordingly, it becomes a central issue to settle for one of such prescriptions in order to perform any sort of detailed study of complexity growth phenomenology. A way to circunvent this problem was pointed out in \cite{Poisson} where it was shown that it is possible to enforce reparametrization invariance of the gravitational action by the addition of an extra counterterm depending on the expansion of codimension-two sections along the null boundaries
\begin{equation}
\label{expansioncounterterm}
\Theta = \partial_\lambda \,\log\,\sqrt{\gamma}
\;. 
\end{equation}
We shall refer to this addition as the  expansion counterterm:
\begin{equation}
\label{Ict}
I_{\Theta} = \sum\limits_i\dfrac{\text{sgn}(N_i)}{8\pi G} \int\limits_{N_i}  \text{d}\lambda\, \text{d}^{d}x \;\sqrt{\gamma}\; \Theta \;\log(\ell_\Theta\, \cdot |\Theta|).
\end{equation}

The appearance of the new length scale $\ell_\Theta$ is interesting. It represents a qualitatively new feature of AC complexity which, as we will see, activates itself  precisely in cases where the entropy has a dynamical behaviour and the effective Hilbert space supporting the singularity  changes its dimension. The presence of $I_\Theta$ has been regarded as necessary to  guarantee the positivity of complexity \cite{Ross} as well as the correct black hole complexity dynamics from collapsing geometries and the verification of the switchback effect \cite{VaidyaI, VaidyaII, WhichAction}. Its precise meaning in microscopic treatments inspired by the notions of circuit complexity remains however quite mysterious (cf. \cite{JeffersonMyers, Myersmixed, Myerscoherent, MyersTFD }). 

In the following we will adopt such prescription for our terminal complexity, meaning that our definition \eqref{terminaldef} is to be understood with
\begin{equation}\label{def}
I[{\cal W}_u] =  \tilde{I} [{\cal W}_u] + I_\Theta [{\cal W}_u]\;,
\end{equation}
with $\tilde{I}$ and $I_\Theta$ given by \eqref{masterformula} and \eqref{Ict} respectively. A careful evaluation of the boundary terms and counterterm will in fact be crucial for our purposes, since we are precisely interested in situations with non-trivial null-expansion.

\section{The local component of terminal complexity}
\label{sec:local}

\noindent

As emphasized in the introduction, a remarkable property of the AC complexity prescription is the occurrence of a quantitatively important  contribution coming directly from the singularity through the evaluation of the YGH term. Despite the fact that its validity is tied to that of the effective theory description in a strong gravity regime, this contribution is argued to be of vital importance when computing the AC complexity, and seems to be the only object that is known so far in the holographic dictionary providing an order one contribution to a finite quantity which is calculated in the vicinity of a singularity.

The YGH contribution is local and formally extensive over the singular set ${\cal S}^*$. However, the volume form is not generally defined at ${\cal S}^*$, which makes the notion of `extensivity' non-trivial.  In order to elucidate this point, let us parametrize the near-terminal metric by  a Gaussian normal  coordinate $\tau$. This foliates  the near-terminal  spacetime into spacelike surfaces $\Sigma_\tau$, according to the proper-time distance to $\CS^*$. In defining a metric on the $\Sigma_\tau$ slices, we extract a conventional power of the proper time according to the ansatz
\begin{equation}\label{locs}
\dd s^2 = -\dd{\tau}^2 + (\tau H)^{2\gamma/d} \, \dd\Sigma_\tau^{\;2}\;,
\end{equation}
where $H^{-1}$ sets the characteristic scale for the expansion away from  the terminal set. In general,  the $d$-dimensional  metric  $\dd\Sigma_\tau^2$  does not have a smooth limit  as $\tau \rightarrow 0^+$, but we may choose the conventional exponent $\gamma$
in such a way that its volume form   does have a smooth limit. We shall actually assume that this volume form  is analytic in $\tau$, since this will be a property of all examples we study (it would be interesting to assess the generality of this assumption.) We will refer to such notion of volume for ${\CS}^*$ as the `comoving volume' of the terminal set and 
denote its measure as $\dd {\rm Vol}_c $.

In this notation, the YGH term in the action is  computed as
\begin{equation}\label{ygh}
\tilde{I}[{\CS}^*]_{\rm YGH}= {1\over 8\pi G} \lim_{{\tau}\rightarrow 0^+} \partial_{\tau} \big[(H{\tau})^{\gamma} \,{\rm Vol}_c [\Sigma_\tau]\big]\;.
\end{equation}
 Picking the term proportional to the comoving volume ${\rm Vol}_c [{\cal S}^*]$ of the singular set,  we find that the YGH term vanishes for $\gamma >1$ and is infinite for $\gamma <1$, except perhaps the case $\gamma =0$ where the answer depends on the possible occurrence of logarithmic terms in the terminal expansion near $\tau=0$. The most interesting case is $\gamma =1$, for which one defines a nontrivial  `comoving complexity density' at the singular set, given by $H/8\pi G$. 
 
 The black hole singularity has $\gamma =1$ and thus presents a purely local contribution to complexity. In fact, this feature appears to be quite general. At spherically symmetric black-hole singularities we have a vanishing ${\bf S}^{d-1}$ and an expanding `radial' direction. Hence, the metric is locally of the Kasner form, i.e. 
 \begin{equation}\label{km}
 \dd s^2 = -\dd \tau^2 + \sum_{j=1}^{d} (H\tau)^{2p_j } \, \dd \sigma_j^2\;,
 \end{equation}
 with a particular choice of Kasner parameters $p_c = \frac{2}{{d}}$ for  $d-1$ `crunching' directions and $p_r = -1+2/d$ for the `ripping' direction. More generally, the Kasner parameters are restricted to satisfy 
 the sum rules $\sum_j p_j  = \sum_j p_j^2 =1$ and any such metric can be put in the form (\ref{locs}) with $\gamma =1$, with
 `comoving' metric
  \begin{equation}\label{comov}
 \dd \Sigma_\tau^2 = \sum_j (\tau H)^{2p_j  - 2/d} \,\dd \sigma_j^2
 \;.
 \end{equation}
  In particular, it has a smooth comoving volume form,  
 \begin{equation}
 \dd{\rm Vol}_c [ \Sigma_\tau] = \wedge_{j=1}^d \dd \sigma_j
 \end{equation}
  as a simple consequence of the sum rule $\sum_j p_j =1$.

   The $\gamma=1$ property and the resulting non-vanishing `complexity density' persist if we let the Kasner parameters depend smoothly  on the `longitudinal' $\sigma_j$. In fact, the  classic results of ref. \cite{BKL,BKL2,BKL3} (BKL) indicate that such a `generalized Kasner'  metrics 
 furnish a good local approximation of the near-singular region (after a slight generalization involving local rescalings and frame rotations.)

 \subsection{Local terminal complexity and coarse-graining}

\noindent

\begin{figure}[t]
$$\includegraphics[width=8cm]{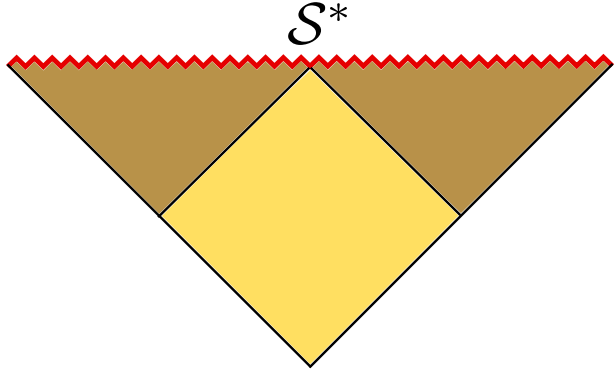} $$   
\begin{center}
\caption{\emph{ When the bulk action is dominated by a negative cosmological constant, the bulk contribution to the AC terminal complexity of  $\CS^*$ is subextensive. While the YGH contribution is extensive over $\CS^*$, the bulk contribution is more negative for the larger set (yellow) than it is for
the union of the smaller sets (brown).  }\label{fig:subextensive}}
 \end{center}
\end{figure}

\begin{figure}[t]
$$\includegraphics[width=8cm]{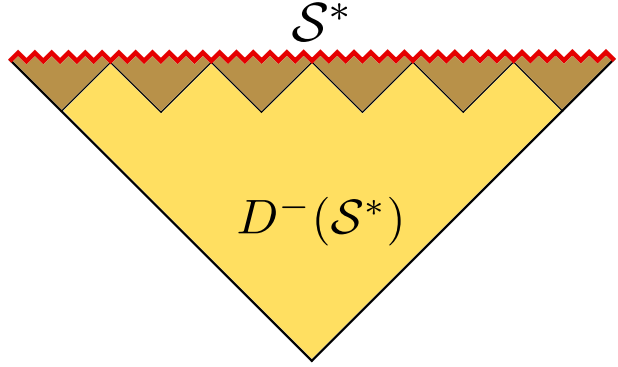} $$   
\begin{center}
\caption{\emph{ When the bulk volume remains sufficiently bounded in the vicinity of the terminal surface, the local complexity of ${\CS}^*$ results from the limit of an extreme coarse graining.  }\label{fig:ultralocal}}
 \end{center}
\end{figure}

The remarkable properties of the local YGH contribution  beg the question of whether we may be able to isolate this term in more physical terms. A natural strategy in this case is to focus on the extensivity of the local contribution, a property not shared by the full AC complexity. As the evaluation of the null boundary contributions generally requires a complete knowledge of the metric, we take in this section a more conservative point of view, and study the pieces of AC complexity that are purely geometrical, i.e. that do not depend on formal `gauge' choices of parametrization or the addition of extra counterterms. We define thus this `geometric complexity' and denote it as $\CC_g$ as the Einstein-Hilbert and YGH contributions to the action on $D^-(\CS^*)$
\begin{equation}\label{ac}
\CC_g^*(u)= \tilde{I}[D^-(\CS^*)]_{\rm bulk} + \tilde{I}[{\cal S}^*]_{\rm YGH}\;.
\end{equation}
In order to illustrate the point on the extensivity of this quantity, we can do so by focusing on the simpler case of vacuum solutions. A vacuum solution is a $(d+1)$-dimensional Einstein manifold whose metric satisfies
\begin{equation}\label{eme}
R_{\mu\nu} = {2\Lambda \over d-1} g_{\mu\nu}\;,
\end{equation}
with cosmological constant $\Lambda$ and no matter degrees of freedom. The bulk contribution to the on-shell action is then
proportional to the spacetime volume
 \begin{equation}\label{bulkc}
 I[X]_{\rm bulk} = {1\over 16\pi G} \int_X (R-2\Lambda) = {\Lambda \over 4\pi G (d-1)} \,{\rm Vol} [X]\;.
 \end{equation}
Assuming a $\gamma=1$ singular set ${\cal S}^*$  with non-vanishing complexity density, we have a full geometric complexity given formally by
\begin{equation}\label{fullq}
\CC_g[{\cal S}^*] =  {\Lambda \over 4\pi G (d-1)} {\rm Vol}\left[D^- ({\cal S}^*)\right] + {H \over 8\pi G} {\rm Vol}_c [{\cal S}^*] \;.
\end{equation}
While the YGH term is extensive along the comoving volume of ${\cal S}^*$, the bulk contribution is extensive in the full spacetime volume of the past domain of dependence. Considering the case $\Lambda <0$, as corresponds to states in an AdS/CFT context, we have a negative-definite bulk contribution, leading to a `subextensivity' property of the full quasilocal complexity. Indeed, under a
coarse-graining of the singular set ${\cal S}^* = \cup_i {\cal S}^*_i$ as indicated in Figure \ref{fig:subextensive}, the expression (\ref{fullq}) satisfies 
\begin{equation}\label{sube}
\CC_g[{\cal S}^*] = \CC_g \left[\cup_i {\cal S}^*_i \right] \leq \sum_i \CC_g[{\cal S}^*_i]\;.
\end{equation}
The inequality is reversed (corresponding to superextensivity) for vacuum singularities in $\Lambda >0$ spaces. The deviation from
extensivity  would disappear if the bulk contributions were to become negligible, a situation we may expect in the limit of extreme coarse graining, illustrated in Figure \ref{fig:ultralocal}. 

\begin{figure}[t]
$$\includegraphics[width=8cm]{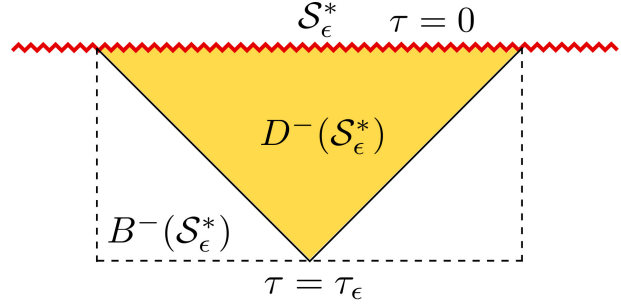} $$   
\begin{center}
\caption{\emph{ Comparison between the $\epsilon$-box and the past domain of dependence of ${\cal S}^*_\epsilon$. }\label{fig:box}}
 \end{center}
\end{figure}

In this case, the limit of infinite coarse-graining does isolate the YGH term. To be more precise, we require that the bulk contributions be consistently smaller than the YGH contribution for small subsets of ${\cal S}^*$. We can check this explicitly for $\gamma =1$ vacuum singularities  described by (\ref{fullq}) and admitting a local Kasner description. Let us consider a fine partition of the singular set by subsets ${\cal S}^*_\epsilon$ with comoving volume of order $\epsilon^d$. The condition for the bulk contribution to be negligible for small sets is that
\begin{equation}\label{epd}
{{\rm Vol}\left[D^- ({\cal S}^*_\epsilon)\right] \over {\rm Vol}_c \left[{\cal S}^*_\epsilon\right]} \sim \epsilon^{\,a}
\;,\end{equation}
with $a >0$. 
Instead of computing the volume of the past domain of dependence, $ D^- ({\cal S}^*_\epsilon)$,  it is easier to compute the volume of the larger set $B^- ({\cal S}^*_\epsilon)$, which `boxes' it
in the standard coordinate frame. If $\tau_\epsilon$ is the maximal value of the $\tau$ coordinate in $D^- ({\cal S}^*_\epsilon)$, the
$\epsilon$-box is defined by the full $\tau\leq \tau_\epsilon$ subset with given comoving coordinates covering ${\cal S}^*$, c.f. Figure (\ref{fig:box}). 
Evidently, ${\rm Vol}\left[B^-({\cal S}^*_\epsilon)\right] \geq {\rm Vol}\left[D^- ({\cal S}^*_\epsilon)\right]$, so that it is enough to establish the condition (\ref{epd}) for $B^- ({\cal S}^*_\epsilon)$.

In order to construct explicitly $B^- ({\cal S}^*_\epsilon)$  for the Kasner metric (\ref{km}) we define ${\cal S}^*_\epsilon$ to be a  $d$-dimensional cube  in the $\sigma$ coordinates with common extent $\Delta \sigma_j = \epsilon$. Its comoving volume is ${\rm Vol}_c \left[{\cal S}^*_\epsilon\right] = \epsilon^d$ and the past domain of dependence, $D^- ({\cal S}^*_\epsilon)$, is a trapezoid with base ${\cal S}^*_\epsilon$ and a  ridge with the topology of a $(d-1)$-dimensional cube, determined by the intersection of light rays in the spacetime plane with faster past-convergence. For any direction $\sigma_j$ we can define a corresponding conformal time coordinate $\eta_j$ such that light rays propagate with unit slope in the $(\eta_j, \sigma_j)$ plane. The explicit relation between $\eta_j$ and the proper time is
\begin{equation}
(1-p_j) H \eta_j = (H\tau)^{1-p_j}
\;,
\end{equation}
where $p_j$ is the Kasner exponent in the direction $\sigma_j$. Light rays whose $\sigma_j$ separation is $\epsilon$ at  $\tau=0$ converge in the past at $\tau_\epsilon^{(j)}$ given by
\begin{equation}
H\tau_\epsilon^{(j)} = \left((1-p_j) {H\epsilon \over 2}\right)^{1\over 1-p_j}\;.
\end{equation}
  Thus, the past domain of dependence of the full ${\cal S}^*_\epsilon$ set is determined by the smallest $\tau_\epsilon^{(j)}$ or, equivalently, by the {\it largest} Kasner exponent which we denote by $p_c$:
\begin{equation}\label{taue}
H\tau_\epsilon = \left((1-p_c){H\epsilon \over 2}\right)^{1 \over 1-p_c}\;.
\end{equation}
 With these ingredients we can compute the volume
  of the $\epsilon$-box as
  \begin{equation}\label{vole}
  {\rm Vol}\left[B^- ({\cal S}^*_\epsilon)\right] = \int_0^{\tau_\epsilon} d\tau \,(H\tau) \, \epsilon^d = {1\over 2H} \,(\tau_\epsilon H)^2 \,\epsilon^d\;,
  \end{equation}
  and verify (\ref{epd}) with $a = 2(1-p_c)^{-1}$. 
  
For   solutions with matter degrees of freedom we need to check that the Lagrangian ${\cal L}_m$ is not too singular. For instance, we may study the  FRW terminal with metric 
\begin{equation}\label{frw}
ds^2 = -d\tau^2 + a(\tau)^2 \,d\Sigma^2\;
\end{equation}

which presents a singularity at $\tau=0$. By construction, the non-singular comoving volume is just given by the volume of the homogeneous and isotropic surfaces $\Sigma$, so that the complexity exponent  $\gamma$ can be read off from the short-time asymptotics of the scale factor $a(\tau)$. Since FRW metrics require non-trivial matter degrees of freedom, we follow standard practice and model them as a perfect fluid with squared speed of sound equal to $w = p/\rho$, where $p$ denotes the pressure and $\rho$ the energy density. Then, we have the standard solution $\rho \;a^{d(1+w)} = $ constant, which leads to 
$a(\tau) \sim \tau^{2 /d(1+w)}$ or, equivalently 
\begin{equation}
\gamma_{\rm FRW} = {2 \over 1+w}\;.
\end{equation}

Approximating the action dimensionally as the volume integral of the energy density $\rho \sim \tau^{-2}$, we estimate
\begin{equation}\label{aces}
I\left[B^- ({\cal S}^*_\epsilon)\right]_{\rm bulk} \propto \epsilon^{\,d} \int_0^{\tau_\epsilon} d\tau \,\tau^\gamma {1\over \tau^2} \sim \epsilon^d \,\tau_\epsilon^{\gamma-1}\;.
\end{equation}

\begin{table}
\label{tablaFLRW}
\begin{center}
\begin{tabular}{|c|c|c|c|c|c|}
\hline
  & \cellcolor{gray!25} AdS & \cellcolor{gray!25} Milne &\cellcolor{gray!25} Matter & \cellcolor{gray!25}Radiation & \cellcolor{gray!25}Stiff matter\\
\hline
 \cellcolor{gray!25} $\omega$ &  $-1$ & $2/d -1$ & $0$ & $1/d$ &		$1$ \\
\hline
\cellcolor{gray!25} $\gamma$ &  $\infty$ & $d$ & $2$ 		& $2d/(d+1)$ & $1$		 \\
\hline
\cellcolor{gray!25} $a(\tau)$ & $\sim e^{H\tau}$ &  $\sim \tau $ &  $\sim \tau^{2/d}$  & $\sim \tau^{2/(d+1)}$ & 	$\sim \tau^{1/d}$	\\
\hline
\end{tabular}
\end{center}
\caption{\textit{Scalings for different cosmological solutions as classified by the equation of state of a perfect fluid.} }
\end{table}

The condition for the coarse-graining procedure to be well-defined is now
\begin{equation}\label{genco}
{I\left[B^- ({\cal S}^*_\epsilon)\right]_{\rm bulk} \over {\rm Vol}_c [{\cal S}^*_\epsilon]} \sim \epsilon^{\,a}\;, \;\;\;a>0\;.
\end{equation}
The physical condition that  the matter equation of state remains  strictly below the stiff limit, $w<1$, implies that $\gamma >1$ and thus (\ref{genco}) is satisfied 
 provided $\tau_\epsilon$ scales with a positive power of $\epsilon$. This
happens for any solution which {\it decelerates} away from the singularity, since the FRW conformal time is given by 
\begin{equation}\label{frwco}
H\eta = \dfrac{d}{d-\gamma}\left(H \tau\right)^{d-\gamma \over d}\;.
\end{equation}
It is precisely for decelerating singularities that we have $\gamma < d$ and $\tau_\epsilon \sim \epsilon^{d\over d-\gamma}$
scaling with a positive power of $\epsilon$, leading to an automatically well-defined coarse-graining limit.

\begin{figure}[t]
$$\includegraphics[width=9cm]{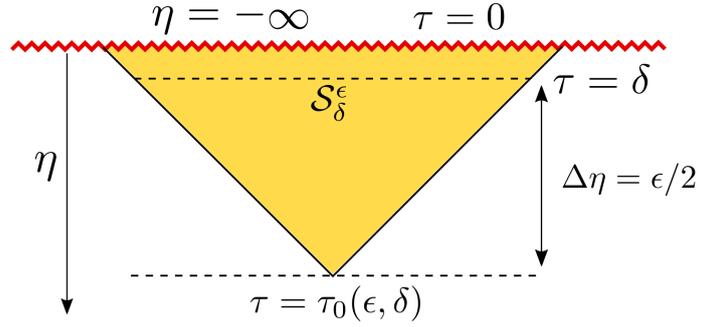} $$   
\begin{center}
\caption{\emph{The regularized terminal ${\cal S}^\delta_\epsilon$ and its past domain of dependence.  }\label{fig:frwacc}}
 \end{center}
\end{figure}

The situation is less clear for FRW metrics that {\it accelerate} away from the singularity, corresponding to $\gamma \geq d$. Now the FRW conformal time plummets to $-\infty$ as $\tau\rightarrow 0^+$. The problem in this case is that $D^- ({\cal S}^*_\epsilon)$ is not itself well defined, as any past light cone emanating from $\tau =0$ and converging at a finite value $\tau_0$ subtends an infinite comoving volume at the terminal surface. To address this point we regularize the terminal surface by bringing it slightly before the singularity at $\tau =\delta$, as indicated in Figure (\ref{fig:frwacc}).  In other words, we compute the past domain of dependence for a small, $\epsilon$-sized subset of $\Sigma_\delta$ rather than ${\cal S}^*$. Let us denote this set ${\cal S}^\delta_\epsilon$. Its past  domain of dependence, $D^- ({\cal S}^\delta_\epsilon)$, has an earliest proper time which is a function of both $\epsilon$ and $\delta$, 
\begin{equation}
\tau_0 (\epsilon,\delta)  = {1\over H} \left((H\delta)^{d-\gamma \over d} + {d-\gamma \over d} {H\epsilon \over 2}\right)^{d \over d-\gamma}\;.
\end{equation}
For $\gamma >d$, this quantity  vanishes linearly in $\delta$ as the terminal time cutoff is removed  at fixed $\epsilon$. 
Hence, when we repeat the estimate 
(\ref{aces}) we find that 
\begin{equation}\label{newaces}
I\left[B^-({\cal S}^\delta_\epsilon)\right]_{\rm bulk} \propto \epsilon^{\,d} \int_{\delta}^{\tau_0} d\tau\, \tau^\gamma \,{1\over \tau^2} \sim \epsilon^d (\tau_0^{\,\gamma -1} - \delta^{\,\gamma-1}) \longrightarrow 0\;,
\end{equation}
as $\delta \rightarrow 0$ at fixed $\epsilon$, since both terms vanish in the limit. Therefore, the bulk contribution vanishes when we remove the regularization at fixed comoving volume, even before we take $\epsilon \rightarrow 0$.

As we have seen thus, the YGH piece of the action can be isolated as a local contribution to the complexity under an extreme coarse-graining procedure. In this respect, the relation of the YGH term  to the full quasilocal  complexity  is analogous to the relation between the  classical thermodynamic entropy, obtained through coarse-graining, and the exact von Neumann entropy of a quantum many-body state.


\chapter{Entropic locking of terminal complexity}
\label{ch:locking}

\noindent
The persistence of complexity growth well after the entropy of the black hole has stabilized is considered to be a fundamental property. More precisely, since the conjectured bound on complexity is non-perturbative in $1/S$, the bulk expansion parameter, the total AC/VC complexity accumulated by an eternal black hole is infinite when computed in leading orders, in correspondence to the infinite volume and action of the black hole interior.

Away from the benchmark example of a static black hole, however, neither the positive growth of complexity nor its decoupling from the entropy dynamics are self evident at all, begging the question of whether these features are to hold for more generic solutions enjoying non-trivial dynamics. As the black hole possess a constant entropy, we may in fact interpret the terminal AC as a measure of the purely infrared contribution to complexity, including only those degrees of freedom which are actually involved in the holographic emergence of the black hole interior. Considering time-dependent UV/IR thresholds in AdS/CFT constructions, we may generalize this setup to situations with dynamical entropy, interpreting the IR Hilbert space as enjoying a time-dependent dimension and associating a quasilocal terminal complexity to them.

\begin{figure}[h]
\begin{center}
\includegraphics[width=6cm]{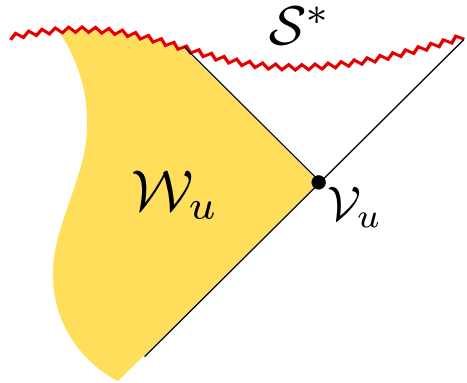} 
\caption{\emph{ The WdW patch $\CW_u$ (in yellow), parametrized by a null coordinate on the boundary of the singularity's past domain of dependence. Entropy is measured by the volume of ${\cal V}_u$ and complexity is measured by the action of ${\cal W}_u$.
This construction is modified in the obvious manner under time reversal.}}
\label{fig:terminalcomplexity}
\end{center}
\end{figure}

More precisely, given a singular spacetime of the type shown in Figure \ref{fig:terminalcomplexity},  with a terminal singularity ${\cal S}^*$ and a horizon bounding the past domain of dependence $D^-({\cal S}^*)$, we can ask how the accumulated complexity compares with the entropy. The formal definitions laid down in section \ref{sec:quasilocal} allow us to perform a sharp comparison, exploring the phenomenology trough a set of qualitatively different dynamical examples in which ${\rm Area}({\cal V}_u)$ varies strongly as $u\rightarrow u_*$ on approaching the singularity. Our results indicate that the rate of complexity growth is generically dominated by the finite-size effects in the Hilbert space, namely the rate of variation of the entropy, rather than the standard process of `entanglement weaving' which leads to linear complexity growth in black holes.

More specifically, this `entropic locking' of complexity is found in two different situations  which illustrate qualitatively different dynamics as classified by the behaviour of its entropy. Our choices are motivated by the ability to compute exactly the complexity on terminal WdW patches using the specific formula \eqref{terminaldef},  but also by our interest in exposing as much as possible the contrast between entropy and complexity when referred to cosmological singularities. One case corresponds to expanding bubbles of Coleman-de Luccia  type, engineered in concrete AdS/CFT scenarios.  Here both the terminal AC complexity and the entropy diverge at the singular locus and we are interested in the relative rates of divergence. The second example is a portion of  the Kasner spacetime, which is known to locally approximate any spacelike singularity  in GR. In this case the entropy vanishes and the terminal AC complexity approaches a constant. 

Our main result is that, as the singularity is approached, $u \rightarrow u_*$, in both classes of examples,  the  terminal AC growth is completely controlled by that of the entropy  $S$ through a law of the form
\begin{equation}
\label{law}
 \delta { \cal C^*} =  a\, \delta S+\dots  \,
\;,
\end{equation}
with $a$ a non-universal constant. Here, $S={\rm Vol} ({\cal V}_u)/4G$ and the dots stand for  $u$-independent contributions or subleading terms as $u\rightarrow u_*$.  More specifically, for the case of expanding bubbles we find the sign of the $a$ coefficient depending on the counterterm scale $\ell_\Theta$, implying that a positive rate of complexity growth  actually requires  picking a sufficiently large value of this scale, as measured in units of the AdS radius of curvature. In the Kasner case, the positivity of the coefficient $a$ is guaranteed by the weaker condition $\LL \gg \ell_{\text{Planck}}$ yielding a decreasing complexity as the effective Hilbert space is reduced.

Before embarking in our tour of examples, we would like to comment briefly on the relation to previous work. In \cite{BarbonRabinoComplexity}  the VC complexity was estimated for a number of cosmological singularities which are naturally embedded into concrete AdS/CFT constructions. In these examples it was found that a regularized version of the VC complexity was monotonically {\it decreasing} on approaching the singularity, in contrast with our statement here for the cuasilocal complexity. A similar behavior was obtained for the AC ansatz in the same examples by \cite{rabinoac}. The reason for this apparent discrepancy is simply that the full complexity computed in \cite{BarbonRabinoComplexity} is dominated by UV contributions to the VC ansatz, and these are highly dependent on the particular details of the embedding into asymptotically AdS geometries. For instance, some of the examples are based on singular CFT metrics which shrink to zero size, and others involve expanding domain walls in the bulk. In the first case it is natural that the UV contribution to complexity should have a negative derivative in time, as corresponds to a shrinking Hilbert space on the full CFT. In the second case, a time-dependent conversion between UV and IR degrees of freedom is introduced in the CFT by switching on a relevant operator with a time-dependent coupling, and the c-theorem explains why the UV again dominates the balance.    Therefore, there is no contradiction since the two monotonicity statements refer to different quantities. The positive monotonicity of the quasilocal complexity defined here (by restriction of the AC/VC ansatz to the interior of $D^- ({\cal S}^*)$,) is compatible with the negative monotonicity of the full complexity, particularly when  the latter is dominated by a strong UV time-dependence.

\section{Vacuum terminals with divergent entropy}

\noindent

The primary example is that of a singularity inside an expanding bubble embedded in an ambient AdS spacetime. Such solutions look like standard crunching cosmologies of FLRW type, where the singularity eventually crunches the whole AdS spacetime in a finite time as measured by the asymptotic global time. The boundary of the bubble has an acceleration horizon which serves as the boundary of the past causal domain $D^- ({\cal S}^*)$. As a result, the entropy of this crunch singularity is infinite.

\begin{figure}[h]
$$\includegraphics[width=8cm]{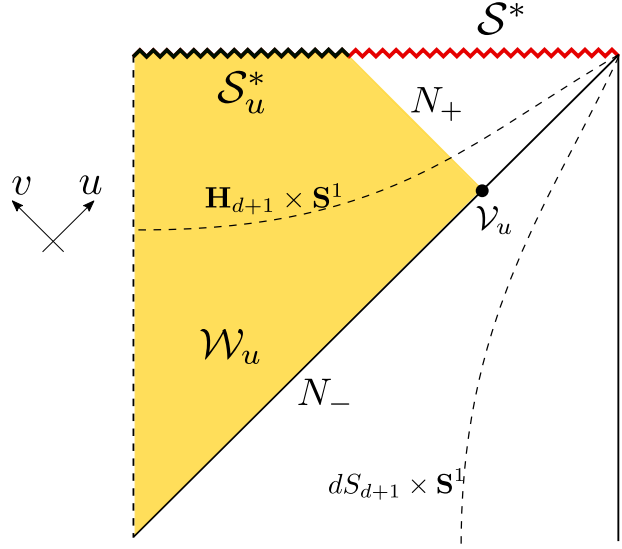} $$   
\begin{center}
\caption{\emph{ The causal structure of the topological crunch singularity and its associated WdW patch. }
\label{fig:tc}}
 \end{center}
\end{figure}

In order to rely on analytic methods,\footnote{For a mereley cosmetic reason, we adopt only in this section the convention that $d$ always stands for the dimensionality of $\CV_u$, i.e. we will work in AdS$_{d+2}$ spacetimes.} we first consider the `topological cruch' model (cf. \cite{Banados, maldapim, BarbonRabinoCrunches}),   which describes a time-dependent compactification of pure AdS$_{d+3}$ with topology AdS$_{d+2} \times {\bf S}^1$, where the ${\bf S}^1$ shrinks to zero size in finite boundary time, producing a spacelike singularity in the interior as shown in Figure \ref{fig:tc}. A holographic interpretation of this model uses a CFT$_{d+2}$ on a spatial manifold with topology $ {\bf S}^{d} \times {\bf S}^1$, where  the sphere is static and the circle shrinks to zero size in finite time. A conformally related description is that of the same CFT on a fixed-size circle, times a de Sitter spacetime.

In the FLRW patch of the $AdS_{d+2}$ the metric for this model is given by
\begin{equation}
\text{d}s^2 = -\text{d}\tilde{t}^{\,2} +  \sin^2 (\tilde{t}\,) \,\text{d}{\bf H}_{d+1}^2 + \cos^2 (\tilde{t}\,) \,\text{d}\phi^2 \, ,
\end{equation}
where we have set the AdS scale to unity and also consider that to be the curvature scale of the boundary metric. The element $d{\bf H}_{d+1}^2$ stands for the unit metric on the $(d+1)$-dimensional Euclidean hyperboloid ,whereas the coordinate $\phi$ is the angle that parametrizes the compact circle ${\bf S}^1$. The time $\tilde{t}$ foliates the cosmology with sections of topology ${\bf H}_{d+1} \times {\bf S}^1$, producing a singularity $\mathcal{S}^*$ when the circle shrinks to zero size at time $\tilde{t}_*=\pi/2$.

In order to perform the required computations, it is useful to define null coordinates on $D^- ({\cal S}^*)$ as follows. First, we introduce a radial coordinate $\chi$ over ${\bf H}_d$ and a conformal time variable $\eta =2\tan^{-1} \left( e^{\tilde{t}/\ell}\right)$ over the AdS$_{d+1}$ factor, so that the metric is written in the form
 \begin{equation}\label{mc}
 ds^2 =\sin^2 (\tilde{t}\,) \left[ -d\eta^2 + d\chi^2 + \sinh^2 (\chi) \,d\Omega_{d}^2\right] +  \cos^2 (\tilde{t}\,) d\phi^2\;.
\end{equation}
Next, we  introduce null compact Kruskal coordinates  
\begin{eqnarray}
\tan u=  e^{\eta + \chi} \;, \qquad \tan v = e^{\eta - \chi} \, ,
\end{eqnarray}
for which we get the full analytic continuation of the metric in the form

\begin{equation}
\label{topmetric}
\text{d}s^2 =  \sec ^2(u-v) \left[ -4 \,\text{d}u \text{d}v + \sin ^2(u-v) \,\text{d}\Omega^2_{d} + \cos^2 (u+v)\, \text{d} \phi^2 \right].
\end{equation}

Considering now the set of nested WdW patches $\CW_u$ labelled by the null coordinate $u$, we are ready to calculate the different contributions to the on shell action from the prescription \eqref{masterformula}. As we will only care about asymptotic behaviours and not the exact full action, it will suffice to consider stripes $\delta \CW_u$ of thickness $\delta u$ for which computations render sensibly simpler results. The bulk action piece of such stripe is given by

\begin{eqnarray}
I_{\text{bulk}}[\delta \CW_u] &=& -\dfrac{(d+2)}{8 \pi G} \int\limits_{\delta \CW_u} \text{d}^{d+3}x \, \sqrt{-g} \\
&=& -\dfrac{2(d+2)V_\Sigma}{8 \pi G} \int\limits^{u+\delta u}_{u} \text{d}u  \int\limits_0^{\pi/2-u} \text{d}v \; \dfrac{\tan^d(u-v)}{\cos^3(u-v)}\cos(u+v)\,.
\end{eqnarray}
where $V_\Sigma$ stands for the area of the ${\bf S}^{d}\times {\bf S }^1$ manifold. Performing this integral and expanding around $u\sim u_*=\pi/2$ we can obtain the asymptotic limit for the bulk action growth

\begin{equation}
\dfrac{\text{d} }{\text{d}u} I_{\text{bulk}}[ \CW_u] \approx -\dfrac{V_\Sigma}{8 \pi G} \,\dfrac{2(d+2^{-d-1})}{d+1}\left( \dfrac{1}{u_*-u} \right)^{d+1}.
\end{equation}

In order to calculate the codimension-one boundary terms, we will choose to parametrize all null boundaries affinely, so that the extrinsic curvature vanishes and such contributions are identically zero. The only non-trivial YGH contribution will be that of the spacelike boundary at the singularity, which is located at $\tau \equiv v+u =\pi /2$. This term is given by
\begin{equation}
I_{\text{YGH}}[ \CW_u] = -\dfrac{1}{8 \pi G} \int\limits_{\CS^*_u} \text{d}^{d+2}x \,\sqrt{\gamma}\,K \, ,
\end{equation} 
where $\gamma$ is the induced metric on $\CS^*_u$ and the integrand is calculated using $\sqrt{\gamma} \, K = g_{\tau \tau}^{-1/2} \, \partial_\tau \sqrt{\gamma}$. Performing the integral again for the slab of thickness $\delta u$ we get the growth rate and its late time limit
\begin{equation}
\dfrac{\text{d} }{\text{d}u}I_{\text{YGH}}[ \CW_u]  = \dfrac{V_\Sigma}{8 \pi G} \dfrac{\tan^{d+1}(2u)}{\sin^{}(2u)}\approx \dfrac{V_\Sigma}{8 \pi G} 2^{-d-1}\left( \dfrac{1}{u_*-u} \right)^{d+1}.
\end{equation}
The only codimension-one contribution that is now left is that of the expansion counterterms \eqref{Ict}. Let us consider thus the WdW null boundaries, which will be given by constant $u,v$ hypersurfaces. For instance, we may start with the past boundary, given by the surface $N_- \equiv (u,0,\Omega_0, \phi_0)$, where the coordinate $u$ here will parametrize the geodesic. Introducing this curve into the geodesic equation, however, we can see that such parametrization is not affine, but rather has the following surface gravity
\begin{equation}
\kappa(u) = 2 \tan u \,.
\end{equation}
Following the standard procedure, we can find an affine parameter $\lambda_-$ from the relation
\begin{equation}
\dfrac{\text{d} \lambda_-}{\text{d}u} = \exp{\int\limits_0^u\kappa(\sigma) \, \text{d} \sigma}\, ,
\end{equation}
which for our case yields
\begin{equation}
\lambda_- = \dfrac{1}{\alpha_-} \tan u\,,
\end{equation}
and we have introduced a constant $\alpha_-$ that parametrizes the freedom to shift the affine parameter. From \eqref{topmetric} we can extract the determinant of the induced transverse metric as well as its expansion and express them in terms of $\lambda_-$
\begin{eqnarray}
\sqrt{\gamma} &=& (\alpha_-\lambda_-)^{d}\, ,\\
\Theta &=&  \dfrac{d}{\lambda_-}\, .
\end{eqnarray}
Feeding this into the counterterm definition \eqref{Ict} we get the contribution to the action
\begin{eqnarray}
\label{counterU}
I^-_{\Theta} &=& \dfrac{V_\Sigma}{8 \pi G} \int\limits_0^{\frac{1}{\alpha_-}\tan u}  \text{d}\lambda_- \, \alpha_- \,(\alpha_-\lambda_-)^{d-1}d \, \log \left(\LL \dfrac{d}{\lambda_-} \right) \\  
&=&\dfrac{V_\Sigma}{8 \pi d G} \tan ^{d}u \, (d \, \log (d \, \LL \, \alpha_- \cot u)+1)\,.
\end{eqnarray}
A similar procedure gives us the affine parameter for the future null boundary $N_+$
\begin{equation}
\lambda_+ = \dfrac{1}{\alpha_+} \tan (u-v)\, ,
\end{equation}
which yields
\begin{equation}
\sqrt{\gamma} = (\alpha_+   \lambda_+ )^{d} (\alpha_+  \lambda_+  \sin (2 u)+\cos (2 u)),
\end{equation}
\begin{equation}
\Theta = \frac{1}{\lambda_+ }\left(d+1-\frac{1}{\alpha_+  \lambda_+  \tan (2 u)+1} \right),
\end{equation}
and we can calculate the corresponding counterterm \footnote{Although this integral is analytically solvable, the result is rather cumbersome and not particularly illuminating. We omit therefore such explicit expression since we will only care about its late time expansion.}

\begin{eqnarray}
\label{counterV}
\scriptsize
I^+_{\Theta} &=&  -\dfrac{V_\Sigma}{8 \pi G} \int\limits^{\frac{1}{\alpha_+} \tan(2u-\frac{\pi}{2})}_{\frac{1}{\alpha_+} \tan u} \text{d}\lambda_+ \, \text{d}^{d+1}x \, \sqrt{\gamma} \,\Theta \,\log(\LL \,|\Theta|).
\end{eqnarray}
Finally, we must also calculate the contribution from the codimension-two joints of $\CW_u$. As the joint $N_+ \cap \CS^*_u $ has vanishing volume, the only one that will produce a non-trivial contribution is $\CV_u = N_- \cap N_+$. Following the rules on section \ref{sec:actionprescription}, the contribution from this joint is given by  
\begin{equation}
\label{Sjoints}
I_{\CV_u} =  -\dfrac{1}{8 \pi G} \oint\limits_{\CV_u} \text{d}^{d+1}x \, \sqrt{\sigma} \, \log | \tfrac{1}{2}k_+ \cdot k_- |\,  ,
\end{equation}
where $k_+$ and $k_-$ are respectively the null tangent vectors of the future and past boundaries. We choose these vectors to be
\begin{eqnarray}
k_+ &=& \alpha_+ (dT+dX)\,,\\
k_- &=& \alpha_- (dT-dX)\,,
\end{eqnarray}
so that they satisfy $k_{\pm} \cdot \partial_T=\alpha_{\pm}$ and $\alpha_{\pm}$ are normalization constants. Substituting these values into \eqref{Sjoints} we get the contribution of the joint
\begin{eqnarray}
\label{jointint}
I_{\CV_u} &=& -\dfrac{1}{8 \pi G} \log \left(\alpha_+ \, \alpha_-   \dfrac{\cos^2u}{2}\right) \oint \text{d}^{d+1}x \, \sqrt{\sigma}  \\ &=& \nonumber -\dfrac{V_\Sigma }{8 \pi G}\log \left(\alpha_+ \, \alpha_-   \dfrac{\cos^2u}{2}\right)  \tan^{d} u\, .
\end{eqnarray}

We observe that the dependence on $\alpha_\pm$ cancels out when \eqref{counterU}, \eqref{counterV} and \eqref{jointint} are added up, and accordingly we can set them to 1 in these expressions. Collecting all results for a late time expansion $u \sim u_*$, we get the following  behaviour for the contributions
\begin{eqnarray}
I_{\text{bulk}} &\approx & -\dfrac{S}{\pi d} \left( \dfrac{d+2^{-d-1}}{d+1}\right)\, , \label{topbulk}\\ 
I_{\text{YGH}} &\approx & \dfrac{S}{\pi d} \times 2^{-d-1}\, ,\label{topYGH} \\
I^-_{\Theta}  &\approx & \dfrac{S}{2\pi d}\left[ 1+ \log\left(\dfrac{V_\Sigma}{4G} \right)+d\log\left(d\, \LL\right)-\log S\right], \\
I^+_{\Theta}  &\approx & \dfrac{S}{2\pi d} \left[ f(d) + \log\left(\dfrac{V_\Sigma}{4G} \right)+d\log\left(d\,\LL\right)-\log S \right], \\
I_{\CV_u}  &\approx & \dfrac{S}{2 \pi d} \left[d\log(2)- 2 \log\left(\dfrac{V_\Sigma}{4G} \right)  +2 \log S \right], 
\end{eqnarray}
where we are dropping terms in fractional powers of the entropy $S$ along the horizon, defined as  
\begin{eqnarray}
\label{entropytop}
S &=& \dfrac{1}{4 G} \oint  \text{d}^{d+1}x \, \sqrt{\sigma}  \\ &=& \nonumber \dfrac{V_\Sigma }{4 G} \tan^{d} u \approx \dfrac{V_\Sigma}{4 G}\left( \dfrac{1}{u_*-u} \right)^{d},
\end{eqnarray}
and keeping only the leading and next-to-leading contributions to the action. The coefficient $f(d)$ is an $\CO(1)$ positive constant given by
$$\scriptstyle  (d+1)f(d)=  \left(d^2+d-1\right) \, _2F_1\left(1,d;d+1;2+\frac{2}{d}\right)-d^2+2 (d+1) d \left(\coth ^{-1}(d+1)-\, _2F_1\left(1,d+1;d+2;2+\frac{2}{d}\right)\right)+2^{-d} \, _2F_1\left(1,d;d+1;1+\frac{1}{d}\right).$$

After adding up all contributions, we can see that the $S\log S$ divergence cancels out, yielding a total late-time complexity dynamics linear in the entropy

\begin{equation}
 \CC^* \approx   a\, S
\end{equation}
with
\begin{equation}
a= \dfrac{1}{\pi d} \left[\dfrac{-d-2^{-d-1}}{d+1} +  2^{-d-1} + \dfrac{1}{2} \left(1+f(d)+d\log (2) \right) +d\log \left( d\, \LL\right) \right].
\end{equation}

As we see, the complexity growth is fully controlled by that of the entropy, diverging as measured by any null coordinate along the horizon. The sign of such growth however will depend on the coefficient, which is essentially controlled by the size of $\LL$,  yielding a positive rate when $\LL \gtrsim 1$ for any dimension.

\begin{figure}[h]
$$\includegraphics[width=5cm]{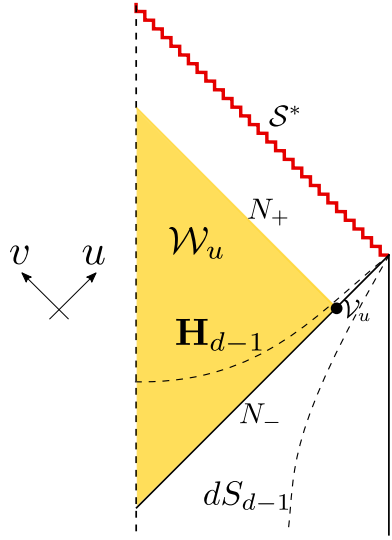} $$   
\begin{center}
\caption{\emph{ The idealized  state generating a null singularity by the collision of a thin-walled bubble with dS$_{d+1}$ worldvolume. Notice that WdW patches anchored at the horizon do not touch the singularity. }\label{fig:tw}}
\label{fig:thinwall}
\end{center}
\end{figure}

It would be interesting to generalize this result to more general solutions with expanding horizons. The need to consider scalar fields with non-trivial potentials generally prevents us from a completely analytic treatment. However, we can offer evidence that the result found is quite robust by examining a similar situation in the so-called thin-wall approximation. Suppose that the bubble has a very narrow outer shell, so that we can describe it as a thin wall expanding into AdS$_{d+2}$, with a de Sitter induced metric. In this case the singularity can be regarded as a null future-directed surface emerging from the boundary  impact time at $t=t_*=\pi $  (see Figure \ref{fig:thinwall}). 

The formal analysis is very similar to that of the topological crunch with the difference that the shrinking ${\bf S}^1$ is not present now. The bulk metric can be obtained therefore simply removing the $\text{d} \phi^2$ factor in \eqref{topmetric} 
\begin{equation}
\label{thinmetric}
\text{d}s^2 =  \sec ^2(u-v) \left[ -4 \, \text{d}u \text{d}v + \sin ^2(u-v) \, \text{d}\Omega^2_{d} \right],
\end{equation}
and the calculations follow very easily from the previous ones. In effect, the bulk contribution is obtained as

\begin{eqnarray}
I_{\text{bulk}}[\CW_u] &=&  -\dfrac{(d+1)V_\Omega}{8 \pi G} \int\limits^{u}_{0} \text{d}u' \int\limits^{u'}_{0} \text{d}v \; \dfrac{\tan^d(u'-v)}{\cos^3(u'-v)} \approx \dfrac{V_\Omega}{8 \pi G d} \left( \dfrac{1}{u_*-u} \right)^{d}.
\end{eqnarray}
where as usual we have performed an expansion around $u \sim u_*$ in the last equality. As the timelike surface at $u=v$ has vanishing induced metric, all codimension-one boundaries of this WdW patch vanish when the parametrization is taken to be affine along the null ones $N_{\pm}$. We need however to compute the counterterms for the later. For both boundaries we get the quantities \footnote{We omit here the normalization constants $\alpha_\pm$ as its cancellation is analogous to the previous case}
\begin{eqnarray}
\lambda_\pm &=&  \tan (u-v)\,, \\
\sqrt{\gamma}_{\pm} &=& (\lambda_\pm)^{d}\,, \\
\Theta_\pm &=& \dfrac{d-1}{\lambda_\pm}\,.
\end{eqnarray}
And the counterterms are very similar to \eqref{counterU}
\begin{eqnarray}
\label{TWct}
I^-_{\Theta} = I^+_{\Theta} &=& \dfrac{V_\Omega}{8 \pi G} \int\limits_0^{\tan u}  \text{d}\lambda \,  \lambda^{d-1} d \, \log \left(\LL\, \dfrac{d}{\lambda} \right) \\  
&=&\dfrac{ V_\Omega}{8 \pi d G} \tan^{d}u \, (d \, \log (d \, \LL \cot u)+1), \\
 &\approx& \dfrac{S}{2\pi d} \left(1+\log \left( \dfrac{V_\Omega}{4G}\right)+ d\log \left( d \,\LL\right) -\log S \right) ,
\end{eqnarray}
where we substituted again the entropy along the horizon, which remains identical as in the previous example \eqref{entropytop}. The joint contribution will also have the same form as in the previous example, therefore cancelling again the $S \log S$ leading divergence in $I_{\Theta}^{\pm}$ for the asymptotic limit $u\sim u_*$. Adding up all pieces of the action we may obtain the total complexity
\begin{eqnarray}
\CC^*\approx  a \,S \,,
\end{eqnarray}
with
\begin{eqnarray}
a= \dfrac{1}{\pi d} \left[-1+\dfrac{1}{2}d\log(2)  + d\log \left( d \,\LL \right)\right],
\end{eqnarray}
an expression that tells us again that the leading divergence is guaranteed to be positive as long as the counterterm scale satisfies $\LL \gtrsim 1$.

\begin{figure}[h]
$$\includegraphics[width=4cm]{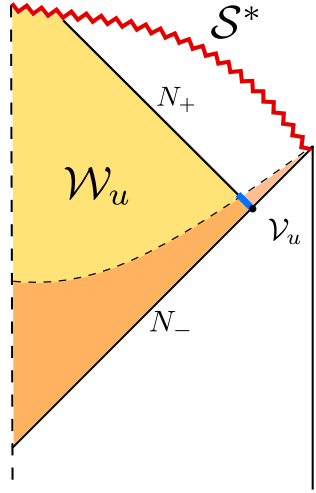} $$   
\begin{center}
\caption{\emph{ Maximally extended FLRW solution embedded in AdS space. The early-times approximation $a(\tilde{t}\,)\approx \tilde{t}$ guarantees that the geometry in the vicinity of $N_-$ (shaded region) is the same as in the pure AdS solutions, implying that $I^-_\Theta$ and $I_{\mathcal{V}_u}$ are identical to those of the vacuum cases. As the same is true for the near-horizon piece of $I^+_\Theta$ (in blue), the logarithmic divergence is generically cancelled provided no new dominating contributions arise from the deep interior piece of $I^+_\Theta$.}}
\label{fig:thickwall}
\end{center}
\end{figure}

As both the latter examples come from spacetimes that are locally AdS, one might be suspicious about the seemingly miraculous cancellation of the logarithmic divergence. Nonetheless, there are reasons to think that such cancellation is generic for more realistic solutions involving a backreacting scalar field that drives a homogeneous cosmology behind the horizon. Indeed, for any negatively curved FLRW solution to enjoy a smooth horizon, the behaviour of the scale factor around the time origin must be of the form $a(\tilde{t}\,) \approx \tilde{t}$. This scaling coincides with the structure present in the near-horizon region from the previous examples and therefore will yield the very same contributions for the $I_\Theta^-$ and $I_{\CV_u}$ terms. The future null counterterm $I^+_\Theta$ on the other hand, will pick up information from the full cosmological solution which in turn will depend on the details of the particular model. In the two analytic examples above, however, it is easy to see that the integrals \eqref{counterV} and \eqref{TWct} are strongly dominated by the near-horizon region, the one responsible of the logarithmic divergence. The far interior, on the other hand, contributes at most linearly to the complexity. As the cosmological solution must be that of the thin wall approximation for a neighbourhood around the horizon, the logarithmic contribution from $I^+_\Theta$ will remain identical and accordingly will produce the generic cancellation of the leading divergence (cf. Figure \ref{fig:thickwall}). This of course does not prevent a possible restoration of higher than linear divergences in both the bulk and  $I^+_\Theta$ contributions when the full cosmological dynamics is taken into account. The estimation of such effects is beyond the scope of this work.

\section{A vacuum terminal with vanishing entropy}

In this section we consider an example in which the entropy of the singularity, as defined by the volume of codimension-two sets ${\cal V}_u$, has precisely the opposite behaviour to the expanding bubble models,  namely it vanishes at the singular locus. Consider the  Kasner metric in $d+2$ dimensions 
\begin{equation}
\label{kasner}
\dd s^2 = -\dd t^2 + \sum_{j=1}^{d+1} (H\,t)^{2p_j} \dd x_j^2\;,
\end{equation}
which conforms the standard vacuum solution with zero cosmological constant and $\mathbb{R}^{d}$ symmetry. Here, $H$ is an inverse-length setting a characteristic value for the expansion away from  the terminal set. Rather than a merely formal solution, the Kasner metric provides a local approximation for `small portions' ($H\Delta x_j \ll 1$) of more general singularities (cf. \cite{BKL, BKL2,BKL3,libro}), making it an object of special interest in the study of terminal spacetimes. Einstein equations impose the coefficients $p_i$ to be restricted to satisfy $\sum_i p_i = \sum_i p_i^2 =1$, where at least one of the exponents $p_i$ must be negative, indicating that at least one direction stretches as one approaches the singularity. The volume of codimension-two surfaces sitting at some constant value of $t$  always vanishes as $|t|$ in the  $t\rightarrow 0$ limit. 

In order to adapt the discussion to the symmetries of the Kasner metric, we  pick one spatial coordinate, say $x_1$, and define the singular set to be a `slab' ${\cal S}^* = \left[-{u_* \over 2} , {u_*\over 2} \right] \times {\bf R}^{d-1}$ for some finite $u_*$ measuring the $x_1$ coordinate length of the finite interval. The set of WdW patches indicated in Figure \ref{fig:kasner} intersect ${\cal S}^*$ along the nested  `slabs' ${\cal S}^*_u =   \left[-{u \over 2} , {u\over 2} \right] \times {\bf R}^{d-1}$. 

In this construction, we regard the holographic data as specified on ${\cal V}_u = \{-{u\over 2}\} \times {\bf R}^{d-1} \cup  \{{u\over 2}\} \times {\bf R}^{d-1}$, and the $x_1$ coordinate takes the role of `holographic' emergent direction. The fact that the holographic data lies on disconnected spaces, interpolated by the `bulk' WdW patches, makes this construction similar to the standard eternal AdS black hole and its dual product CFTs \cite{Maldacenaeternal}, with the crucial difference that here the entropy density 
\begin{equation}
{\tilde S} = {1\over V_{{\bf R}^{d-1}}} {{\rm Vol} ({\cal V}_u )\over 4G} \, ,
\end{equation}
vanishes as  the singularity is approached in the limit $u\rightarrow u_*$.

Let us pick spacetime units so that the `Hubble rate' $H=1$, and pass to conformal coordinates 
\begin{equation}
\label{kasnermetric}
\text{d}s^2 = \left((1-p_1) \tau \right)^{\frac{2p_1}{1-p_1}} \left(-\text{d}\tau^2 + \text{d}x_1^2 +  \sum\limits_{i=2}^{d+1}\left((1-p_1)\tau\right)^{\frac{2(p_i-p_1)}{1-p_1}} \text{d}x_i^2 \right),
\end{equation} 
where $\tau$ is the conformal time in the $(t, x_1)$ plane. As in this case we are dealing with a Ricci flat solution, the bulk contribution of $\CW_u$ will be trivially zero. It will suffice thus to compute the expansion counterterms, joints and the YGH contribution from the singular locus. For the latter, it is easy to see from \eqref{kasnermetric} that $\sqrt{\gamma}\,K=1$, implying the very simple contribution
\begin{equation}
I_{\text{YGH}}= \dfrac{V_{{\bf R}^{d-1}}}{4\pi G}\, u \,,
\end{equation}
where $V_{{\bf R}^{d-1}}$ is the volume of the non-compact directions, appearing here in the sense that we may define a finite complexity density $\widetilde{\CC}=\CC /V_{{\bf R}^{d-1}}$. In order to compute the counterterms, we may define as usual the null coordinates by $\tau, x_1 = \frac{1}{2} (u\pm v)$. Starting with the past boundary $N_- = (u,v_0,\vec x_{i,0}),$ we find that its surface gravity is given by

\begin{figure}[h]
$$\includegraphics[width=9cm]{kasner.pdf} $$   
\begin{center}
\caption{\emph{ WdW patch for a Kasner slab ${\cal S}^*_u = \left[-{u \over 2} , {u\over 2} \right] \times {\bf R}^{d-1}$. }}
\label{fig:kasner}
\end{center}
\end{figure} 

\begin{equation}
\kappa = -\frac{2 p_1}{(p_1-1) (u+v_0)}\, ,
\end{equation}
and following the standard procedure we get the correct affine parameter
\begin{equation}
\lambda_-(u) =  \dfrac{1}{\alpha_-}(u+v_0)^{1/\delta}\, ,
\end{equation}
where we defined for simplicity $\delta= \frac{1-p_1}{1+p_1}$. We can calculate now the induced volume element and the expansion
\begin{eqnarray}
\sqrt{\gamma} &=& -\dfrac{\delta}{1+\delta}(\alpha_- \lambda_-)^\delta\,, \\
\Theta &=& \dfrac{\delta}{\lambda_-}\,.
\end{eqnarray}
Feeding it into the counterterm action we get
\begin{eqnarray}
I_{\Theta}^- &=& \dfrac{ V_{{\bf R}^{d-1}}}{8 \pi G} \int\limits_{\lambda_-(u_0)}^{\lambda_-(u)} \text{d} \lambda_- \, \dfrac{\delta^{2}}{\delta+1} \,\alpha_-^\delta \, \lambda^{\delta-1} \, \log\left(\dfrac{\LL \,\delta}{\lambda_-} \right) \\  
&=& -\dfrac{ V_{{\bf R}^{d-1}}}{8 \pi G} \dfrac{1}{\delta+1}(u+v_0)\left[   1+ \delta \log \left(\delta \, \LL \alpha_- \right) -\log \left(u+v_0 \right)  \right]  - (u \leftrightarrow u_0) .\nonumber
\end{eqnarray}

Repeating the procedure above for the future null boundary $N_+$ we can see that the counterterm contribution is given by an almost identical expression where we must exchange the roles $u, u_0 \leftrightarrow u$ and $v_{0} \leftrightarrow v_{0}, v_*$ with $v_*=-v_0=-u_0=u_*$ the value at the singularity.
\begin{eqnarray}
I_{\Theta}^+ &=& -\dfrac{ V_{{\bf R}^{d-1}}}{8 \pi G} \int\limits^{\lambda_+(v_*)}_{\lambda_+(v_0)} \text{d} \lambda_+ \, \dfrac{\delta^{}}{\delta+1} \,\alpha_+^\delta \, \lambda_+^{\delta-1} \, \log\left(\dfrac{\LL \,\delta}{\lambda_+} \right) \\  
&=& -\dfrac{ V_{{\bf R}^{d-1}}}{8 \pi G} \dfrac{1}{\delta+1}(u+v_0) \left[  1+ \delta\,\log \left( \delta \LL \alpha_+  \right) -\log \left(u+v_0 \right) \right] .\nonumber
\end{eqnarray}
Now, we must calculate the contribution from the joints of these surfaces. Choosing the normal vectors
\begin{eqnarray}
k_+ &=& \alpha_+ (d\tau+dx_1)\,,\\
k_- &=& \alpha_- (d\tau-dx_1)\,,
\end{eqnarray}
we get the contribution
\begin{equation}
I_{\CV_u}= \dfrac{ V_{{\bf R}^{d-1}}}{4 \pi G} \dfrac{\delta}{\delta+1}\left(u+v_0\right) \log \left( \alpha_+ \, \alpha_- \,\left((u+v_0)\dfrac{\delta}{\delta+1} \right)^{\frac{\delta-1}{\delta}} \right)\,,
\end{equation}
and similarly for the south tip
\begin{equation}
I_{\southcorner}= -\dfrac{ V_{{\bf R}^{d-1}}}{8 \pi G} \dfrac{\delta}{\delta+1}\left(u_0+v_0\right) \log \left( \alpha_+ \, \alpha_- \,\left((u_0+v_0)\dfrac{\delta}{\delta+1} \right)^{\frac{\delta-1}{\delta}} \right)\,,
\end{equation}
and as we see, the dependence on $\alpha_{\pm}$ will cancel with that of the counterterms above. The entropy along the horizon in this case will be

\begin{equation}
S= -\dfrac{2\delta}{\delta+1} \dfrac{V_{{\bf R}^{d-1}}}{4G}(u+v_0)\, ,
\end{equation}
so we may rewrite again everything as a function of the entropy
\begin{eqnarray}
\label{kasnerYGH}
I_{\text{YGH}} &=& \dfrac{\delta+1}{2\pi \delta} (S_0/2-S)\, ,\\
2(I_{\Theta}^+ +I_{\Theta}^-)&=&\dfrac{S}{\pi \delta} \left[ 1+\log\left( \frac{\delta}{\delta+1}\, \frac{V_{{\bf R}^{d-1}}}{2G}\right) +\delta \log( \delta \,\LL )-\log S \right] - \dfrac{1}{2} \left( S \leftrightarrow S_0 \right)\, , \\
I_{\CV_u} &=& \dfrac{S}{\pi\delta} \dfrac{\delta-1}{2}\left[ \log\left(\dfrac{V_{{\bf R}^{d-1}}}{2G}\right) -\log S\right]\, ,\\
I_{\southcorner} &=& \dfrac{S_0}{ \pi\delta} \dfrac{\delta-1}{4}\left[ -\log\left(\dfrac{V_{{\bf R}^{d-1}}}{2G}\right) +\log S_0\right]\,,\label{kasnertip}
\end{eqnarray}
where $S_0$ is a constant that stands for the entropy evaluated at the south tip of the WdW patch $(u_0, v_0$). As we see, we recover the same structure for the counterterm as in the expanding case \eqref{TWct} and the black hole (cf. \cite{Myerstdep}) but in which the constant $\delta$ seems to play now the role of the codimension-2. Interestingly such effective dimension can in fact recover the value $\delta = d$ by considering the `holographic coordinate' $x_1$ to be the ripping direction in the most isotropic case, i.e. $p_1 = \frac{1-d}{d+1}$ and $p_{i \neq 1}=\frac{2}{d+1} $. Such values for the Kasner exponents correspond furthermore to the near-singularity approximation of the Schwarzschild black hole metric.

From equations \eqref{kasnerYGH} to \eqref{kasnertip}, we see that we obtain a log-linear locking for the total complexity density in the Kasner solution. Particularly, we can extract a growth rate of the form
\begin{eqnarray}
\delta \widetilde{\CC}^* &=&  \left( a\, + b\log \tilde{S} \right) \delta \tilde{S}\, ,
\end{eqnarray}
with
\begin{eqnarray}
a &=&\dfrac{1}{\pi \delta} \left[ -\delta + \dfrac{\delta+1}{2} \log \left(\dfrac{1}{2G}\right) +\log\left( \dfrac{\delta}{\delta+1} \right)+ \delta\log(\delta \, \LL )  \right]\, ,\\
b &=& -\dfrac{\delta+1}{2 \pi \delta}\,  ,
\end{eqnarray}
where we have restored the  entropy density $\tilde{S}=S/V_{{\bf R}^{d-1}}$. As we can see, the terminal behaviour of the complexity is now controlled by the balance of the linear and logarithmic pieces, in turn depending on the UV/IR character of the scale $\LL$. Within the validity of the Kasner approximation, however, we are forced to consider entropies that do not exceed the size of the effective Kasner patch, i.e. $S \ll G^{-1}$, implying that the term in $\sim \log G$ becomes always dominant at late times. Since $\delta$ is a strictly positive quantity, this linear dominance cannot be contrarested by any choice of $\LL$ as doing so would require to pick transplanckian values of the counterterm, taking us out of the realm of the effective theory action. Accordingly, we obtain the generic late time behaviour
\begin{equation}
\delta \widetilde{\CC}^* \approx a\,  \delta \tilde{S}\, .
\end{equation}
where $a$ is a positive quantity by the considerations above. For the case of the Kasner singularity, we therefore see that complexity is a decreasing quantity, an effect that signals the dominance of the strong shrinking of the effective Hilbert space over the natural tendency of complexity to grow in systems at equilibrium. 

\newpage\null\thispagestyle{empty}\newpage

\chapter{Epilogue: holographic complexity and the arrow of time }
\label{ch:arrow}

\noindent

As we have seen in the previous chapters, the local contributions to complexity that are picked up by the boundary contribution at the singularity have qualitatively different behaviour depending on the nature of the solution. If we regard the generalized Kasner behaviour as `generic' we may be tempted to say that all spacelike singularities tend to have a non-vanishing  local complexity density. On the other hand, there are important examples of singularities whose YGH contribution vanishes, as is the case of those  occurring in standard FLRW spacetimes. 

To bring this simple point home, we can apply \eqref{ygh} to the standard FLRW metric
\begin{equation}
ds^2 = -d\tau^2 + a(\tau)^2 \,d\Sigma^2\;,
\end{equation}
with a singularity at $\tau=0$. As mentioned in the previous section, standard solutions to the Einstein equations for a perfect fluid with equation of state $p=\omega \rho$ yield terminal metrics of the form \eqref{locs} with  $\gamma = {2 \over 1+w}$. This result implies that the only FLRW singularity with a finite complexity density is the slightly unphysical case with `stiff matter', $w=1$, leading to $\gamma=1$. On the other hand, a formally infinite contribution to the complexity density, associated to $\gamma <1$, would require $w>1$  in the FLRW context, i.e. a violation of  the energy conditions on the matter degrees of freedom. 

Imposing the physical condition that the matter is strictly below the `stiff' limit, $w<1$, we have $\gamma >1$, implying  a {\it vanishing} local complexity. Hence, we find that  `ordered' singularities of FLRW type have a vanishing local contribution to holographic complexity whereas more generic (Kasner) ones do not. This feature suggests that holographic complexity might be a suitable diagnostic tool to classify GR singularities according to their nature, an effort which goes back to \cite{Penrose, misner, BKL, BKL2, BKL3 }.

\begin{figure}[t]
\begin{center}
\includegraphics[height=6cm]{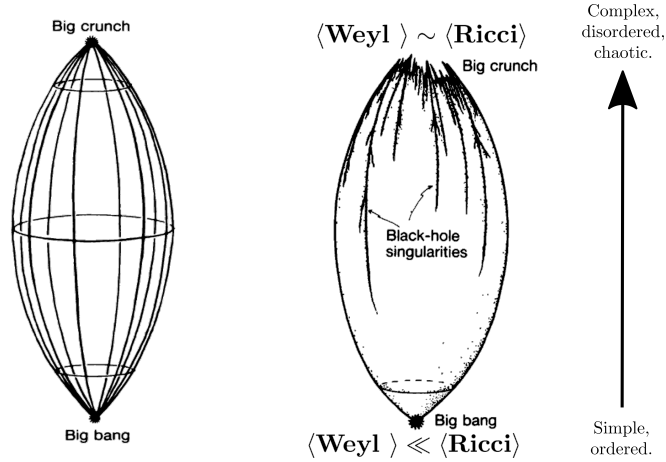}
\end{center}
\caption{\emph{Schematic picture representing Penrose's argument. In the left, a perfectly homogeneous bouncing cosmology. In the right, a more realistic cosmology undergoing a crunch with scattered black holes. }}
\label{fig:penrosecriterion}
\end{figure}

In \cite{Penrose} Penrose gave a local criterion for the complexity of a singularity. The basic observation was that `ordered' singularities, like the ones in FLRW models, have vanishing Weyl curvature, whereas more generic ones, such as those arising in gravitational collapse, have a generically divergent Weyl tensor.  Penrose argued that the Weyl criterion would be associated to a large gravitational entropy flowing into the singularity, a suggestion based on the heuristic picture of a generic cosmological crunch, full of chaotic black hole collisions (see Figure \ref{fig:penrosecriterion}). Since black holes are known to carry entropy, a corresponding notion of entropy may be assigned to the union of all singularities enclosed by the colliding black holes.

As the Weyl tensor does not seem to be directly related with any known measure of entropy in the context of AdS/CFT, one basic observation of this section is that holographic complexity, rather than entropy, might provide a more natural language in the seek for a classification of singularities, a picture supported by the behaviour of the YGH contribution to AC complexity in the Kasner and FLRW spacetimes.

As it turns out, however, there is an interesting twist to this story. According to the classic  BKL analysis \cite{BKL,BKL2,BKL3}, the vicinity of a generic spacelike singularity is not quite described by a single generalized Kasner metric, but rather an oscillating  regime where a series of `epochs' succeed one another, each epoch being locally described by a generalized Kasner solution of the type (\ref{km}). The values of the Kasner parameters, $p_j$, change from one epoch to the next one in a deterministic but chaotic manner \footnote{Since the proposal of the BKL hypothesis, some objections have been raised (e.g. \cite{BarrowTipler}) that question the validity of the approximation. Numerical simulations of gravitational collapse (cf. \cite{Singularitynumerics1, Singularitynumerics2} ) however have shown agreement with the BKL picture.}. In addition, the frame determining the special coordinates $\sigma_j$ in (\ref{km}) undergoes a rotation, and furthermore the induced volume form at fixed $\tau$ is rescaled by a finite factor which we may absorb in the dimensionful expansion parameter $H$. As a result, the geometry in the $n$-th epoch is well approximated by 
\begin{equation}\label{nth}
ds^2 |_{(n)} = -d\tau^2 + (H_n \tau)^{2/d} \;d^2\Sigma_\tau^{(n)}\;,
\end{equation}
where $d^2 \Sigma_\tau^{(n)}$ is a rotated version of (\ref{comov}) with Kasner parameters $p_j^{(n)}$. All epochs are described
by $\gamma=1$ metrics but, crucially, they have slightly different parameters $p_j^{(n)}$ and $H_n$.  In particular, the substitution rule for the expansion parameter follows the recurrence relation
\begin{equation}\label{eva}
H_{n+1} = (2p_r^{(n)} + 1) \,H_n\;,
\end{equation}
where $p_r^{(n)} <0$ is always the `ripping' parameter of the $n$-th epoch. Since $2p_r^{(n)} + 1 < 1$ for all $n$, the series of $H_n$ is monotonically decreasing. 

\begin{figure}[h]
$$\includegraphics[width=5cm]{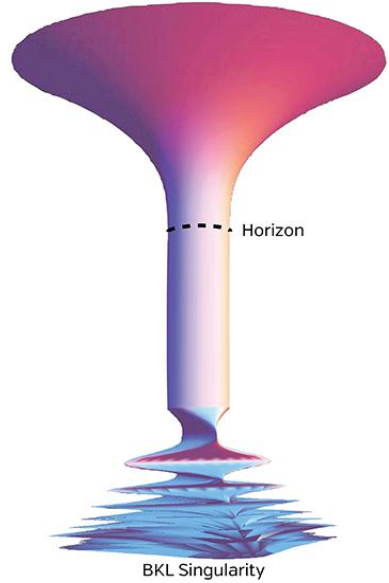} $$   
\begin{center}
\caption{\emph{ Schematic picture of a BKL singularity formed in a generic anisotropic black hole collapse. }
\label{fig:BKLsingularity}}
\end{center}
\end{figure}

If we compute the YGH contribution to complexity by placing a regulating surface and taking the limit, the result of the complexity density is determined by the limit of $H_n$. Namely it is proportional to 
\begin{equation}\label{infp}
 \prod_{\rm epochs} (2p_r +1)\;.
 \end{equation}
 
According to the analysis of \cite{BKL,BKL2,BKL3} the truly generic singularity features an infinite number of Kasner epochs. In this situation  the  product (\ref{infp}), featuring  an infinite set of numbers in the open interval $(0,1)$,  is bound to vanish for almost all singularities as the induced volume on the successive Kasner slices becomes diluted as the singularity is approached. Ultimately, a cutoff at Planck time from the singularity must be imposed, implying that the complexity computed by this ansatz would have a finite suppression factor determined by the number of epochs taking place until Planck time. An estimation of the the number of Kasner epochs or the physical scale parameter $H$ for realistic cosmologies depends on the dynamical details of the setup \footnote{For realistic cosmologies such as the one of our universe, it has been argued that the BKL analysis undergoes only an order 1 number of Kasner epochs \cite{Henneaux}.}, limiting thus our ability to quantify generically the order of the suppression factor \eqref{infp}.

\section*{Conclusions and outlook}
\addcontentsline{toc}{chapter}{Conclusions and outlook}
\noindent

In this part we have introduced quasilocal notions of AC complexity for terminal sets such as spacelike singularities in General Relativity. The basic idea is to build WdW patches restricted to the past causal domain of the singular set. Holographic data are associated to codimension-two surfaces on corresponding horizons, assigning a notion of entropy to a singularity by looking at the area of these codimension-two sets. 

Within our `terminal complexity' definition, we have studied solutions with non-trivial entropy dynamics, showing the strong impact on complexity dynamics of a varying Hilbert space already at the level of the leading bulk approximation. In particular, we study exact solutions with divergent entropy, given by Coleman-de Luccia type solutions, and also patches of the  Kasner spacetime with vanishing entropy at the singularity, representing the local description of generic singularities in GR. We find the remarkable result that in these two very different situations, the terminal AC approaches a unified linear form in terms of the entropy
\begin{equation}
\delta \CC^*  \approx a \, \delta S  + \dots \,,
\end{equation}
up to $u$-independent constants and subleading terms.  Although the detailed form of the coefficient $a$  depends on the particular solution, some general properties are to be noticed. In particular, we find the general behavior
\begin{equation}
a = a_1 + a_2\,\log \LL\;,
\end{equation}
in appropriate units\footnote{ These are given by the AdS radius of curvature for the expanding bubble examples, and by the inverse Hubble scale $H^{-1}$ for the Kasner example.}.  Both $a_1$ and $a_2$ are strictly positive constants for all dynamical scenarios, therefore relegating negative values of $a$ only to those choices of $\LL$ that correspond to UV scales ($\LL \ll 1$) in the case of the expanding scenarios or even transplanckian ($\LL \ll \ell_{\text{Planck}}$) for the local case. Remarkably, this class of solutions comprise the most radical example of sensitivity of complexity to $\LL$ as they are the only known ones for which the late-time dynamics is qualitatively affected by the size of this scale. If one is to believe that the monotonic growth must be a generic property of terminal complexity in the expanding cases, the result above forces us to consider $\LL$ as an IR scale of the same order or lower than the the lowest scale present in the CFT, i.e. the curvature radius of the boundary metric.

In brief, we observe that an `entropic locking' of AC complexity arises when the entropy has strong dynamics near a spacelike singularity. On general grounds, we can imagine that the complexity grows linearly within a fixed Hilbert space, but it may have more complicated dynamics when the effective dimensionality of the Hilbert space, of order ${\rm exp}(S)$, changes abruptly with time. This was the situation found in \cite{BarbonRabinoComplexity} in cases where the complexity was dominated by strong time-dependence of UV degrees of freedom. In the situations described in this paper, we are only concentrating in IR sectors, in the holographic sense, but again the effective Hilbert spaces have strongly time-dependent dimensionality and this phenomenon dominates the rate of change of complexity.

Next, our results shed some light to the old program of classifying GR singularities according to their inherent complexity,(cf. \cite{Penrose, misner, BKL, BKL2, BKL3 }), which acquires an interesting outlook when combined with holographic ideas, providing a suitable language for these hypotheses. 

In particular we find that the  YGH term evaluated at the singularity defines a `complexity density' which serves as a holographic version of the Weyl curvature criterion by Penrose. We show that this contribution can be isolated  from the terminal AC complexity by a coarse-graining procedure, and we explicitly check that this quantity sets apart `simple' singularities, such as the one at a FRW crunch, from `complex' ones, such as the generic black hole singularity.  

 Nonetheless, within the local description of generic singularities, as presented in the classic BKL analysis, our ansatz assigns  a vanishing  complexity density to the formal infinite sequences of chaotic Kasner `epochs'. Since these chaotic structures are generic in the light of the BKL analysis, we would conclude that the  local complexity density of generic spacelike singularities is zero. We find this phenomenon puzzling, and consider that it begs the question of the proper interpretation of the scale $H$ and the order of the supression factor for realistic cosmologies within an effective GR approach.

\addcontentsline{toc}{chapter}{References} 


\cleardoublepage
\bibliographystyle{utphys}
\bibliography{biblio}

\end{document}